%% file: main.tex
\documentclass[a4paper,11pt]{article}
\pdfoutput=1
\usepackage{jcappub}
\usepackage{aas_macros}

\usepackage{bm}
\usepackage{color}
\usepackage{xcolor}
\usepackage{xspace}
\usepackage{comment}

\usepackage{tabularx}
\def\beq{\begin{eqnarray}}
\def\eeq{\end{eqnarray}}

\usepackage{soul,xcolor}
\setstcolor{red}

\usepackage{physics}
\usepackage{ifthen}
\usepackage{url}

\usepackage{graphicx}% Include figure files
\usepackage{dcolumn}% Align table columns on decimal point
\usepackage{bm}% bold math
\usepackage{subfigure}
\usepackage[nointegrals]{wasysym}
\usepackage{verbatim}
\usepackage{color}
\usepackage{amsmath}
\usepackage[utf8]{inputenc}
\usepackage{orcidlink}

\usepackage{enumitem,amssymb}
\newlist{todolist}{itemize}{2}
\setlist[todolist]{label=$\square$}
\usepackage{pifont}

\newcommand{\fg}{f}
\newcommand{\fil}{h}

\usepackage[normalem]{ulem}
\usepackage{xcolor}

\title{Estimation and mitigation of foregrounds in projected kSZ velocity reconstruction}

\input{authors_cev.tex}

\emailAdd{ce425@cam.ac.uk}

\abstract{The kSZ effect has recently emerged as a powerful probe for precision cosmology through its ability to reconstruct the large-scale velocity field. In particular, the kSZ-reconstructed velocity--galaxy cross-correlation is sensitive to signatures of primordial non-Gaussianity through its imprint on the galaxy bias. The kSZ velocity reconstruction is performed using small-scale information from CMB temperature and galaxy overdensities. As the sensitivity of these measurements improves, systematic effects such as extragalactic foreground contamination present in CMB maps become increasingly important. We present a study of foreground biases to the kSZ-reconstructed velocity--galaxy cross-correlation. We derive the relevant foreground contributions from the thermal Sunyaev-Zel'dovich effect and the cosmic infrared background, modeling them using a halo model description of the dominant one- and two-halo terms. We compare our analytic predictions to measurements obtained using ACT DR6 temperature maps and DESI Legacy Imaging Survey galaxies, finding qualitative agreement. We introduce a parity-odd estimator constructed from antisymmetric combinations of tomographic velocity--galaxy correlations and show analytically that, under the Limber approximation, this estimator entirely cancels the foreground contamination while preserving the full cosmological signal without loss of signal-to-noise. Finally, we apply this parity-odd estimator to the data combination mentioned above and show that the fit to the velocity--galaxy correlation is dramatically improved compared to the analysis without mitigation; our estimator detects the signal at 11$\sigma$, with an amplitude consistent with recent studies.}
 
\begin{document}
\maketitle
\setcounter{tocdepth}{1}

\tableofcontents

\section{Introduction}

Cosmic microwave background (CMB) observations have entered an era in which secondary anisotropies, generated by the interaction of CMB photons with the intervening matter, have become central science targets of current and future surveys \citep{SimonsObservatory:2018koc,thesimonsobservatoryacollaboration2025simonsobservatorysciencegoals,sehgal2019cmbhdultradeephighresolutionmillimeterwave,2011PASP..123..568C,2011ApJS..194...41S}. Since these secondary signals predominantly appear on small angular scales, high-resolution CMB experiments are now positioned to probe in detail the imprint of large-scale structure (LSS) on the CMB. This opens a unique window into the study of the late-time universe through CMB observations, and provides a complementary tracer to traditional LSS surveys.

One promising secondary is the kinetic Sunyaev–Zel’dovich (kSZ) effect \citep{sunyaev,zeldovich,Sunyaev:1980nv}. The kSZ signal arises from the Doppler shift experienced by CMB photons that Thomson scatter off free electrons with non-zero peculiar velocities. Photons scattering off electrons moving away from (towards) us are redshifted (blueshifted), generating a characteristic small-scale signal that is modulated by the large-scale radial velocity field. This effect correlates the small-scale CMB temperature fluctuations $T_S$ with the small-scale electron density field $\delta_S^e$: $\left<T_S \delta_S^e\right>$.\footnote{Throughout this work we use a subscript $S$ to indicate a small-scale mode and a subscript $L$ to indicate a large-scale mode.} While this correlation vanishes on average, it becomes non-zero in small patches with fixed non-zero velocity. The kSZ amplitude depends on the distribution of the ionized gas as well as the large-scale velocity field. Over the past few years, kSZ measurements have leveraged these two dependencies to extract information on both the baryon distributions in halos and the velocity field as a cosmological probe (e.g.,~\cite{Hand2012,Planck2013_kSZ,Soergel2016,Hill2016,ACTPol:2015teu,Kusiak2021,Tanimura2021,Calafut2021,Hadzhiyska:2025mvt,Ried_ksz_25,krywonos2024constraintscosmologylambdacdmkinetic,Lague2024,Hotinli:2025tul,lai_ksz_25,McCarthy_kSZ2_25,Roper2025,Hadzhiyska_ksz_baryon_24,Hadzhiyska:2025egz,Gong:2025ffw}). Various techniques have been proposed to measure the kSZ effect in a ``tomographic'' fashion, whereby a galaxy survey recovers the three-dimensional information from a two-dimensional CMB measurement (e.g.,~\cite{2009arXiv0903.2845H,Shao:2010md,Deutsch_ksz_17,ksz_bispectrum}).

In this work, we are interested in the use of the kSZ effect to reconstruct the projected radial-velocity field \cite{zhang_v,2017JCAP...02..040T,Deutsch_ksz_17,ksz_bispectrum}. This field is a fundamental cosmological observable: it probes the growth of structure, is sensitive to the underlying matter distribution, and encodes information about the statistics of primordial fluctuations. The kSZ velocity signal (and, more generally, its cross-correlation with the large-scale galaxy density) is also sensitive to beyond-$\Lambda$CDM effects including non-Gaussianity parameters~\cite{Munchmeyer_ksz_forecast_18,Adshead:2024paa}, compensated isocurvature perturbations~\cite{Hotinli:2019wdp,Kumar:2022bly,krywonos2024constraintscosmologylambdacdmkinetic}, modified gravity~\cite{2019PhRvD.100h3522P,AnilKumar:2022flx,Patki:2025oyp},  dark energy~\cite{Adolff:2025fli}, general relativistic effects in galaxy surveys~\cite{2019JCAP...10..024C}, and other large-scale anomalies in the standard cosmological model~\cite{Cayuso:2019hen} (particularly in combination with polarized SZ tomography~\cite{Kamionkowski:1997na,Deutsch_ksz_17}). 

In particular, at the current level of precision, the reconstructed velocity field can be cross-correlated with biased tracers of the LSS to constrain primordial non-Gaussianity (PNG) of the local type, parametrized by $f_{\text{NL}}$ \cite{Munchmeyer_ksz_forecast_18,2019JCAP...10..024C,tishue2025kszopticaldepthdegeneracy}. The presence of local PNG induces a characteristic scale-dependent bias in LSS tracers that becomes most significant on the largest scales \cite{Dalal:2007cu}, making the velocity--galaxy cross-correlation $P_{vg}(k)$, or its two-dimensional projection $C_L^{v g}$, a powerful probe of this effect. Moreover, cross-correlation measurements with unbiased tracers, such as the velocity field, can additionally bypass the sample-variance limit through sample-variance cancellation \citep{Seljak_2009}. Recent analyses have demonstrated the feasibility of these measurements, and the field has quickly moved from reporting upper limits~\citep{Bloch:2024kpn} and first measurements of the reconstructed velocity field~\cite{McCarthy_ksz_2024,Lague2024} to high-significance measurements~\cite{Hotinli:2025tul,lai_ksz_25,McCarthy_kSZ2_25}, including cosmological constraints~\citep{krywonos2024constraintscosmologylambdacdmkinetic,Lague2024, Hotinli:2025tul,McCarthy_kSZ2_25}. 
These developments mark the transition of kSZ velocity reconstruction analyses from first measurements to a competitive probe for cosmology.

To reconstruct the velocity field, the kSZ-induced mode coupling between small-scale CMB temperature fluctuations and the electron field is used, with galaxies serving as a proxy for the electron distribution. In the presence of a long-wavelength velocity mode, the small-scale correlation $\left<T_S \delta_S^g \right>$ is modulated across the sky by the large-scale radial velocity $v_r$. Several approaches have been developed to extract the velocity information from this correlator \cite{zhang_v,2017JCAP...02..040T,Deutsch_ksz_17,ksz_bispectrum,2020PhRvD.102d3520M,Contreras:2022zdz,2023JCAP...02..051C,Kvasiuk_ksz_23}. Here, we study this reconstruction projected in different redshift bins, using the two-dimensional quadratic estimator introduced in Ref.~\cite{Deutsch_ksz_17}, implemented in the flat-sky limit (following Ref.~\cite{Kvasiuk_ksz_23}), which reconstructs the kSZ velocity by averaging over optimally weighted pairs of the small-scale temperature and galaxy overdensity, such that $\hat{v}_r \sim \left<T_S \delta_S^g \right>$. To compute these mode couplings analytically, we build upon the \texttt{CosmoBLENDER}\footnote{\url{https://github.com/abaleato/CosmoBLENDER/tree/ksz_velocity}} framework \cite{baleato_2025}, originally developed to model foreground contamination in CMB lensing quadratic estimators (e.g.,~\cite{2003PhRvD..67h3002O}).  

As the precision of these measurements improves, systematic effects that were previously subdominant will become increasingly important. In particular, extragalactic foregrounds present in CMB temperature maps can bias the signal, since they trace the same underlying matter density field as the galaxy field entering the velocity-reconstruction estimator. Among these foregrounds, the dominant sources of extragalactic contamination are expected to come from the thermal Sunyaev–Zel’dovich (tSZ) effect \cite{sunyaev,zeldovich}, generated from the inverse Compton scattering of CMB photons off hot, free electrons in galaxy clusters and groups, and the cosmic infrared background (CIB), originated from dusty, star-forming galaxies~\cite{Puget:1996fx,Planck:2013wqd}. The presence of both these fields in our temperature maps can bias the velocity reconstruction by generating additional mode couplings between the small-scale CMB temperature and the galaxy field. In fact, evidence for such contamination has already been seen in data: in recent measurements of $C_L^{\hat{v} g}$, Ref.~\citep{McCarthy_kSZ2_25} reported a residual systematic signal attributed to foreground contamination present in the reconstructed velocity field, and Ref.~\cite{Hotinli:2025tul} saw evidence for contamination in $P_{vg}(k)$ at 90\,GHz.

{In this work, we focus on the foreground contamination to the kSZ velocity--galaxy cross-correlation $C_L^{\hat{v} g}$, which contains a contribution of the form $\left<\delta_L^g \fg_S\delta_S^g\right>$ induced by foregrounds ($\fg$), and is generally non-zero due to non-Gaussianity induced by non-linear structure formation. However, foregrounds correlated with the small-scale galaxy field can also affect the large-scale galaxy--kSZ velocity cross-correlation $\left<\delta^g_L \hat{v}_r\right>$ through a mean-field contribution sourced by their Gaussian, isotropic correlation. For an isotropic survey (covering the full sky), this leads to a non-zero estimator monopole, but this can leak into higher multipoles in the presence of a non-trivial survey geometry. This effect adds \textit{variance}, as opposed to a systematic bias to $\left<\delta^g_L \hat{v}_r\right>$. This was investigated at the map level of $\hat{v}_r$ in Ref.~\cite{2023JCAP...02..051C} and found to be small and localized to regions close to the edge of the mask. The effect on observables like $\left<\delta^g_L \hat{v}_r\right>$ and the kSZ velocity auto-correlation $\left<\hat{v}_r \hat{v}_r\right>$ (which gains a systematic bias) can be controlled through careful calibration of the ``mean-field'' contribution, analogous to CMB lensing~\cite{Smith:2007rg}. In this work, we do not explore this mean-field contribution and instead focus on the bias term $\left<\delta_L^g \fg_S\delta_S^g\right>$ presented above.}

One of the main motivations for the analytic approach adopted in this work is our focus on the largest scales of the velocity--galaxy cross-correlation, which are most relevant for $f_{\text{NL}}$ constraints. Previous work on foreground contamination to similar estimators, such as CMB lensing, extensively studied the impact of foregrounds to the signal on simulations. However, extending this approach to kSZ velocity reconstruction is particularly challenging, as it would require simulations that can describe the small-scale galaxy--foreground correlations while spanning large enough volumes to capture accurately the large-scale modes of interest. Given these challenges, an analytic approach may provide a more straightforward framework to study these effects. {Throughout this work, we restrict our analysis to the large-scale regime of the relevant spectra, where most of the signal-to-noise is concentrated.}

The aim of this work is to present a dedicated analysis of the contributions from extragalactic foreground contamination to the projected kSZ velocity reconstruction.  We begin by identifying the relevant squeezed bispectra that contribute to this bias (see Sec.~\ref{sec:2}).  We also explore foreground contributions to the velocity in the response formalism, showing that they must be proportional to the matter density field (Sec.~\ref{sec:response}). Then, we model their signal using the halo-model framework (Sec.~\ref{sec:halo_model}), before showing our results in Sec. \ref{sec:results}, where we explore the dependence of these biases with frequency and redshift (Sec.~\ref{sec:freq_redshift}), benchmark our predictions against data (Sec.~\ref{sec:data_comparison}), and compare its contribution to the cosmological signal (Sec.~\ref{sec:neighbouring}). Finally, we propose a mitigation strategy (Sec.~\ref{sec:mitigation}) based on the symmetry properties of the estimator, and show explicitly that it removes the foreground contribution entirely while preserving the full signal-to-noise. We present its first implementation to data (Sec.~\ref{sec:mitigation_data}) and show that it allows us to achieve a good fit to a $C_L^{\hat vg}$ measurement.

\section{kSZ velocity reconstruction and foreground bispectra} \label{sec:2}

The kSZ contribution to the CMB temperature anisotropy $\Delta T^{\mathrm{kSZ}}(\hat{\boldsymbol{n}})$ can be written as the line-of-sight projection of the radial velocity $v_r$ and electron number density $n_e$ \citep{sunyaev,Sunyaev:1980nv}:
\begin{equation}
    \frac{\Delta T^{\text{kSZ}}}{T_{\mathrm{CMB}}}(\hat{\boldsymbol{n}}) = - \sigma_{\text{T}} \int d \chi \, a(\chi) \; v_r(\chi, \hat{\boldsymbol{n}}) n_e(\chi, \hat{\boldsymbol{n}}) \; ,
    \label{eq:kSZ_def}
\end{equation}
where  $\sigma_{\text{T}}$ is the Thomson cross-section, $a(\chi)$ the scale factor at comoving distance $\chi$, $\hat{\boldsymbol{n}}$~denotes the line-of-sight direction, $v_r \equiv \hat{\boldsymbol{n}}\cdot\boldsymbol{v}$ is the radial component of the peculiar velocity~$\boldsymbol{v}$, and $T_{\mathrm{CMB}}$ is the CMB temperature monopole (measured by COBE-FIRAS to be~$T=2.726\, \mathrm{K}$~\cite{Fixsen:1996nj,Mather:1998gm,Fixsen:2009ug}). The kSZ anisotropy can be written as the integrated differential optical depth $\frac{d\tau}{d\chi}$ weighted by the (negative) radial velocity, where the  optical depth $ \tau$ is given by
\begin{equation}
    \tau(\boldsymbol{\hat{n}}) = \int d\chi \,a(\chi) \,\sigma_T \,n_e(\chi,\boldsymbol{\hat{n}}).
\end{equation}

In this section, we review the projected-velocity reconstruction formalism in the flat-sky limit and derive the foreground-induced contributions to the cross-correlation of the reconstructed velocity and the galaxy overdensity. We consider a setup where the galaxies in a survey can be binned tomographically into redshift bins labelled by Greek indices $\alpha$.

\subsection{kSZ velocity reconstruction estimator} \label{sec:QE_estimator}

In the flat-sky approximation, the quadratic estimator for the projected radial velocity field in redshift bin $\alpha$ can be written as \citep{Kvasiuk_ksz_23}
\begin{equation}
\hat{v}^{\alpha}_r(\boldsymbol{L}) = \lambda(L) \int d^{2}\boldsymbol{\ell}\;
   \fil^{\alpha}(\boldsymbol{\ell},\boldsymbol{L-\ell})\,
   T^{\text{obs}}(\boldsymbol{\ell})\,
   g^{\alpha}(\boldsymbol{L-\ell})\, \; ,
   \label{eq:QE}
\end{equation}
where $T^{\text{obs}}(\boldsymbol{\ell})$ is the Fourier transform of the observed CMB temperature, including the lensed primary CMB and kSZ signals, as well as foregrounds and instrumental noise. We denote its angular power spectrum by $C_\ell^{TT}$. The field $g^{\alpha}(\boldsymbol{\ell})$ corresponds to the projected galaxy overdensity in redshift bin $\alpha$; the precise definition for this projection is given in Eq.~\eqref{eq:g_projection}. Its auto-spectrum, including shot noise, is denoted by $C^{g^{\alpha}g^{\alpha}}_{\ell}$. 

The mode-coupling filters $\fil^{\alpha}$ are chosen to minimize the variance of the estimator (first derived on the full-sky in Ref.~\cite{Deutsch_ksz_17}, and written in the flat-sky approximation in Ref.~\cite{Kvasiuk_ksz_23}) and are given by
\begin{equation}
    \fil^{\alpha}(\boldsymbol{\ell},\boldsymbol{L-\ell}) 
    = 
    \frac{C^{\tau g^{\alpha}}_{|\boldsymbol{L-\ell}|}}
    {C^{TT}_{\ell} \, C^{g^{\alpha}g^{\alpha}}_{|\boldsymbol{L-\ell}|}} \; ,
    \label{eq:filters}
\end{equation}
where $C_\ell^{\tau g^{\alpha}}$ denotes the optical depth--galaxy overdensity cross-spectrum for the galaxies in bin $\alpha$. In practice, this quantity is either computed analytically assuming a framework such as the halo model, or measured from simulations.

The normalization of the estimator $\lambda(L)$ is given by:
\begin{equation}
    \lambda(L) = -\left[ \int d^2 \boldsymbol{\ell} \frac{ (C^{\tau g^{\alpha}}_{|\boldsymbol{L-\ell}|})^2}{C^{TT}_{\ell} C^{g^{\alpha}g^{\alpha}}_{|\boldsymbol{L-\ell}|}} \; \right]^{-1} .
\end{equation}
This normalization is chosen to enforce an unbiased estimator, i.e., $\langle\hat{v}^\alpha_r(\boldsymbol{L})\rangle = v^\alpha_r$ when the fiducial $C_\ell^{\tau g^\alpha}$ matches the truth.
In practice, any mis-estimation of this power spectrum is often accounted for by marginalizing over a nuisance parameter referred to as the ``velocity bias'' and labelled by $b_v$.\footnote{We emphasize that the ``velocity bias'' $b_v$ is unrelated to any large-scale bias of any objects with respect to dark matter, and could alternatively be referred to as an ``optical depth bias''.} On large scales this parameter is scale-independent\footnote{The scale independence of $b_v$ has been demonstrated in $N$-body simulations in~\cite{Giri:2020pkk} and Gaussian simulations in~\cite{2023JCAP...02..051C}.}, so mis-modeling $C_\ell^{\tau g^{\alpha}}$ primarily affects the overall amplitude and does not bias analyses targeting strongly scale-dependent signals such as the scale-dependent bias induced by $f_{\mathrm{NL}}$. The velocity bias is defined in the flat-sky formalism as 
\begin{align}
b_v^\alpha &= \lim_{L\rightarrow 0}
\left( -\lambda(L) \int d^2\boldsymbol{\ell}\, \fil^\alpha(\boldsymbol{\ell},\boldsymbol{L}-\boldsymbol{\ell}) \left[C^{\tau g}_{|\boldsymbol{L}-\boldsymbol{\ell}|}\right]^\text{true}\right) \nonumber \\
&=
\left[
\int d\ell\,\ell\,
\frac{
\left(C_\ell^{\tau g^\alpha}\right)^2
}{
C_\ell^{TT}\,C_\ell^{g^\alpha g^\alpha}
}\right]^{-1}
\int d\ell\,\ell\,
\frac{
C_\ell^{\tau g^\alpha}\,
\left[C_\ell^{\tau g^\alpha}\right]^{\mathrm{true}}
}{
C_\ell^{TT}\,C_\ell^{g^\alpha g^\alpha}
}\, .
\end{align}
For incorrect $C_\ell^{\tau g^\alpha}$, the velocity bias leads to the mis-normalization
$\langle\hat v^\alpha_r(\boldsymbol{L})\rangle = b_v^\alpha v^\alpha_r(\boldsymbol{L})$ on large scales.

\subsection{The foreground bispectrum}\label{sec:bispectrum}

In this work, we are interested in studying the cross-correlation of the reconstructed velocity and galaxy overdensity: $\left< \hat{v}_r^\alpha(\boldsymbol{L}) g^{\beta}(\boldsymbol{L}')\right>$. This cross-correlation can then be computed in our current formalism as:
\begin{equation}
    \left< \hat{v}_r^\alpha(\boldsymbol{L}) g^{\beta}_L(\boldsymbol{L}^\prime)\right> =  \lambda(L) \int d^{2}\boldsymbol{\ell}\;
   \fil^{\alpha}(\boldsymbol{\ell},\boldsymbol{L-\ell})\,
   \left< T^{\text{obs}}(\boldsymbol{\ell})\,
   g^{\alpha}_S(\boldsymbol{L-\ell})\,
   g^{\beta}_L(\boldsymbol{L}^\prime) \right>  \; ,
   \label{eq:first_bis}
\end{equation}
where we have again used the subscripts $S$ and $L$ to distinguish the small-scale galaxy field entering the reconstruction ($g^{\alpha}_S$) and the large-scale galaxies used in the cross-correlation ($g^{\beta}_L$).

We decompose the observed temperature field as
\begin{equation}
    T^{\text{obs}} = T^{\mathrm{kSZ}} +T^{\mathrm{CMB
    }} + \fg + n ,
    \label{eq:T_decomp}
\end{equation}
where: $T^{\mathrm{kSZ}}$ is the kSZ temperature signal;  $T^{\mathrm{CMB}}$ is the lensed primary CMB; $\fg$ refers to contributions from extragalactic foregrounds; and $n$ is the instrumental noise, assumed to be Gaussian and uncorrelated with the galaxy field. For the purposes of kSZ measurements, $T^{\mathrm{CMB}}$  also acts as uncorrelated Gaussian noise, and so we will drop it going forward and incorporate it into $n$.

We can then write Eq.~\eqref{eq:first_bis} as
\begin{equation}
\begin{split}
\left< \hat{v}_r^\alpha(\boldsymbol{L}) g^{\beta}_L(\boldsymbol{L}^\prime)\right> =  \lambda(L) \int d^{2}\boldsymbol{\ell}\;
\fil^{\alpha}(\boldsymbol{\ell},\boldsymbol{L-\ell})\,\left[
\left< T^{\mathrm{kSZ}}(\boldsymbol{\ell})\,
g^{\alpha}_S(\boldsymbol{L-\ell})\,
g^{\beta}_L(\boldsymbol{L}^\prime) \right>  \right. \\
\left. +\left< \fg(\boldsymbol{\ell})\,
g^{\alpha}_S(\boldsymbol{L-\ell})\,
g^{\beta}_L(\boldsymbol{L}^\prime) \right>+\left< n(\boldsymbol{\ell})\,
g^{\alpha}_S(\boldsymbol{L-\ell})\,
g^{\beta}_L(\boldsymbol{L}^\prime) \right> \right] \; .
\end{split}
\end{equation}
The first term corresponds to the cosmological signal of interest $C_L^{v^\alpha g^\beta}$, while the last term vanishes. In a Gaussian Universe the second term would also vanish. However, extragalactic foregrounds trace the underlying large-scale structure, and the non-linear evolution of structure formation generates a non-zero bispectrum between the foreground field and the galaxy density. As a result, the correlator $\left< \fg(\boldsymbol{\ell})\,g^{\alpha}_S(\boldsymbol{L-\ell})\,g^{\beta}_L(\boldsymbol{L}^\prime) \right>$ is non-zero and leads to a bias in the reconstructed velocity--galaxy cross-correlation. 

Therefore, by defining
\begin{equation}
    \left< \hat{v}_r^\alpha(\boldsymbol{L}) g^{\beta}(\boldsymbol{L}^\prime)\right> \equiv   (2\pi)^2 \delta^{(2)}(\boldsymbol{L}+\boldsymbol{L}^{\prime})\left(C_L^{\hat{v}^\alpha g^\beta} + \Delta C_L^{\hat{v}^\alpha g^\beta} \right),
    \label{eq:angular_cls}
\end{equation}
we identify the foreground-induced contribution to the velocity–galaxy cross-correlation---which is the quantity of interest in this work---with 
\begin{equation}
     \Delta C_L^{\hat{v}^\alpha g^\beta} \equiv \lambda(L) \int d^{2}\boldsymbol{\ell}\;
   \fil^{\alpha}(\boldsymbol{\ell},\boldsymbol{L-\ell})\,
   B^{\text{2D}}_{\fg,g^\alpha_S,g^\beta_L}(\boldsymbol{\ell},\boldsymbol{L-\ell},\boldsymbol{-L}) \; ,
   \label{eq:def_Delta_Cl}
\end{equation}
where the projected angular bispectrum is defined as
\begin{equation}
    \left<
    \fg(\boldsymbol{\ell}_1)\,
    g_S^\alpha(\boldsymbol{\ell}_2)\,
    g_L^\beta(\boldsymbol{\ell}_3)
    \right>
    =
    (2\pi)^2
    \delta^{(2)}(\boldsymbol{\ell}_1+\boldsymbol{\ell}_2+\boldsymbol{\ell}_3)\,
    B^{\text{2D}}_{\fg,g^\alpha_S,g^\beta_L}(\boldsymbol{\ell}_1,\boldsymbol{\ell}_2,\boldsymbol{\ell}_3) \; .
\end{equation}
Then, if the foregrounds are correlated with the galaxies used in the measurement, and given the non-Gaussianity induced by non-linear evolution---which generates a non-zero matter bispectrum correlation---a bias is induced in the reconstructed velocity–galaxy cross-correlation. Correlations between galaxies and the foreground components considered in this work (tSZ and CIB) are expected and have indeed been detected and analyzed at high significance in data (see, e.g.,~\cite{2021PhRvD.103f3513S, Serra:2014pva,Liu:2025zqo}, or most recently \cite{LaPosta:2026cek} for tSZ cross-correlations and~\cite{Jego:2022eqo,Chiang:2025ujk} for CIB cross correlations), as well as gravity-induced non-Gaussianity measured through, e.g., the galaxy bispectrum \cite{Scoccimarro:2000sn}, or CIB/tSZ non-Gaussian statistics \cite{Planck:2013wqd,Crawford:2013uka}.

\subsection{Bispectrum projection} \label{sec:projection}

To relate the object of interest $B^{\text{2D}}_{\fg,g^\alpha_S,g^\beta_L}(\boldsymbol{\ell}_1,\boldsymbol{\ell}_2,\boldsymbol{\ell}_3)$ to its corresponding three-dimensional counterpart we first need to define the projected observables entering our calculation. The projected galaxy overdensity in redshift bin $\alpha$ is:
\begin{equation}
    g^\alpha(\hat{\boldsymbol{n}})
    =
    \int dz\,
    \frac{d\chi}{dz}\,
    W_g^\alpha(z)\,
    \delta_g(\hat{\boldsymbol{n}},z) \; ,
    \label{eq:g_projection}
\end{equation}
where $\delta_g(\hat{\boldsymbol{n}},z)$ is the three-dimensional galaxy overdensity field and
\begin{equation}
    W_g^\alpha(z)
    =
    H(z)\,
    \frac{dN^\alpha}{dz}
    \label{eq:g_kernel}
\end{equation}
is the redshift kernel of the galaxy sample in bin $\alpha$, with $H(z)$ being the Hubble parameter, $c$ the speed of light, and $dN^{\alpha}/dz$ the normalized redshift distribution of the galaxy sample in redshift bin $\alpha$. 

For the foreground leg of our correlator, the projected Compton-$y$ field describing the tSZ is
\begin{equation}
    y(\hat{\boldsymbol{n}})
    =
    \int dz\,
    \frac{d\chi}{dz}\,
    a(z)\,
    y_{\text{3D}}(\hat{\boldsymbol{n}},z) \; ,
\end{equation}
where $y_{\text{3D}}(\hat{\boldsymbol{n}},z)$ is the 3D Compton-$y$ field which will be described in detail in Sec.~\ref{sec:tsz_modeling} (Eq.~\ref{eq:y_to_Pe}). The projected CIB intensity $ I_\nu$ at a frequency $\nu$ is
\begin{equation}
    I_\nu(\hat{\boldsymbol{n}}) = \int dz \frac{d\chi}{dz} a(z) j_\nu(\hat{\boldsymbol{n}},z) \; ,
    \label{eq:CIB_I_projection}
\end{equation}
where $j_\nu$ is the comoving infrared emissivity and its parametrization in the halo-model framework will be presented in Sec.~\ref{sec:cib_modeling}. Since, given our notation, the remaining redshift dependence of both our foreground tracers corresponds to the scale factor, we can define a common foreground redshift kernel: $W_\fg(\chi) = a(\chi)$.

Combining these definitions, we relate the projected angular bispectrum to its corresponding three-dimensional bispectrum using the Limber approximation~\cite{1953ApJ...117..134L,Lemos:2017arq}:
\begin{equation}
    B^{\text{2D}}_{\fg,g^\alpha_S,g^\beta_L}
    (\boldsymbol{\ell}_1,\boldsymbol{\ell}_2,\boldsymbol{\ell}_3)
    =
    \int dz\,
    \frac{d\chi}{dz}\,
    \frac{W_\fg(z)\, W_{g_S}^\alpha(z)\, W_{g_L}^\beta(z)}
         {\chi^4}
    \,
    B_{\fg,g_S,g_L}
    \!\left(
        \frac{\boldsymbol{\ell}_1}{\chi},
        \frac{\boldsymbol{\ell}_2}{\chi},
        \frac{\boldsymbol{\ell}_3}{\chi};
        z
    \right) ,
    \label{eq:projection}
\end{equation}
where $B_{\fg,g_S,g_L}\!\left(\boldsymbol{k_1},\boldsymbol{k_1},\boldsymbol{k_1};z \right)$ is the 3D foreground--galaxy--galaxy bispectrum that we will characterize using the halo model in Sec. \ref{sec:halo_model}. These projections assume a spatially-flat geometry but can be easily generalized to a curved background.

\section{Response formalism intuition} \label{sec:response}

Before presenting the halo-model calculations of the relevant bispectra, it is useful to develop some intuition for the foreground-induced biases using the response formalism \citep{Barreira:2017sqa,Barreira:2017fjz}. First, we note that foreground contamination generates an extra contribution to the reconstructed velocity. Directly using the $T^{\text{obs}}$ decomposition presented in Eq.~\eqref{eq:T_decomp}, we see that foreground fields $f$ generate an extra term to the velocity estimator~(Eq.~\ref{eq:QE}) of the form 
\begin{equation}
 \hat{v}^{\alpha}_{r,\fg}(\boldsymbol{L})
= \lambda(L)
\int d^{2}\boldsymbol{\ell}\;
\fil^{\alpha}(\boldsymbol{\ell},\boldsymbol{L}-\boldsymbol{\ell})\, \fg(\boldsymbol{\ell}) \;
g^{\alpha}(\boldsymbol{L}-\boldsymbol{\ell})\; .
\label{eq:vel_response}
\end{equation}

Since we are interested in the large-scale behavior of this quantity where $L \ll \ell$, it is natural to interpret the object of interest as depending on the behavior of the short-scale modes in the presence of a long-wavelength perturbation. In this limit, the average over short-wavelength modes of Eq.~\eqref{eq:vel_response} is controlled by the response of the small-scale two-point function $ \left< \fg(\boldsymbol{\ell}) g^{\alpha}(\boldsymbol{L}-\boldsymbol{\ell})\right>_S$ (where, here, the subscript $S$ indicates that the average is only taken over short modes), to long-wavelength gravitational observables. In the absence of primordial non-Gaussianity, the leading observables are second derivatives of the gravitational potential, namely the matter overdensity $\delta_m$ and the tidal field $K_{ij}$ (which in turn can be expressed as a function of $\delta_m$ in Fourier space) \citep{2015JCAP...11..007S,2015JCAP...07..030M}. Therefore, to make this intuition more quantitative, we need to understand how the short-scale correlator $\langle \fg(\boldsymbol{\ell}) g^{\alpha}(\boldsymbol{L}-\boldsymbol{\ell}) \rangle_S$ behaves in the presence of a long-wavelength matter density mode. We will first consider its three-dimensional counterpart, $\langle \fg_{\text{3D}}(\boldsymbol{k}) \delta_g(\boldsymbol{K}-\boldsymbol{k}) \rangle_S$, before projecting to the observed geometry; in this section, we will assume that the two- and three-dimensional variables are related in the flat-sky, plane-parallel approximation via the projections $f(\hat{\mathbf{n}})=\int d \chi W_f(\chi) f_{\text{3D}}(\hat{\mathbf{n}} \chi, \chi)$ and $g^\alpha(\hat{\mathbf{n}})=\int d \chi W^\alpha_g(\chi) \delta_g(\hat{\mathbf{n}} \chi, \chi)$ for foreground and galaxy fields, respectively (and $W$ indicates the relevant projection kernel). 

The three-dimensional correlator is now not statistically homogeneous, and instead homogeneity is broken locally by the presence of a long-wavelength mode, so that short modes become correlated in the presence of a long mode $\delta_m(\boldsymbol{K})$. The off-diagonal correlator (i.e., the correlator with $\boldsymbol{K}\neq0$) in the limit $|\boldsymbol{K}|\ll |\boldsymbol{k}|$ can be evaluated by applying the response formalism presented in Ref.~\cite{Barreira:2017sqa}, noting that the spectra and response functions will be appropriate to the three-dimensional foreground--galaxy cross-spectrum $P^{fg}(k)$ instead of the matter power spectrum. For now, we will neglect any additional stochastic noise term, uncorrelated with the long-wavelength mode, that may enter into $ \hat{v}^{\alpha}_{r,\fg}$, but we will restore it later. This implies that at leading order in the separation of scales and in the long-wavelength mode (assuming throughout this section that $\boldsymbol{K}\neq0$) \cite{Barreira:2017sqa}
\begin{align}
    \left< \fg_{\mathrm{3D}}(\boldsymbol{k})\,\delta_g(\boldsymbol{K}-\boldsymbol{k})\right>_S &\approx \mathcal{R}(k,\mu_{\boldsymbol{k},\boldsymbol{K}}) P^{fg}(k) \delta_m(\boldsymbol{K}) \nonumber \\
    &=\left[
R_1(k)
+ R_K(k)
\left(\mu_{\boldsymbol{k},\boldsymbol{K}}^2-\frac{1}{3}\right)
\right]P^{\fg\,g}(k)~
\delta_m(\boldsymbol{K}) \; ,
\label{refMC}
\end{align}
where $\mu_{\boldsymbol{k},\boldsymbol{K}} \equiv \hat{\boldsymbol{k}}\!\cdot\!\hat{\boldsymbol{K}}$ and $\mathcal{R}$ is the total response function. Here the second equality just writes the total response $\mathcal{R}$ in terms of $R_1(k)$ and $R_K(k)$, which denote the isotropic and tidal response functions, respectively. The intuition underlying this expression is that the position-dependent foreground--galaxy cross-power spectrum (which is probed by the correlator in the limit $K\ll k$) can depend on only two long-wavelength quantities, the density field and the tidal field, and these dependencies are encoded by functions $R_1(k)$ and $R_K(k)$.
While the isotropic and tidal response functions can be explicitly computed for the matter power spectrum \cite{Barreira:2017sqa}, here they describe the response of the position dependent foreground--galaxy cross-spectrum $P^{fg}(k)$ to long-wavelength modes, which has not yet been well-characterized; despite this, the response formalism can also be used to describe the modulation of $P^{fg}(k)$ and other small-scale LSS spectra \cite{2021JCAP...05..069V} by long-wavelength modes. Finally, note that we have identified the argument of the matter density field modulating the response with $\boldsymbol{K}$, since, in the end, our analysis aims only to explain correctly foreground effects on cross-spectra of $\hat v_r$ with external LSS tracers $C$, which are rigorously described via the bispectrum $\left< \fg_{\mathrm{3D}}(\boldsymbol{k})\,\delta_g(\boldsymbol{K}-\boldsymbol{k})C_{\mathrm{3D}}(\boldsymbol{p})\right>$; as described in Ref.~\cite{Barreira:2017sqa}, the momentum-conserving delta function ensures that $\boldsymbol{K}$ is the argument of $\delta_m$. 

We can now project this expression for the off-diagonal three-dimensional correlator to evaluate the foreground contribution to the velocity estimator. Using the definitions of the projected fields above we obtain
\begin{equation}
\langle f(\boldsymbol{\ell})\, \delta_g(\boldsymbol{L}-\boldsymbol{\ell})\rangle_S
= \int d\chi_1\, d\chi_2\;
  \frac{W_{f}(\chi_1)}{\chi_1^2}\;
  \frac{W^\alpha_g(\chi_2)}{\chi_2^2}\;
  \left\langle
    \widetilde{f}_{\mathrm{3D}\perp}\!\left(\frac{\boldsymbol{\ell}}{\chi_1},\, \chi_1\right)\;
    \tilde{\delta}_{g\perp}\!\left(\frac{\boldsymbol{L}-\boldsymbol{\ell}}{\chi_2},\, \chi_2\right)
  \right\rangle_S\; ,
\end{equation}
where the tilde and perpendicular symbols used in $\widetilde{f}_{\mathrm{3D}\perp}$, $\tilde{\delta}_{g\perp}$ indicate a partial Fourier transform operating only in the $\boldsymbol{k}_\perp$ plane, perpendicular to the line-of-sight direction. 
Fourier transforming also in the parallel direction and inserting our expression for the three-dimensional correlator from Eq.~\eqref{refMC} we obtain
\begin{align}
  \left<\hat{v}^{\alpha}_{r,\fg}(\boldsymbol{L})\right>_S
  &\approx \lambda_L \int d^2\boldsymbol{\ell}\; h^\alpha_{\boldsymbol{\ell},\boldsymbol{L}}
     \int d\chi_1\, d\chi_2\;
     \frac{W_{f}(\chi_1)}{\chi_1^2}\,
     \frac{W^\alpha_g(\chi_2)}{\chi_2^2}
     \notag\\
  &\qquad\times\;
  \int \frac{dk_{\parallel_1}}{2\pi}\, \frac{dk_{\parallel_2}}{2\pi}\;
     \mathcal{R}(k_1,\, \mu_{\boldsymbol{k_1},\boldsymbol{k_1}+\boldsymbol{k_2}})\, P^{fg}(k_1)
     \delta_m\bigl(
       \boldsymbol{k}_{\perp 1} + \boldsymbol{k}_{\perp 2},\;
       k_{\parallel 1} + k_{\parallel 2}
     \bigr)\;
     e^{i(k_{\parallel 1}\chi_1 + k_{\parallel 2}\chi_2)} \; ,
  \label{eq:full_expansion}
\end{align}
where
$  \boldsymbol{k}_1
  = \left(\boldsymbol{k}_{\perp 1} =\frac{\boldsymbol{\ell}}{\chi_1},\, k_{\parallel 1}\right)$,
  $\boldsymbol{k}_2
  = \left(\boldsymbol{k}_{\perp 2}=\frac{\boldsymbol{L} - \boldsymbol{\ell}}{\chi_2},\, k_{\parallel 2}\right)$, and we have used the abbreviated notation $h^\alpha_{\boldsymbol{\ell},\boldsymbol{L}} \equiv\fil^{\alpha}(\boldsymbol{\ell},\boldsymbol{L-\ell})$ and $\lambda_L \equiv \lambda(L)$.
This result implies that (under our squeezed-limit approximation) foregrounds produce an additional term in the velocity reconstruction that is dependent on the long-wavelength matter field $\delta_m\bigl(
       \boldsymbol{k}_{\perp 1} + \boldsymbol{k}_{\perp 2},\;
       k_{\parallel 1} + k_{\parallel 2}
     \bigr)$.

Going forward, we will derive an expression that gives the correct result when cross-correlated with another long-wavelength projected field within the Limber approximation: $\left<\hat{v}^{\alpha}_{r,\fg}(\boldsymbol{L})C(\boldsymbol{L}')\right>$. For this purpose, we may assume all $k_\parallel=0$, whenever useful, and drop the integral over $\chi_2$ (along with one of the $k_\parallel$ integrals), setting $\chi_2=\chi_1$. We then obtain the following result:
\begin{align}
\left<\hat{v}^{\alpha}_{r,\fg}(\boldsymbol{L})\right>_S &\approx  \lambda_L \int d^2\boldsymbol{\ell}\; h^\alpha_{\boldsymbol{\ell},\boldsymbol{L}} \int d\chi\;
\frac{W_{f}(\chi)}{\chi^2}\,
\frac{W^\alpha_g(\chi)}{\chi^2}\;
\mathcal{R}\!\left(\frac{\ell}{\chi},\, \mu_{\boldsymbol{\ell}, \boldsymbol{L}}\right)
P^{fg}\!\left(\frac{\ell}{\chi}\right)
\int \frac{dk_\parallel}{2\pi}\;
{\delta}_m\!\left(\frac{\boldsymbol{L}}{\chi},\, k_\parallel\right)
e^{ik_\parallel \chi} \nonumber \\
&\approx \lambda_L \int d^2\boldsymbol{\ell}\; h^\alpha_{\boldsymbol{\ell},\boldsymbol{L}}
\int d\chi\;
\frac{W_{f}(\chi)}{\chi^2}\,
\frac{W^\alpha_g(\chi)}{\chi^2}\;
\mathcal{R}\!\left(\frac{\ell}{\chi},\, \mu_{\boldsymbol{\ell}, \boldsymbol{L}}\right)
P^{fg}\!\left(\frac{\ell}{\chi}\right)\;
\tilde{\delta}_{m}{}_{\perp}\!\left(\frac{\boldsymbol{L}}{\chi},\, \chi\right) \; ,
\end{align}
where $\mu_{\boldsymbol{\ell},\boldsymbol{L}} \equiv \hat{\boldsymbol{\ell}}\cdot\hat{\boldsymbol{L}}$. Hence,
\begin{equation}
\left<\hat{v}^{\alpha}_{r,\fg}(\boldsymbol{L})\right>_S \approx \int d\chi\;
  \frac{W_{f}(\chi)}{\chi^2}\,
  \frac{W^\alpha_g(\chi)}{\chi^2}\;
  b^\alpha_{fg}(\chi,\boldsymbol{L})\;
  \tilde{\delta}_{m}{}_{\perp}\!\left(\frac{\boldsymbol{L}}{\chi},\, \chi\right) \; ,
\end{equation}
where we have defined $b^\alpha_{fg}(\chi,\boldsymbol{L})\equiv \lambda_L \int d^2\boldsymbol{\ell}\; h^\alpha_{\boldsymbol{\ell},\boldsymbol{L}}\; \mathcal{R}\!\left(\frac{\ell}{\chi},\mu_{\boldsymbol{l},\boldsymbol{L}}\right) P^{fg}\!\left(\frac{\ell}{\chi}\right)$.

Under the separation of scales $L \ll \ell$, we can simplify our expression for the bias $b^\alpha_{fg}$. In this limit, the power spectra entering the estimator filters can be evaluated at $|\boldsymbol{L}-\boldsymbol{\ell}|\sim \ell$. Furthermore, the only angular dependence inside the $\boldsymbol{\ell}$-integral arises through $\mu_{\boldsymbol{\ell},\boldsymbol{L}}^2$, whose azimuthal average satisfies $\langle \mu_{\boldsymbol{\ell},\boldsymbol{L}}^2 \rangle_\varphi = 1/2$. Performing the angular integral therefore reduces the expression to a purely radial integral, as can be seen in the following expression (where we have inserted the definitions of $h^\alpha_{\boldsymbol{\ell},\boldsymbol{L}}$ and $\mathcal{R})$: 
\begin{align}
b^\alpha_{fg}(\chi,\boldsymbol{L})&\equiv \lambda_L \int d^2\boldsymbol{\ell}\; \frac{C_{|\boldsymbol{L}-\boldsymbol{\ell}|}^{\tau g^\alpha}}
{C_\ell^{TT} C_{|\boldsymbol{L}-\boldsymbol{\ell}|}^{g^\alpha g^\alpha}} \; \left[
R_1\left(\frac{\ell}{\chi}\right)
+ R_K\left(\frac{\ell}{\chi}\right)
\left(\mu_{\boldsymbol{\ell},\boldsymbol{L}}^2-\frac{1}{3}\right)
\right] P^{fg}\!\left(\frac{\ell}{\chi}\right)\nonumber\\
&\approx
2\pi\lambda_L
\int d\ell\;\ell\;
\frac{C_\ell^{\tau g^\alpha}}
{C_\ell^{TT} C_\ell^{g^\alpha g^\alpha}} 
\left[
R_1\left(\frac{\ell}{\chi}\right)
+ \frac{1}{6} R_K\left(\frac{\ell}{\chi}\right)
\right]
P^{fg}\!\left(\frac{\ell}{\chi}\right).
\label{scaleindFGbias_complicated}
\end{align}
Crucially, noting also that the normalization $\lambda_L$ is approximately constant in the low-$L$ regime we are interested in, this expression is independent of $L$ at leading order in $L/\ell$. To a good approximation, on the large scales of interest, we may hence assume that $b^\alpha_{fg}(\chi,\boldsymbol{L})=b^\alpha_{fg}(\chi)$. 

It follows that the foreground contamination to the velocity estimator takes the form of a new projected tracer of the long-wavelength density field: 
\begin{equation}
\left<\hat{v}^{\alpha}_{r,\fg} (\hat{\mathbf{n}})\right>_S \approx \int d\chi\;
  \frac{W_{f}(\chi)  W^\alpha_g(\chi)}{\chi^2}
\,
  b^\alpha_{fg}(\chi)\; \delta_m(\chi \hat{\mathbf{n}}, \chi) \; .
  \label{finalApproxEq}
\end{equation}
This tracer has a redshift kernel given by $[W_{f}(\chi)  W^\alpha_g(\chi)]/\chi^2$ and a scale-independent effective bias that encodes the response of the small-scale foreground--galaxy power spectrum to long-wavelength modes. This bias is expected to be slowly varying within a redshift bin so it can be approximated as constant within each bin, $b^\alpha_{fg}(\chi)\approx b^{\alpha}_{fg}$.

Thus far, we have neglected any stochastic noise contribution which arises from small-scale fluctuations that are uncorrelated with the long wavelength mode; to more accurately model $\hat{v}^{\alpha}_{r,\fg} (\hat{\mathbf{n}})$, these can be straightforwardly included by modifying the right-hand side of Eq.~$\eqref{finalApproxEq}$ via $b^\alpha_{fg}(\chi)\; \delta_m(\chi \hat{\mathbf{n}}, \chi) \rightarrow  b^\alpha_{fg}(\chi)\; \delta_m(\chi \hat{\mathbf{n}}, \chi) +\, \epsilon$, where $\epsilon$ is a stochastic term uncorrelated with the long-wavelength modes. In cross-correlation with an external LSS tracer (which can be expected to have a correlated stochastic contribution), this stochastic term can be expected to produce only power that is approximately constant on large scales, since all correlation functions involving $\epsilon$ only have support on very small scales.

In the presence of primordial non-Gaussianity of the local type, another additional contribution must be taken into account in the argument above. This arises from the primordial gravitational potential $\Phi$, which in the presence of local-type primordial non-Gaussianity can be written as
\begin{equation}
    \Phi(\boldsymbol{x}) = \phi_G(\boldsymbol{x}) + f_{\text{NL}}\left(\phi_G^2(\boldsymbol{x}) - \left< \phi_G^2 \right> \right),
\end{equation}
where $\phi_G$ denotes its Gaussian component. For $f_{\text{NL}}\neq 0$, this induces a coupling between long- and short-wavelength modes, such that the amplitude of the small-scale power spectrum is modulated by the presence of a long-wavelength \emph{potential} mode.  Within the response formalism, this implies that the small-scale foreground--galaxy power spectrum responds not only to the long-wavelength matter density perturbation and tidal field, but is also proportional to the long-wavelength potential $\Phi_L$ times $f_{\text{NL}}$. This contribution, following the identical argument as for the density field, then produces a bias, which in this case is scale dependent and hence degenerate with the cosmological signal of interest. However, as we show in Sec.~\ref{sec:mitigation}, the mitigation strategy presented here removes all foreground contributions (within the Limber approximation) based on symmetry arguments, and is expected to perform similarly for $\delta_m$ and $\Phi$, because they have the same parity symmetry.

\section{Halo-model description of extragalactic foreground biases} \label{sec:halo_model}

Quantifying the impact of extragalactic foregrounds on kSZ velocity-reconstruction measurements requires the computation of the foreground–galaxy–galaxy bispectrum. Since both the CIB and the tSZ effect are produced nearly entirely by matter contained within halos, the halo model provides a natural framework for modeling these bispectra.

The halo model (see~\cite{2002PhR...372....1C} for a review) has been  used extensively to describe correlations between galaxies, gas, and other large-scale structure tracers across a wide range of scales. It is known to have limitations, including inaccuracies in the transition between the 1-halo and 2-halo regimes. Additionally, the signals of interest---the small-scale galaxy distribution, tSZ signal, and CIB signal---are highly uncertain. However, the halo model remains a powerful tool and one of the few analytic frameworks to describe multi-tracer correlations involving gas and galaxies across scales. As it is not our goal to explain the small-scale galaxy--tSZ or galaxy--CIB cross-correlations as predictions of a cosmological model or of a more general halo-model parametrization, we instead use halo-model predictions as an approximate guide to how large we expect these foreground biases to be, and their scale dependencies. For this purpose, the halo model provides a well-motivated and sufficiently accurate tool. {We note that there are more direct methods of measuring and constraining the properties of our tracers, such as directly measuring and interpreting the cross-correlation between the temperature and the galaxies on small scales (e.g., with measurements like those of~\cite{Liu:2025zqo}).}

Here, we restrict our analysis to the 1-halo and 2-halo contributions to the relevant bispectra. Although a 3-halo term is also present, previous implementations of similar types of bispectra---the ones arising from biases to the lensing reconstruction cross-correlations in Ref.~\cite{baleato_2025}---have shown that the 1-halo and 2-halo terms are sufficient to achieve agreement with simulations, and therefore we neglect the 3-halo contribution in this work. The 1-halo term corresponds to configurations in which all three legs of the bispectrum are sourced within the same halo; the 2-halo term is one in which two legs reside in one halo and the third in another; and the 3-halo term is one with 3 different halos. As we are interested in a bispectrum which has support on two small-scale legs and one large-scale leg, it is not surprising that the 1- and especially the 2-halo terms are dominant. In fact, the leading contribution will come from the 2-halo term in which the small-scale galaxy and foreground modes are located in one halo, while the large-scale galaxy mode is in another.

Therefore, we describe the halo-model bispectrum at redshift $z$ as a sum of the 1-halo and 2-halo terms. The general 1-halo contribution for three tracers $Q_1$, $Q_2$, and $Q_3$ takes the form
\begin{equation}
    B_{Q_1 Q_2 Q_3}^{\text{1h}}
    (\boldsymbol{k}_1,\boldsymbol{k}_2,\boldsymbol{k}_3;z)
    =
    \int dM \, n(M,z) 
    Q_1(k_1,M,z)
    Q_2(k_2,M,z)
    Q_3(k_3,M,z) \; ,
    \label{eq:gen_1h}
\end{equation}
where $n(M,z)$ is the halo mass function, $M$ indicates halo mass, and we have replaced $\boldsymbol{k}_i$ with $k_i\equiv\left|\boldsymbol{k}_i\right|$ by assuming spherical symmetry. Throughout this work we adopt the Sheth--Tormen implementation of $n(M,z)$~\cite{2002MNRAS.329...61S}. The spherically-symmetric profiles $Q_i(k,M,z)$ are the Fourier transforms of the real-space 3D profiles that describe the different tracers entering the bispectrum. These will be described in detail in the following subsections. It is known that, at large enough scales, calculations of the 1-halo term can unphysically dominate over the 2-halo term. Different approaches have been developed to correct for this effect (e.g.,~\cite{2002PhR...372....1C,PhysRevD.101.103522}); here we introduce a damping factor in the style proposed by Ref.~\cite{2015MNRAS.454.1958M}, and also used in Ref.~\cite{baleato_2025}, when calculating the 1-halo contribution. 

The 2-halo contribution can be written as
\begin{equation}
\begin{split}
    B_{Q_1 Q_2 Q_3}^{\text{2h}}
    (\boldsymbol{k}_1,\boldsymbol{k}_2,\boldsymbol{k}_3;z)
    =
    P_{\rm lin}(k_1,z)
    \int dM \, n(M,z) b(M,z)
    Q_1(k_1,M,z) \\
    \times
    \int dM' \, n(M',z) b(M',z)
    Q_2(k_2,M',z)
    Q_3(k_3,M',z)
    + \text{perms}\, ,
    \label{eq:gen_2h}
\end{split}
\end{equation}
where $P_{\rm lin}(k,z)$ is the linear matter power spectrum (which we compute with \texttt{CAMB}~\cite{Lewis:1999bs}) and $b(M,z)$ is the linear halo bias (for which we use the Sheth--Tormen prescription~\cite{Sheth-Torman}), such that the halo--halo power spectrum is given by $P_{\rm hh}(k|M_1,M_2) = b(M_1,z)\, b(M_2,z)\, P_{\rm lin}(k,z)$, which is a good approximation for the scales in which the 2-halo term dominates. We show the explicit expressions of these bispectra used in this work in Appendix \ref{app:bispectra}.

Throughout this analysis, all halo profiles are assumed to be spherically symmetric, so that $Q_i$ depend only on $k_i \equiv |\boldsymbol{k}_i|$. This symmetry significantly simplifies the numerical evaluation of the quadratic estimator integral presented in Eq.~\eqref{eq:QE}. As described in detail in Ref.~\cite{baleato_2025}, the spherical symmetry of the 3D profiles translates into an azimuthal symmetry in the 2D quadratic estimator integral, which enables the use of Gaussian quadratures, accelerating these mode-coupling integrals significantly relative to brute-force, two-dimensional integrations. Given that this mode-coupling integral must be evaluated at each mass and redshift step from Eqs.~\eqref{eq:projection}, \eqref{eq:gen_1h}, and~\eqref{eq:gen_2h}, this computational efficiency is crucial. We refer the reader to Ref.~\cite{baleato_2025} for details of the fast evaluation of Gaussian quadratures.

\subsection{Galaxy HOD} \label{sec:galaxy_HOD}

The halo occupation distribution (HOD) framework  \cite{HOD} provides a prescription to model the probability $P(N \mid M)$ that a halo of mass $M$ hosts $N$ galaxies. In the standard HOD framework, the galaxy population is decomposed into contributions from central and satellite galaxies, with expectation values $\langle N_{\mathrm{cen}} \rangle$ and $\langle N_{\mathrm{sat}} \rangle$ denoting the mean number of central and satellite galaxies, respectively. The number of central galaxies in a halo $N_{\text{cen}}$ is either 0 or 1, while the presence of satellite galaxies is dependent on the existence of a central galaxy, and is usually modeled to scale as a function of the halo mass. Assuming that the number of satellites follows a Poisson distribution given the presence of a central galaxy ($N_{\text{cen}} = 1$) \cite{2005ApJ...633..791Z}, the total mean number of galaxies in a halo is then given by $\langle N_{\mathrm{gal}} \rangle = \langle N_{\mathrm{cen}} \rangle + \langle N_{\mathrm{sat}} \rangle$.

In this work, we model $\langle N_{\mathrm{cen}} \rangle$ and $\langle N_{\mathrm{sat}} \rangle$ using the abundance-matching technique described in Appendix B.3 of \cite{ksz_bispectrum} and implemented in \texttt{hmvec}\footnote{\url{https://github.com/simonsobs/hmvec}}. This procedure defines the expectation values of the numbers of central and satellite galaxies from the stellar mass threshold of the galaxy sample of interest. To match the properties of our survey, we begin with the fiducial angular number density as a function of redshift, convert it into a comoving number density, and infer the corresponding minimum stellar mass threshold. This threshold is then used to specify the halo occupation functions $\langle N_{\mathrm{cen}} \rangle$ and $\langle N_{\mathrm{sat}} \rangle$. In this work, we use the redshift distribution of the full Dark Energy Spectroscopic Instrument (DESI) luminous red galaxies (LRG) Extended sample presented in Ref.~\cite{2023JCAP...11..097Z} to build the HOD.

The prescription to incorporate the HOD of the galaxies into any $N$-point correlation function within the halo model has been extensively studied, and different prescriptions exist in the literature (e.g.,~\cite{2014MNRAS.439..123L}). In this work, we are interested in modeling the foreground--galaxy--galaxy bispectrum, restricting our analysis to the 1-halo and 2-halo contributions. Therefore, we only need to model contributions with either one or two galaxies per halo. We follow the prescription of Ref.~\cite{2014MNRAS.439..123L}, and write down the factorial moments as derived in Appendix B of Ref.~\cite{baleato_2025}. Then, the 3D profiles entering Eqs.~\eqref{eq:gen_1h} and~\eqref{eq:gen_2h} corresponding to the galaxy legs $Q_{g_i}(k_i\mid M,z)$ take the form
\begin{equation}
    Q_{g_1}(k_1,M,z) = \frac{1}{\bar{n}_g} \left<N_{\text{gal}}(M) \right> u_g(k_1 \mid M) \; ,
\end{equation}
when there is only one galaxy per halo, and
\begin{equation}
    Q_{g_1}(k_1,M,z) Q_{g_2}(k_2,M,z) = \frac{1}{\bar{n}^2_g} \left<N_{\text{sat}} \right> \left[ 2 \left<N_{\text{cen}} \right> + \left<N_{\text{cen}} \right> \right] u_g(k_1 \mid M) u_g(k_2 \mid M)  \; ,
\end{equation}
when two galaxies are present in the same halo. Here, $u_g(k \mid M)$ describes how the galaxies are spatially distributed within the halo, and we model it with a Navarro–Frenk–White (NFW) density profile \cite{1996ApJ...462..563N}, and the mean number density of galaxies $\bar{n}_g$ is defined by
\begin{equation}
    \bar{n}_{\text{g}} = \int dM n(M) \left(\langle N_{\mathrm{cen}} \rangle+\langle N_{\mathrm{sat}} \rangle \right) \; .
\end{equation}

\subsection{tSZ profiles}\label{sec:tsz_modeling}

The tSZ effect \cite{sunyaev,zeldovich} arises when CMB photons inverse-Compton scatter off hot ionized electrons---mostly located in the intracluster medium of massive halos---along the line of sight. This scattering transfers energy from the electrons to the photons, shifting photons to higher frequencies and producing a characteristic spectral distortion of the CMB. The resulting temperature fluctuation can be written as
\begin{equation}
    \frac{\Delta T^{\text{tSZ}} (\boldsymbol{\theta},M,z)}{T_{\text{CMB}}}
    =
    g_\nu \, y(\boldsymbol{\theta},M,z) \; ,
\end{equation}
where $g_\nu = x/\tanh(x/2)-4$ encodes the frequency dependence of the tSZ signal, with $x\equiv h\nu/(k_{\rm B}T_{\text{CMB}})$. Here, $k_{\rm B}$ is the Boltzmann constant, $T_{\text{CMB}}$ the mean CMB temperature, and $h$ the Planck constant. The Compton-$y$ parameter is defined as the line-of-sight integral of the electron pressure:
\begin{equation}
    y(\boldsymbol{\theta},M,z)
    \equiv
    \frac{\sigma_{\text{T}}}{m_e c^2}
    \int_{\text{l.o.s.}}
    P_e\!\left(
    \sqrt{l^2 + d_A^2 |\boldsymbol{\theta}|^2},
    M,z
    \right)
    dl \; ,
    \label{eq:y_to_Pe}
\end{equation}
where $P_e$ is the electron pressure profile, $d_A$ is the angular diameter distance, $m_e$ is the electron mass, and $\sigma_{\text{T}}$ is the Thomson cross-section.

To describe the pressure profile as a function of halo mass and redshift we adopt a generalized NFW \cite{1996ApJ...462..563N} parameterization:
\begin{equation}
    P_e(x)
    =
    P_0
    \left(\frac{x}{x_c}\right)^{\gamma}
    \left[1+\left(\frac{x}{x_c}\right)^{\alpha}\right]^{-\beta},
    \quad
    x \equiv r/R_{200} \; .
    \label{eq:e_press_profile}
\end{equation}
Here, $R_{200}$ is the radius enclosing a mean density 200 times the critical density of the Universe $\rho_{\mathrm{crit}}$ at redshift $z$, such that $M_{200} = (4\pi/3)\,200\,\rho_{\mathrm{crit}}(z)\,R_{200}^3$, and $x_c$ is a dimensionless parameter that sets the characteristic scale radius of the profile. Taking the Fourier transform of the quantity defined in Eq.~\eqref{eq:e_press_profile} directly relates to the 3D $Q_i$ profiles that appear in Eqs~\eqref{eq:gen_1h} and~\eqref{eq:gen_2h} for the case of tSZ: $Q_y(k,M,z)$.

In this work we consider two different calibrations of this profile. The first corresponds to the commonly used ``Battaglia'' profile from Ref.~\cite{2012ApJ...758...75B}, which was calibrated using hydrodynamical simulations. In this model the parameters that control the amplitude and shape of the profile vary with halo mass and redshift following:
\begin{equation}
    A
    =
    A_0
    \left(\frac{M_{200}}{10^{14}M_{\odot}}\right)^{\alpha_m}
    (1+z)^{\alpha_z},
    \label{eq:Battaglia_params}
\end{equation}
where $A=\{P_0,\beta,x_c\}$ and the parameters $A_0$, $\alpha_m$, and $\alpha_z$ are fitted separately for each of them. 

More recent analyses based on observational data have suggested deviations from the Battaglia calibration, potentially showing stronger baryonic feedback than what was present in the simulations used to derive the original fit~\cite{DES:2021sgf}. One example is the profile presented in \cite{2021PhRvD.103f3514A}, which we refer to here as the ``Amodeo'' profile. In that work the pressure profile is fitted directly to stacked tSZ and kSZ measurements from the Atacama Cosmology Telescope (ACT) at the positions of constant stellar mass (CMASS) Baryon Oscillation Spectroscopic Survey (BOSS) galaxies \cite{2014ApJS..211...17A}. These galaxies span a redshift range of $0.4 < z < 0.7$ and are predominantly central galaxies residing in halos of characteristic mass $M_{\rm halo} \sim 3\times10^{13} M_\odot$\footnote{Halo mass in this case is quoted as $M_{200}$, at a radius $R_{200}$.}, similar to the galaxies we will be making our comparisons to in Sec.~\ref{sec:data_comparison}, although we do not necessarily expect the two samples to match in detail. In this model the parameters controlling the inner structure of the profile, $\gamma$ and $x_c$, are fixed to the Battaglia values, while $P_0$, $\alpha$, and $\beta$ are fitted directly to the data. Here, the parametrization of $P_0$ and $\beta$ does not introduce an explicit mass or redshift dependence. In Sec.~\ref{sec:data_comparison}, we compare the impact that different pressure profiles have on our bias estimates, finding our predictions to be moderately dependent on the choice of pressure profile parametrization (see Figs~\ref{fig:data_150} and \ref{fig:data_90}).

The tSZ signal is strongly dominated by the most massive clusters (e.g.,~\cite{2002ARA&A..40..643C}). As a result, the high-mass end of the different terms entering the halo model (such as the HOD for the galaxies or the pressure profile itself) play an important role in predicting the amplitude of the tSZ-induced biases. This is due to the steep mass scaling of the tSZ signal, approximately proportional to $M^{5/3}$, which up-weights contributions at high masses. In practice, observational analyses can mitigate the impact of massive clusters by directly masking them\footnote{Alternative mitigation strategies, such as profile hardening \cite{2013MNRAS.431..609N,2014JCAP...03..024O,2020PhRvD.102f3517S}, have become a common standard in other types of reconstruction analyses such as CMB lensing. We do not study the impact of these techniques in this work.} in the CMB maps before running the velocity-reconstruction pipeline. In our analytic calculations, we mimic this procedure by imposing a maximum halo mass cutoff in the mass integrals appearing in Eqs.~\eqref{eq:gen_1h} and~\eqref{eq:gen_2h}. This allows us to obtain estimates of the tSZ-induced biases that are less sensitive to uncertainties in the high-mass tail of our halo-model parametrizations.

\subsection{CIB model}\label{sec:cib_modeling}

The CIB is produced by radiation emitted by dust heated by ultraviolet starlight in star-forming galaxies. A fraction of this radiation is absorbed by dust in the interstellar medium and then re-emitted at infrared wavelengths. Therefore, the cumulative infrared emission from galaxies across the redshift range where star formation occurs can be detected by CMB experiments. The observed CIB intensity $I_\nu$ at frequency $\nu$ was defined as the line-of-sight projection of the comoving infrared emissivity $j_\nu$ in Eq.~\eqref{eq:CIB_I_projection}.

Within the halo-model framework, the CIB can be described by associating the infrared emission of galaxies with their host dark matter halos. Since the CIB originates from intrinsically discrete sources (although most are unresolved), this modeling can be implemented in a way analogous to halo occupation models such as the one presented in Sec.~\ref{sec:galaxy_HOD}. In this approach the comoving emissivity is written as a sum of contributions from central and satellite galaxies, 
\begin{equation}
    j_\nu(z) =
    \int dM\, n(M,z)
    \left[
    f_\nu^{\text{cen}}(M,z)
    +
    f_\nu^{\text{sat}}(M,z)
    \right],
\end{equation}
where $n(M,z)$ is the halo mass function. The contributions from central and satellite galaxies are given by
\begin{align}
    f_\nu^{\text{cen}}(M,z)
    &=
    \frac{1}{4\pi}
    N_{\text{cen}}(M,z)
    L_{(1+z)\nu}(M,z) \; , \\
    f_\nu^{\text{sat}}(M,z)
    &=
    \frac{1}{4\pi}
    \int dm\,
    n_{\text{sub}}(m,z|M)
    L_{(1+z)\nu}(m,z) \; ,
\end{align}
where $N_{\mathrm{cen}}(M,z)$ is the number of central galaxies in the halo, $n_{\text{sub}}(m,z|M)$ denotes the subhalo mass function (e.g.,~\cite{2004ApJ...609...35K}) and $L_{(1+z)\nu}(M,z)$ is the infrared luminosity spectral density of a galaxy hosted in a halo of mass $M$.

The mass, redshift, and frequency dependence of the infrared luminosity have been studied in detail in previous works. Here, we adopt the parametrization introduced in Ref.~\cite{Shang:2011mh}, which was fitted in Ref.~\cite{Planck:2013wqd} to the set of auto- and cross-power spectra measured by \textit{Planck} at 217, 353, 545, and 857\,GHz, together with \textit{IRAS} measurements at 3000\,GHz. A concise summary of this model, together with the values of the best-fit parameters, can also be found in Appendix A of~\cite{baleato_2025} (as well as in Ref.~\cite{2021PhRvD.103j3515M}). To model how the CIB component enters the bispectra presented in Eqs.~\eqref{eq:gen_1h} and~\eqref{eq:gen_2h}, we follow the prescription in Ref.~\cite{baleato_2025}:
\begin{equation}
    Q_I(k,M,z) = u(k,M,z) \left[\frac{f^{\text{cen}}_\nu(M,z)}{u(k,M,z)} + f_\nu^{\text{sat}}(M,z)\right] \; .
    \label{eq:CIB_profile}
\end{equation}

\section{Results} \label{sec:results}

In this section we present the extragalactic foreground bias contributions to the velocity--galaxy cross-correlation. To do so, we consider a configuration designed to resemble current observational datasets. In particular, we use ACT data release 6 (DR6) \cite{AtacamaCosmologyTelescope:2025vnj,AtacamaCosmologyTelescope:2025blo,AtacamaCosmologyTelescope:2025nti} to describe the CMB contribution together with the galaxies from the DESI Legacy Imaging Survey (DESILS)  \cite{2023JCAP...11..097Z,2023AJ....165...58Z,2019AJ....157..168D}, which are used both as the small-scale galaxy field entering the velocity reconstruction and as the large-scale galaxy tracer that is cross-correlated with the velocity. This setup is chosen to match the measurement performed in Ref.~\cite{McCarthy_kSZ2_25}, to which we will compare our predictions in Sec.~\ref{sec:data_comparison}. Although the main results in Ref.~\cite{McCarthy_kSZ2_25} were obtained by analysis of a component-separated map that preserved the CMB blackbody component, we avoid the difficulties associated with tracing the foreground contamination through the component separation process by comparing to analogous measurements performed using the single-frequency maps (as were used for the measurement of Ref.~\cite{lai_ksz_25}). We perform all of our calculations assuming a $\Lambda$CDM cosmology as measured by \textit{Planck} using the \texttt{CamSpec} likelihood (see table A.1 of~\cite{Planck:2018vyg}), and quote our halo masses as virial mass $M_{\text{vir}}$ \cite{1998ApJ...495...80B}. 

To implement this configuration, we choose the filters entering Eq.~\eqref{eq:filters} to be the same spectra as those used in the measurement of Ref.~\cite{McCarthy_kSZ2_25}.\footnote{The analysis of Ref.~\cite{McCarthy_kSZ2_25} is done separately on the North and South patches of the sky, and there are slightly different filters in each patch due to the varying depth properties of the CMB survey and  varying number densities of the galaxy surveys. For purposes of our foreground bias contamination, we choose the filters from the North patch, but we expect the calculation to be extremely similar.} These include the $C_\ell^{TT}$ measured from ACT DR6 temperature maps~\cite{AtacamaCosmologyTelescope:2025vnj} at both 90 and 150\,GHz, the measured galaxy overdensity power spectra in each redshift bin $C_\ell^{g^\alpha g^\alpha}$ (which includes shot noise), and the electron--galaxy correlation $C_\ell^{\tau g}$ obtained from $N$-body simulations (the \texttt{AbacusSummit} light cone simulations \cite{Maksimova_2021,Hadzhiyska_2021}) in which galaxies were populated using an LRG-like HOD and the $\tau$ field was calibrated to the TNG-300 hydrodynamical simulations~\cite{nelson2021illustristngsimulationspublicdata} using the transfer function method of Ref.~\cite{2025arXiv250411794L}. The halo-model description of both the small- and large-scale galaxy fields is constructed from the observed galaxy number density using the abundance-matching technique implemented in \texttt{hmvec} and described in Appendix~B.3 of Ref.~\cite{ksz_bispectrum}. The tomographic redshift bins considered throughout correspond to the (fiducial) 16-bin case from Ref.~\cite{McCarthy_kSZ2_25}, spanning approximately $0.4 < z < 1.0$.

We begin by focusing on the case where the reconstructed velocity field and the galaxy overdensity entering the cross-correlation are in the same redshift bin. In this configuration the cosmological signal is expected to be negligible, since velocities pointing into (out of) an overdensity (underdensity) are equally likely to be pointing towards or away from us, and the velocity-galaxy correlation cancels out on all but the very largest scales.
As a result, this setup provides a particularly clean way to isolate the foreground contamination contribution. We first (Sec. \ref{sec:freq_redshift}) present our foreground calculations in this configuration and study their dependence on observing frequency and tomographic redshift. Later, in Sec.~\ref{sec:data_comparison}, we compare our analytic predictions with measurements from data in these same redshift bins, allowing us to benchmark our modeling against observations directly. Finally, in Sec.~\ref{sec:neighbouring}, we repeat the calculation for reconstructed velocities and galaxies located in neighboring redshift bins (where the cosmological velocity--galaxy cross-correlation signal peaks) and compare them to the predicted velocity--galaxy signal. Note that, as our calculation is in the Limber approximation, we require some overlap between redshift bins for the foreground signal to be non-zero. This is certainly the case for nearby bins in our analysis, given the significant overlap due to the photo-$z$ errors of the underlying catalog.

\subsection{Frequency and redshift dependence} \label{sec:freq_redshift}

The aim of this section is to obtain a first characterization of the different foreground contributions to the velocity--galaxy cross-correlation, namely the squeezed bispectra defined in Eq.~\eqref{eq:def_Delta_Cl} for the case of the tSZ and CIB foregrounds. We will focus on understanding the frequency and redshift dependence of these contributions before comparing our predictions to data.

First, to study the frequency dependence of the foreground contributions we compute the analytic predictions for a single redshift bin centered at $\bar{z} = 0.42$, corresponding to the first bin of the 16-bin case in Ref. \cite{McCarthy_kSZ2_25}. Although labeled ``90\,GHz'', the ACT DR6 maps~\cite{2025JCAP...11..061N} have passbands of finite width. Formally, we should integrate the SEDs of the components over these passbands; as we are not interested in making a precision prediction, we instead compute effective frequencies for each component (following the prescription in Appendix D of Ref.~\cite{2020JCAP...12..047A}) assuming that the CIB follows a modified black body spectrum with spectral index $\beta=1.6$ and dust temperature of $20 \,\mathrm{K}$. In the 150 GHz map, the effective frequencies are close enough to 150 GHz that we perform our calculations at 150 GHz; for the 90 GHz map we instead compute the CIB at 97 GHz and the tSZ frequency response at 95 GHz.

We adopt a virial mass cut of $M_{\text{vir}} = 3 \times 10^{14}\,M_\odot$ in order to mimic the cluster-masking procedure usually performed in data analyses, as explained in Sec.~\ref{sec:tsz_modeling}. In particular, we choose the quoted mass threshold in order to match roughly the ACT DR5 cluster catalogue~\cite{2021ApJS..253....3H} that was masked in the temperature maps used in Ref.~\cite{McCarthy_kSZ2_25}, noting that the exact conversion between signal-to-noise in the $y$-map and virial mass is not trivial. We reconstruct the velocity field using modes up to $\ell_{\text{max}} = 8000$. Throughout this section we model the electron pressure using the Battaglia profile.

\begin{figure}
    \centering
    \includegraphics[width=\linewidth]{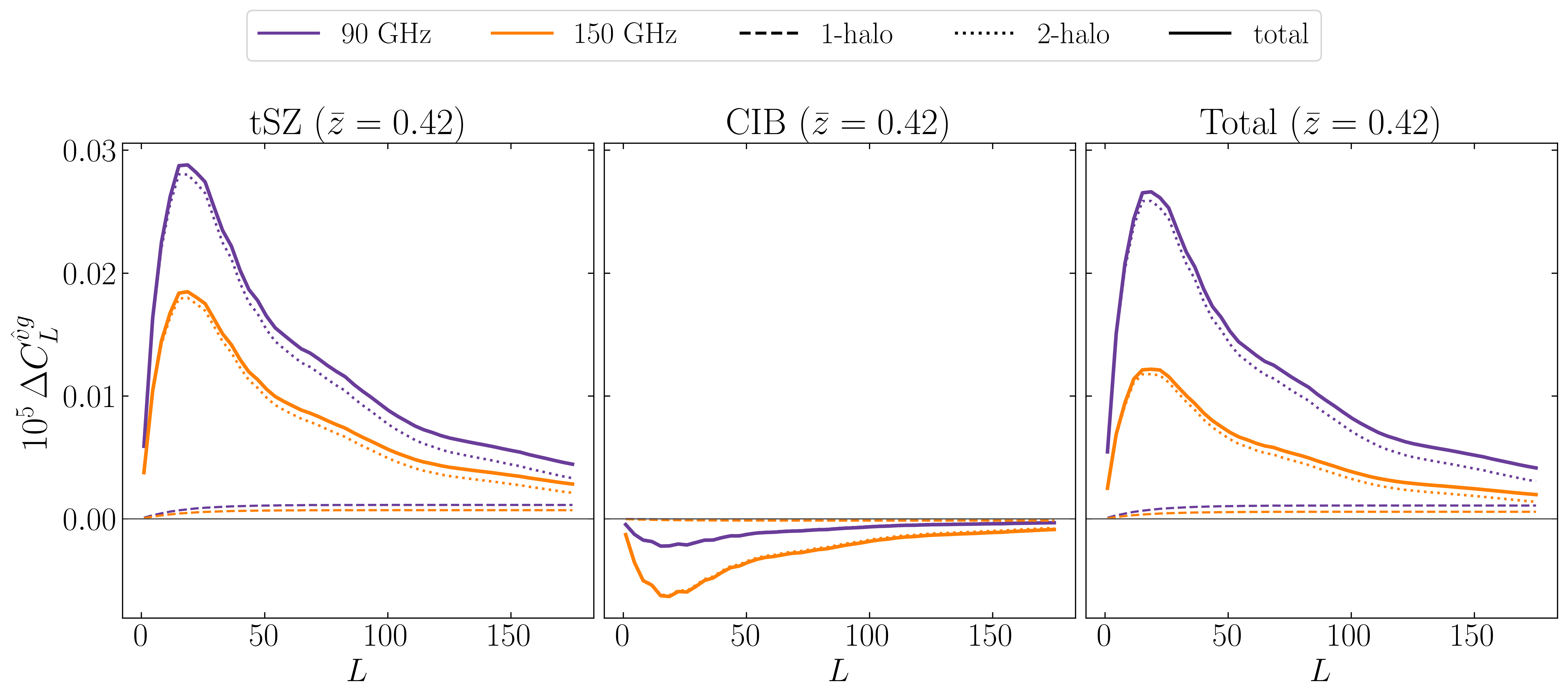}
    \caption{Analytic estimates of the foreground bias to the cross-correlation of kSZ-reconstructed velocity and galaxies for a redshift bin centered at $\bar{z}=0.42$, corresponding to the first bin of the 16-bin case in Ref. \cite{McCarthy_kSZ2_25}. We show the contributions from the 1-halo (dashed) and 2-halo (dotted) terms for tSZ (left), CIB (center), and their sum (right). The velocity field has been reconstructed using modes up to $\ell_{\text{max}} = 8000$ at frequencies of 150\,GHz (orange) and 90\,GHz (purple). The mass cut used in these computations is $M_{\text{vir}} = 3\times 10^{14}M_\odot$. We remind the reader that the ``2-halo'' contribution quantifies correlations between one large-scale galaxy leg in one halo, and two small-scale foreground and galaxy legs in another halo; this is clearly dominant, as expected.}
    \label{fig:main_sing_freq}
\end{figure}

In Fig.~\ref{fig:main_sing_freq} we show the resulting bias contributions to the velocity--galaxy cross-correlation for the tSZ, CIB, and their sum, separating the 1-halo and 2-halo terms. We note that, unlike the case of lensing reconstruction biases, there is no mixed tSZ--CIB contribution here, since the CMB temperature enters only one leg of the quadratic velocity reconstruction. In practice, multi-frequency cleaning techniques such as internal linear combinations (ILC)~\cite{bennett_92} can be used to mitigate the foreground contributions studied here. In fact, our current framework within \texttt{CosmoBLENDER} allows to compute the foreground-bias predictions in an ILC framework using the \texttt{BasicILC}\footnote{\url{https://github.com/EmmanuelSchaan/BasicILC}} code. However, in order to isolate clearly the different foreground contributions, we will only present the analytical calculations for the single-frequency cases.

The first feature we notice in Fig.~\ref{fig:main_sing_freq} is that the tSZ and CIB contributions have opposite signs, leading to a partial cancellation in the total foreground bias. This behavior can be directly understood from the frequency dependence of the two signals. Since there is only one temperature leg entering the bispectrum, the overall sign of the contamination is determined by the frequency dependence of each foreground component, together with the minus sign appearing in the definition of the kSZ anisotropy in Eq.~\eqref{eq:kSZ_def}. The relative amplitudes at $90$ and $150\,\mathrm{GHz}$ also follow the expected trends: the tSZ contribution increases toward $90\,\mathrm{GHz}$, where the tSZ spectral distortion is stronger, while the CIB contribution decreases following its spectral energy distribution.

Another important feature is that, as expected, on the large angular scales relevant for $f_{\mathrm{NL}}$ measurements ($L \lesssim 150$), the foreground contamination is dominated by the 2-halo term. In particular, the major contribution to this squeezed bispectrum can be traced to the term in which the large-scale galaxy leg is located in one halo, while the foreground and small-scale galaxy legs are located in the other. We  do not explicitly compare our foreground predictions with the signal in this same-redshift-bin configuration since the cosmological signal is expected to be extremely small. We will compare our predictions to the signal in Sec. \ref{sec:neighbouring}, where we explore the impact of foregrounds to the neighboring-redshift-bins case, where the cosmological signal peaks. 

At the redshift $\bar{z}=0.42$ adopted in Fig.~\ref{fig:main_sing_freq}, the foreground contamination is dominated by the tSZ contribution. This will remain true for the tomographic bins considered in the following sections and reflects the redshift overlap between the galaxy tracer and the tSZ and CIB. The tSZ signal receives most of its contribution from massive halos at lower redshifts---where there is a bigger overlap with the galaxy sample---whereas the CIB peaks at higher redshifts ($z\sim 2$) during the epoch of peak star formation.

\begin{figure}
    \centering
    \includegraphics[width=\linewidth]{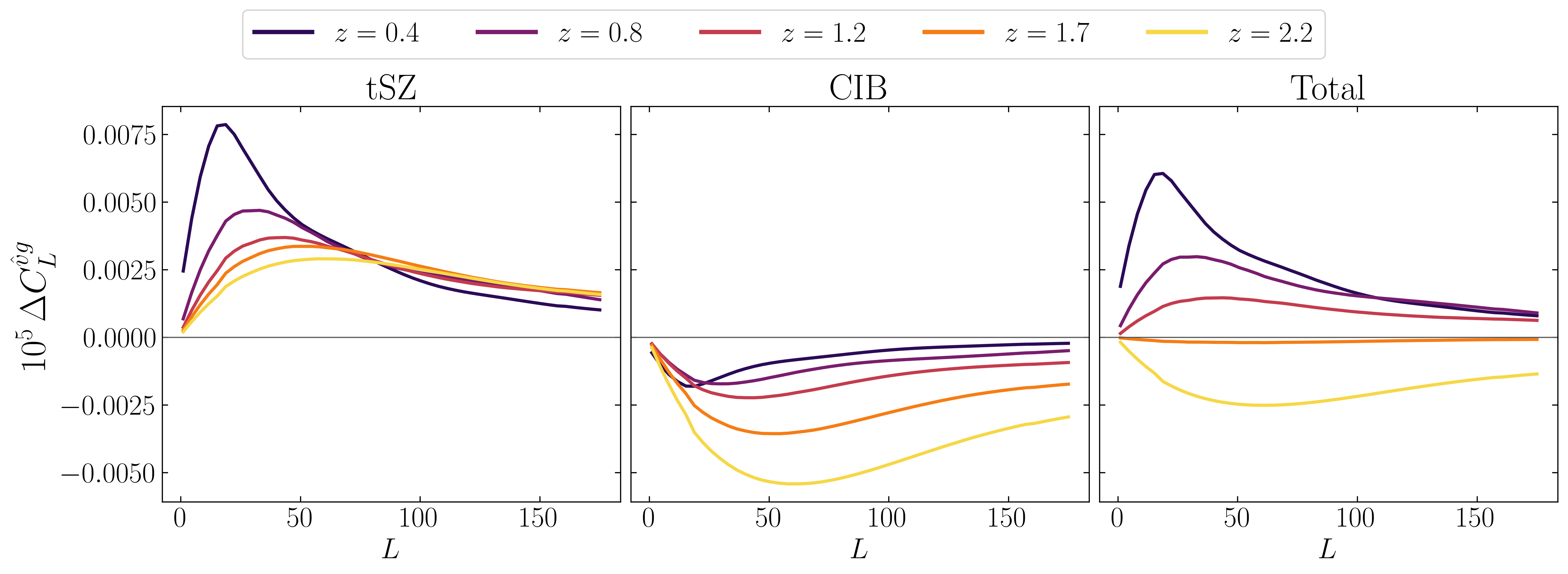}
    \caption{Redshift dependence of the foreground biases to the cross-correlation of kSZ-reconstructed velocity and galaxies at an observing frequency of $150\,\mathrm{GHz}$. The panels show the contributions from tSZ (left), CIB (center), and their sum (right), for tomographic bins centered at $z=[0.4, 0.8, 1.2, 1.7,2.2]$ (dark blue through yellow) with equal widths in comoving distance $\Delta \chi = 300\,\mathrm{Mpc}$. Calculations are done using a mock galaxy tracer with a constant minimum stellar mass threshold of $M_{\text{min}}^* = 10^{11.5}M_\odot$ across redshifts. The velocity reconstruction uses modes up to $\ell_{\text{max}} = 8000$, and the electron pressure profile is modeled using the Battaglia prescription.}
    \label{fig:redshift_dependence}
\end{figure}

It is interesting to show how the tSZ and CIB contributions evolve with redshift explicitly; to do so we compute the foreground bias predictions across a broader redshift range. Since our HOD is not defined at the high redshifts relevant for this exercise, we instead construct a simple mock galaxy tracer by imposing a minimum stellar-mass threshold of $M_{\text{min}}^* = 10^{11.5}M_\odot$ constant in redshift up to $z=5$. This HOD is constructed with the same abundance matching technique described in Sec.~\ref{sec:galaxy_HOD}, as implemented in \texttt{hmvec}. We then define tomographic bins as Gaussian distributions of equal width in comoving distance $\Delta \chi = 300\,\mathrm{Mpc}$ centered at $z = [0.4, 0.8, 1.2, 1.7,2.2]$. To compute the reconstructions analytically we need consistent spectra for the filters $C_\ell^{gg}$ and $C_\ell^{\tau g}$ entering Eq.~\eqref{eq:filters}, so we compute them analytically using the halo model. To compute these spectra we use the mock HOD, and tomographic bins that have just been defined, and describe the optical depth using the Battaglia AGN density profile presented in Ref.~\cite{2016JCAP...08..058B}. The resulting foreground predictions are shown in Fig.~\ref{fig:redshift_dependence}, where we show that the CIB contribution increases with redshift and becomes comparable to, and eventually larger than, the tSZ signal around the epoch of peak star formation ($z \sim 2$), changing the sign of the total foreground bias.

{We note that the Limber approximation is valid for multipoles satisfying $L \gg \bar{\chi}/{\Delta \chi}$, where $\Delta \chi$ is the width of each redshift bin and $\bar{\chi}$ is their average radial distance. For the redshift bins considered here, with $\Delta \chi \sim 300~\mathrm{Mpc}$ and $\bar{\chi}\sim 1500$--$5000\,\mathrm{Mpc}$, this corresponds to $L \approx 5$--$18$. A more precise treatment would require including beyond-Limber corrections to have an accurate estimate of the foreground biases at the lowest multipoles. {While this might be important for $f_{\text{NL}}$ constraints, we do not pursue this further in this work.}}

\subsection{Comparison to data} \label{sec:data_comparison}

In this section we compare our theoretical predictions with measurements obtained using the pipeline presented in Ref.~\cite{McCarthy_kSZ2_25}. For the purpose of characterizing the foreground contributions studied in this work, we run the pipeline on single-frequency temperature maps rather than on the needlet internal linear combination (NILC) \cite{bennett_92, 2009A&A...493..835D} maps used in the original analysis. The goal of this subsection is to benchmark our analytic calculations against observations and help understand the origin of the foreground contamination previously found in this dataset. To that end, we note that the data will have contributions from shot noise resulting from the underlying discrete nature of the CIB sources and the galaxies, so we present the relevant terms in App.~\ref{app:shot_noise} and include them in our halo model in this section. These contributions are found to be small, corresponding to at most 2\% of the presented total signal. We do not  aim to perform a fit of the halo model to the data, as this would require a more careful treatment of the halo-model details and is beyond the scope of this work. Instead, we present predictions that have been calibrated to previous datasets, in different scales, masses and redshifts than the data we are comparing them to. Therefore, this should serve as a qualitative comparison only, to show whether the amplitudes and shapes of our predictions are roughly consistent with the data, rather than aiming to have precise agreement.

\begin{figure}
    \centering
    \includegraphics[width=\linewidth]{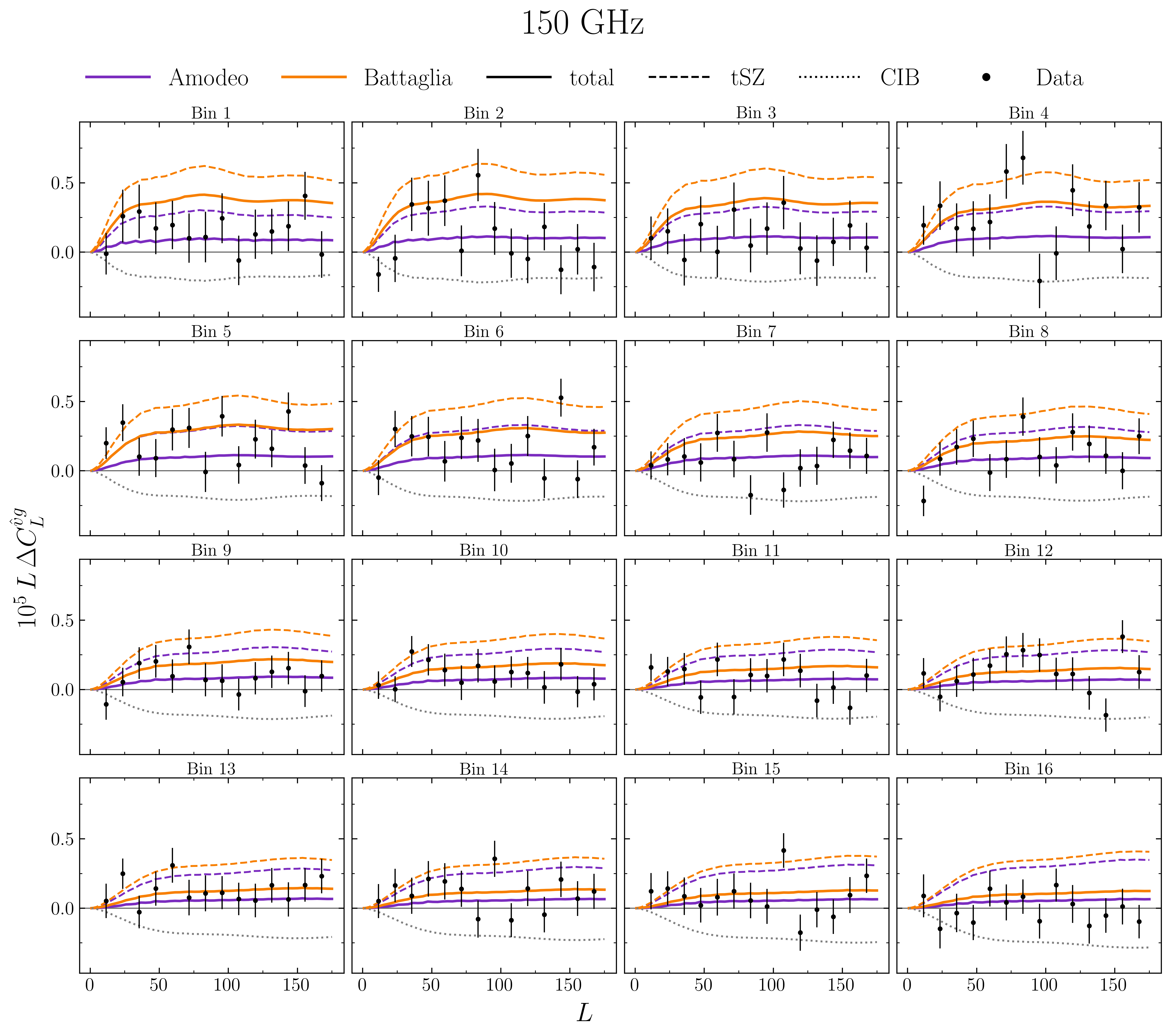}
    \caption{Comparison between our analytic predictions and measurements obtained using ACT DR6 CMB temperature maps and galaxies from DESILS, processed with the same pipeline as in Ref.~\cite{McCarthy_kSZ2_25}, at a frequency of $150\,\mathrm{GHz}$. The predictions shown are not fits to the data, instead we use two parametrizations of the electron pressure profile, the ``Amodeo'' (purple lines) and ``Battaglia'' (orange lines) models introduced in Sec.~\ref{sec:tsz_modeling}, and the DESILS-like HOD and CIB models presented in Secs~\ref{sec:galaxy_HOD} and \ref{sec:cib_modeling}, respectively. We show separately the tSZ (dashed) and CIB (dotted) contributions as well as their sum (solid). The comparison is performed for the same-redshift-bin configuration across the 16 tomographic bins used in Ref.~\cite{McCarthy_kSZ2_25}. The analytic calculations use a higher mass cut of $M_{\text{vir}} = 3 \times 10^{14}\,M_\odot$, chosen to mimic the cluster masking procedure applied to the CMB maps, and the velocity reconstruction uses modes up to $\ell_{\text{max}} = 8000$.}
    \label{fig:data_150}
\end{figure}

\begin{figure}
    \centering
    \includegraphics[width=\linewidth]{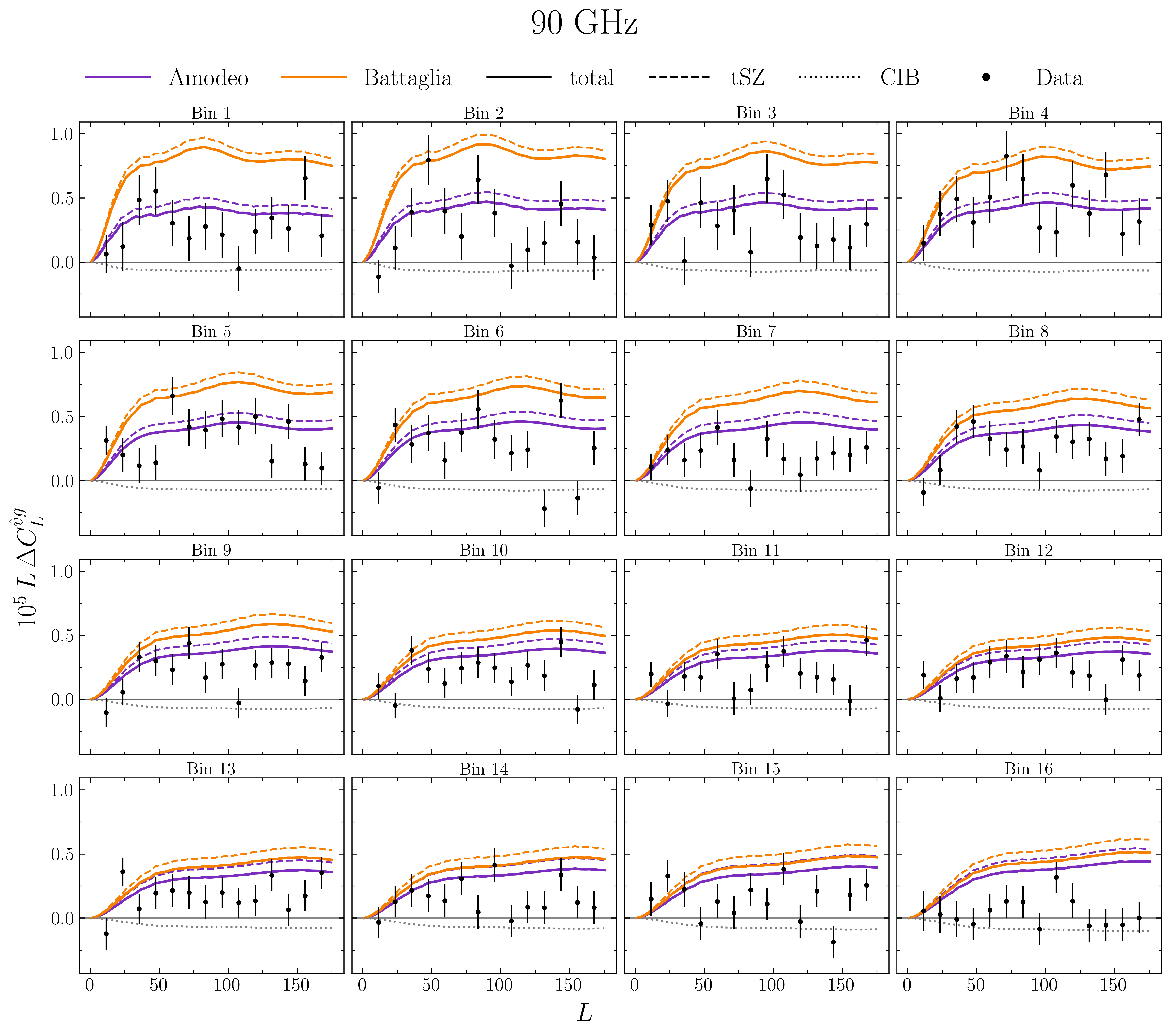}
    \caption{Comparison to data of our analytic calculations using CMB data from ACT DR6 at a single frequency of 90\,GHz and using DESILS as the galaxies. We show our predictions using the same configuration as Fig.~\ref{fig:data_150} but computed at $90\,\text{GHz}$.}
    \label{fig:data_90}
\end{figure}

The velocity-reconstruction pipeline used on the data uses a modifed version of the \texttt{ReCCo} algorithm described in Ref.~\cite{2023JCAP...02..051C}, which implements the real-space version of the quadratic estimator introduced in Ref.~\cite{Deutsch_ksz_17} (these real-space estimators first appeared in Ref.~\cite{2018PhRvD..98f3502C}). This is the full-sky version of the flat-sky quadratic estimator presented in Sec.~\ref{sec:QE_estimator}. The reconstruction is performed tomographically, splitting the DESI Legacy sample into 16 redshift bins spanning $0.4\lesssim z \lesssim 1.0$, and defined by thresholding their photometric redshift, as described in Ref.~\cite{McCarthy_kSZ2_25}. We estimate the true redshift distribution accounting for a Gaussian photo-$z$ error $\sigma(z)/(1+z)=0.027$.\footnote{We note that the resulting estimates of the redshift distributions are much less accurate than those of the full photo-$z$ sample, which is computed using DESI spectroscopy.} The temperature maps used for the reconstruction correspond to the public ACT DR6 co-added single-frequency, source-free maps using observations taken only at night time~\cite{AtacamaCosmologyTelescope:2025vnj}, while the small-scale galaxy field entering the estimator is taken from the DESI DR9 extended photometric LRG sample \cite{2023JCAP...11..097Z,2019AJ....157..168D}. The reconstructed velocity field is then cross-correlated with the large-scale galaxy overdensity estimated from the DESI Legacy DR10 galaxy sample defined with the same redshift bins \cite{2023JCAP...11..097Z,2023AJ....165...58Z}. The measurement is performed separately on the northern and southern galactic hemispheres; after measuring the cross-correlation we co-add the measurements at the bandpower level. The resulting measurements are shown in Figs.~\ref{fig:data_150} and \ref{fig:data_90} for the observed frequencies of $150$ and $90\,\mathrm{GHz}$ respectively.

To compare our analytic predictions to the data we consider the two prescriptions for the electron pressure profile introduced in Sec.~\ref{sec:tsz_modeling}, which we refer to as the ``Battaglia'' and ``Amodeo'' profiles. In Figs~\ref{fig:data_150} and \ref{fig:data_90} we find that our calculations depend moderately on the model used, with the Amodeo profile predicting a lower signal than Battaglia. In the case of $150\,\mathrm{GHz}$ (Fig.~\ref{fig:data_150}) we find both prescriptions to be in reasonable agreement with the data, which we remind the reader only contain foreground contamination and negligible signal coming from the velocity--galaxy correlation in this equal-redshift-bin case. This demonstrates the power of our model to explain and characterize the foreground contamination found in this dataset, despite the uncertainties arising from the details of the halo model.

{We also show the measurements obtained at $90\,\mathrm{GHz}$ (Fig.~\ref{fig:data_90}), where the agreement between the analytic predictions and the data is slightly worse, especially when using the Battaglia pressure profile. At this frequency, the total bias is dominated by the tSZ contribution, making the predictions even more sensitive to the modeling of the electron pressure profile as well as to the high-mass tail of the HOD. In addition, the parameters of the Amodeo profile were calibrated using a different galaxy sample (the BOSS CMASS sample) and were fitted assuming fixed redshift and mass dependencies. As a result, we do not expect this model to match our measurements at high precision, which span a broader redshift range and probe halos of different characteristic masses. Indeed, one can see that the absence of redshift evolution in the Amodeo pressure profile leads to increasing discrepancies with the data toward higher-redshift bins; in particular we note that only the first eight bins of our study overlap in redshift with the BOSS CMASS sample. Another possible source that can contribute to the discrepancy is the absence of a radio-source component in our modeling, which would contribute with opposite sign to the tSZ signal and is expected to be more significant at 90\,GHz than at 150\,GHz.}

{In the $90\,\mathrm{GHz}$ case, where the tSZ contribution is most relevant, the Amodeo profile provides a better qualitative description of the measurements than the Battaglia profile. However, this statement is highly dependent on the details of the HOD and assumptions about the CIB, and a more careful analysis is required to make any claims regarding the actual distribution of electrons around our galaxies. We also remind the reader that the quantity studied here corresponds to the combination of tSZ and CIB contributions to the squeezed bispectrum $\langle \fg\,g\,g\rangle$ introduced in Sec.~\ref{sec:bispectrum}. We note that it would be interesting to explore more direct probes of the pressure profile around the halos, such as measurements of the $\langle y\,g\rangle$ correlation, to provide a direct approach for constraining these profiles.}

\subsection{Neighboring-bins comparison} \label{sec:neighbouring}

Finally, it is worth exploring how much the foreground contribution biases the cosmological signal of interest. The $C_L^{\hat{v}^\alpha g^\beta}$ signal is most significant when the redshift bins $\alpha$ and $\beta$ are separated by a distance close to the velocity coherence length. In this configuration there is a clear direction in which the velocity is pointing and so the velocity--galaxy correlation does not vanish as it does in the same-redshift-bin case. However, there remains significant overlap between the redshift bins, meaning that foreground contribution may not be negligible. Therefore, we study the foreground contribution for the case where $\alpha$ and $\beta$ correspond to neighboring bins, where the cosmological signal is expected to be most significant.

\begin{figure}
    \centering
    \includegraphics[width=\linewidth]{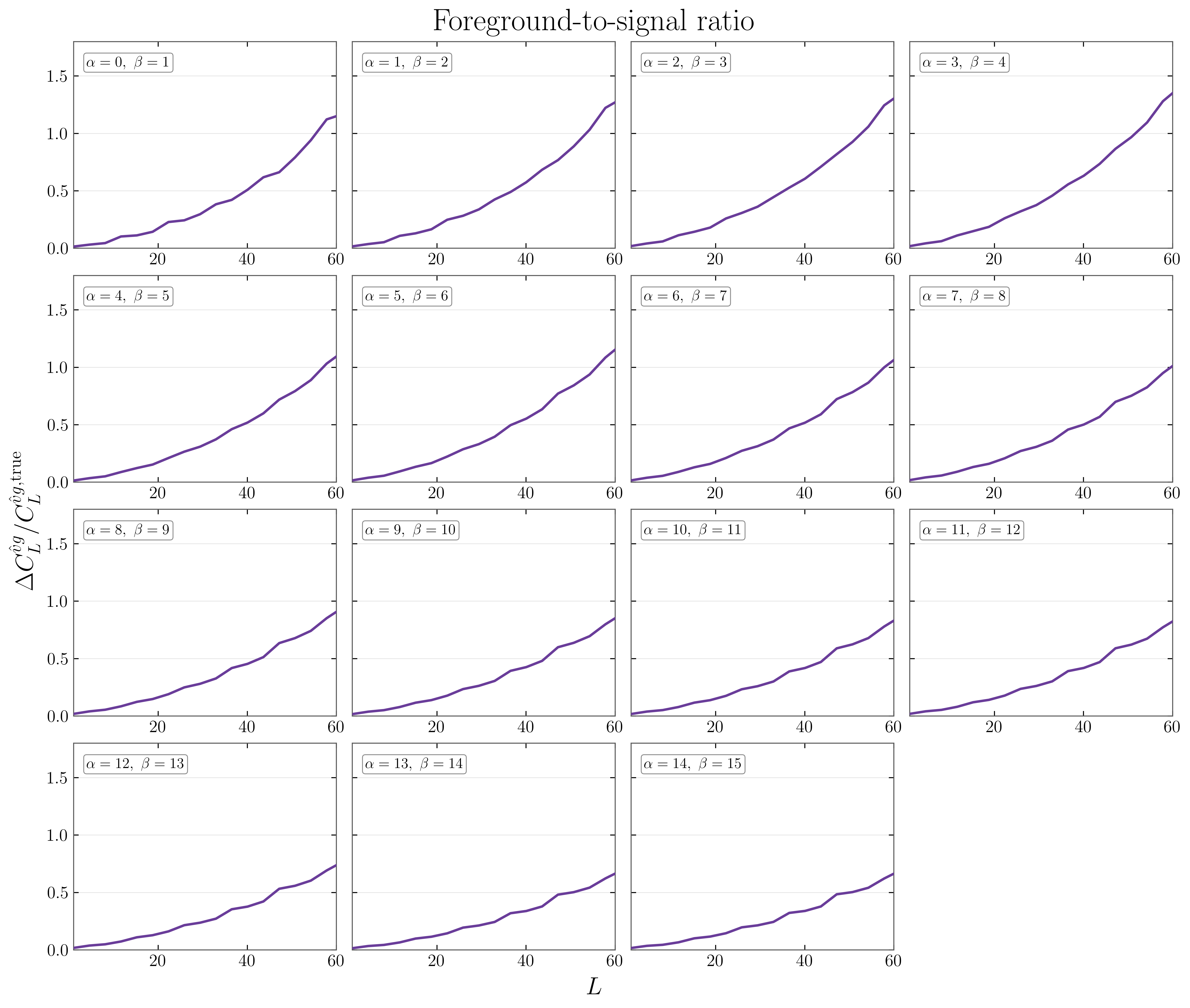}
    \caption{Ratio between the predicted foreground contribution and the expected kSZ velocity–galaxy cross-correlation signal for the neighboring-bin configuration, where the velocity reconstruction is performed in redshift bin $\alpha$ and the galaxy density tracer is in bin $\beta$. All predictions are shown for an observing frequency of $150\,\mathrm{GHz}$. The foreground calculations assume a mass cut of $M_{\text{cut}} = 3 \times 10^{14}\,M_\odot$ to mimic the cluster-masking procedure and use modes up to $\ell_{\text{max}} = 8000$ for the velocity reconstruction. The theoretical signal has been computed using a modified version of \texttt{ReCCO} {with $f_{\text{NL} }= 0$}. We show that, across the scales where the signal-to-noise is concentrated, the foreground contamination can reach the amplitude of the predicted signal.}
    \label{fig:difz_ratio}
\end{figure}

In Fig.~\ref{fig:difz_ratio} we show the ratio of our analytical foreground prediction and the expected velocity--galaxy cross-correlation signal. We compute the theoretical prediction for the signal using a modified version of \texttt{ReCCO} \cite{2023JCAP...02..051C}, which incorporates the beyond-Limber corrections of \cite{2020JCAP...05..010F}, assuming the same fiducial cosmology as our foreground calculations. In this neighboring-bin configuration, {at the relevant scales used for $f_{\text{NL}}$ in Ref.~\cite{McCarthy_kSZ2_25} ($L \lesssim60$), we find that the foreground contribution can become comparable to the signal. We note that this estimate is computed at a single frequency of $150\,\mathrm{GHz}$, whereas the analysis in Ref.~\cite{McCarthy_kSZ2_25} mitigated foreground contributions using NILC maps. Nevertheless, residual foregrounds were found in the data, and we show that the foreground contribution is still significant (although reduced) at the lower multiples, most relevant for $f_{\text{NL}}$ constraints. This highlights the need for robust foreground-mitigation strategies in kSZ velocity-reconstruction analyses.}

\section{Mitigation strategy: parity-odd estimator} \label{sec:mitigation}

Given the clear presence of foreground contamination in the velocity--galaxy cross-correlation, it is crucial to develop mitigation strategies that allow these measurements to be used for precision cosmology. {As noted in Ref.~\cite{ksz_bispectrum}, the kSZ signal and the non-Gaussian foreground signal have different symmetries: the kSZ-induced bispectrum is parity-odd under a reflection along the line of sight, while the foreground-induced bispectrum is parity-even. An estimator that respects this symmetry can isolate the kSZ-induced signal with no loss of signal-to-noise. In the 3D formalism, this has been achieved by measuring the velocity-galaxy power spectrum \textit{dipole}~\cite{Lague2024,Hotinli:2025tul} (with respect to the line of sight direction), which is naturally parity-odd. However, recent measurements of the projected tomographic signal~\cite{lai_ksz_25,McCarthy_kSZ2_25} have measured a full $C_L^{\hat{v}^\alpha g^\beta}$ datavector, which mixes parity-even and parity-odd terms (although fitting a model which is parity odd mostly isolates the parity-odd contribution).}  

In the projected tomographic approach, we suggest exploiting the different symmetry properties of the cosmological signal and the foreground contamination at the estimator level, which cancels the foreground contribution entirely while preserving the cosmological signal. {In Ref.~\cite{McCarthy_kSZ2_25} this was achieved by filtering the large-scale galaxy distribution in order to produce a direct estimate of the large-scale velocity field using the continuity equation (see, e.g.,~\cite{2024PhRvD.109j3533R,2024PhRvD.109j3534H}), which was then used}{ as the large-scale leg of the bispectrum. However, this introduced significant modeling difficulties due to the complex continuity-equation velocity reconstruction process, which would not be present in a $C_L^{\hat vg}$ analysis. Instead, here we explore a simpler parity-odd estimator first suggested in Ref.~\cite{McCarthy_kSZ2_25}:}
\begin{equation}
    C_L^{\hat{v} g,\mathrm{ODD}^{\alpha,\beta}}
    \equiv
    C_L^{\hat{v}^\alpha g^\beta}
    -
    C_L^{\hat{v}^\beta g^\alpha} \; .
\end{equation}
Using the decomposition introduced in Eq.~\eqref{eq:angular_cls}, together with the definition of the foreground contribution in Eq.~\eqref{eq:def_Delta_Cl}, the foreground contribution to this parity-odd combination $\Delta C_L^{\hat{v} g,\mathrm{ODD}^{\alpha,\beta}}$ is proportional to\footnote{This expression is symbolic: there is an implicit integration over $\boldsymbol{\ell}$ in the first line and an additional integral over $z$ in the second.}
\begin{align}
\Delta C_L^{\hat{v} g,\mathrm{ODD}^{\alpha,\beta}} &\propto \fil^\alpha(\boldsymbol{\ell},\boldsymbol{L}-\boldsymbol{\ell})B^{\text{2D}}_{\fg,g^\alpha_S,g^\beta_L}
(\boldsymbol{\ell},\boldsymbol{L}-\boldsymbol{\ell},-\boldsymbol{L})
- \fil^\beta(\boldsymbol{\ell},\boldsymbol{L}-\boldsymbol{\ell})
B^{\text{2D}}_{\fg,g^\beta_S,g^\alpha_L}
    (\boldsymbol{\ell},\boldsymbol{L}-\boldsymbol{\ell},-\boldsymbol{L}) \nonumber \\
 &\propto
    \left[\fil^\alpha(\boldsymbol{\ell},\boldsymbol{L}-\boldsymbol{\ell})W_{g_S}^\alpha(z)\, W_{g_L}^\beta(z)  B_{\fg,g_S,g_L} 
    \!\left(
        \frac{\boldsymbol{\ell}}{\chi},
        \frac{\boldsymbol{L}-\boldsymbol{\ell}}{\chi},
        -\frac{\boldsymbol{L}}{\chi};
        z
    \right) \right.
    - \nonumber \\ &\hspace{0.1\textwidth} \left. \fil^\beta(\boldsymbol{\ell},\boldsymbol{L}-\boldsymbol{\ell}) W_{g_S}^\beta(z)\, W_{g_L}^\alpha(z)B_{\fg,g_S,g_L}
    \!\left(
        \frac{\boldsymbol{\ell}}{\chi},
        \frac{\boldsymbol{L}-\boldsymbol{\ell}}{\chi},
        -\frac{\boldsymbol{L}}{\chi};
        z
    \right) \right]
     \; ,
\end{align}
following Eq.~\eqref{eq:projection}. The two terms cancel under the Limber approximation provided that: (i) the redshift distributions for the galaxies $g_S$ used in the velocity reconstruction are the same as for the galaxies $g_L$ used to correlate with the reconstruction; and (ii) evolution in the galaxy sample $g_S$ is small so that $\fil^\alpha \approx \fil^\beta$. As a result, the foreground contribution to the parity-odd estimator vanishes:
\begin{equation}
    \Delta C_L^{\hat{v} g,\mathrm{ODD}^{\alpha,\beta}} \approx 0 \; .
\end{equation}

The same cancellation does not occur for the cosmological signal. Again neglecting very large-scale effects, the true velocity–galaxy correlation changes sign under the exchange $\alpha \leftrightarrow \beta$. Consequently, the parity-odd combination preserves the signal while removing the foreground contamination:
\begin{equation}
    C_L^{\hat{v} g,\mathrm{ODD}^{\alpha,\beta}}
    \approx
    2\, C_L^{\hat{v}^\alpha g^\beta,\mathrm{true}} \; ;
\end{equation}
or, in the presence of $b_v$:
\begin{equation}
    C_L^{\hat{v} g,\mathrm{ODD}^{\alpha,\beta}}
    =
     b_v^\alpha C_L^{\hat{v}^\alpha g^\beta,\mathrm{true}}-  b_v^\beta C_L^{\hat{v}^\beta g^\alpha,\mathrm{true}}\; .
\end{equation}
We therefore obtain an estimator that suppresses the extragalactic foreground contribution while retaining the full cosmological signal without any loss of signal-to-noise. In the following we demonstrate how adopting this parity-odd estimator can significantly improve the fit to $C_L^{\hat vg}$ data presented in~Ref.~\cite{McCarthy_kSZ2_25}.

\subsection{Demonstration on ACT DR6+DESILS} \label{sec:mitigation_data}

We now demonstrate the efficacy of fitting  $C_L^{\hat{v} g,\mathrm{ODD}^{\alpha,\beta}}$  to data instead of the full $C_\ell^{\hat v^\alpha g^\beta}$ dataset. We use very similar datasets and processing  to that described in Ref.~\cite{McCarthy_kSZ2_25}, with some exceptions;  we refer the reader to that reference for in-depth details. In particular, we again perform kSZ velocity reconstruction on the combination of the small-scale DESI DR9 extended photometric LRG sample~\cite{2023JCAP...11..097Z} and the small-scale ACT DR6 black-body NILC map~\cite{2024PhRvD.109f3530C,2025JCAP...11..061N} using the pipeline described in~\cite{McCarthy_ksz_2024,McCarthy_kSZ2_25} and implemented in our modified version of \texttt{ReCCO}~\cite{2023JCAP...02..051C}. In all cases we perform the kSZ velocity reconstruction and cross correlation with the large-scale galaxies separately in the north and south galactic hemispheres, and co-add the measurements on both hemispheres at the bandpower level.

In Ref.~\cite{McCarthy_kSZ2_25}, foregrounds were mitigated by performing continuity-equation velocity reconstruction on the large-scale galaxy field to create a continuity-equation estimate of the velocity field in each redshift bin $\hat v^{\mathrm{cont}}$; no evidence was found for any foreground contamination, and the best-fit model had a good fit to the data. In contrast, when the kSZ velocity--galaxy cross power $C_L^{\hat v^g}$  was analyzed, evidence for foreground contamination was found {(despite the NILC foreground cleaning adopted for the CMB maps)} and the authors could not find a best-fit model with a good fit to the data, even when looking at subsets of the data that removed the most contaminated parts (i.e., the $\alpha=\beta$ diagonal). We will now demonstrate that using $C_L^{\hat{v} g,\mathrm{ODD}^{\alpha,\beta}}$  alleviates this problem.

\begin{figure}\centering
\includegraphics[width=\textwidth]{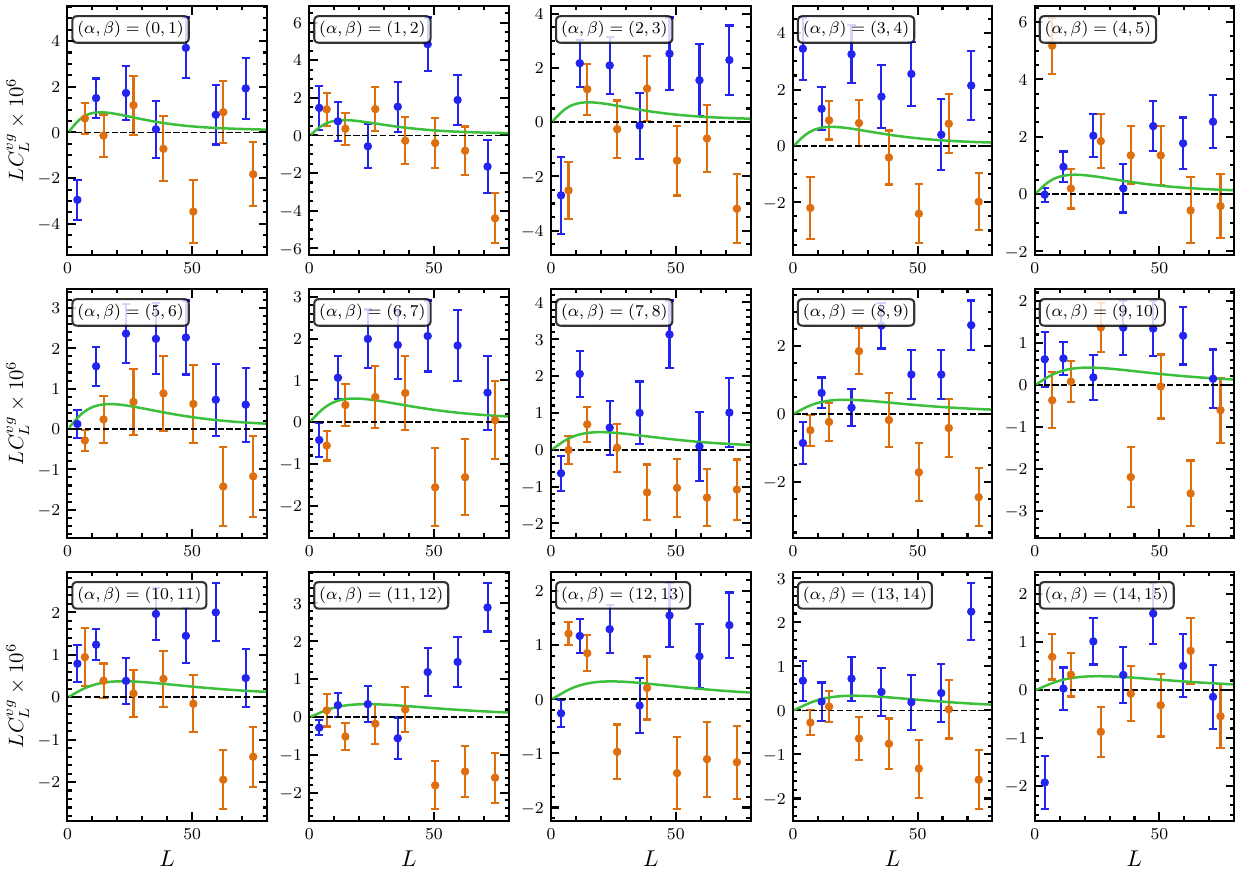}
\includegraphics[width=0.7\textwidth]{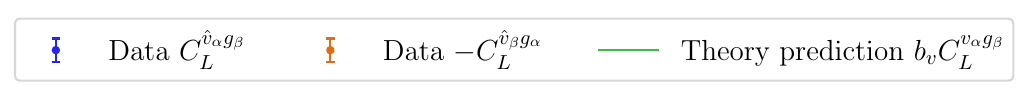}
\caption{The $C_L^{\hat v g}$ datapoints measured between neighbouring bins, $|\alpha-\beta|=1.$ We plot $C_L^{\hat{v}_\alpha g_\beta}$ (blue points) and $-C_L^{\hat{v}_\beta g_\alpha}$ (orange points) on the same plot as these should be extremely close to each other (if foreground contribution is negligible). We slightly offset the points in multipoles to enhance visibility. We only show the theory curve for $C_L^{v_\alpha g_\beta}$ (green) and not $-C_L^{v_\beta g_\alpha}$ as they are essentially overlapping. For $b_v$, we use the best-fit from {Ref.}~\cite{McCarthy_kSZ2_25}, i.e.,~$b_v=0.33$.}\label{fig:clvg_data_new}
\end{figure}

We cross-correlate our reconstructed kSZ velocity with overdensity maps estimated from the large-scale DR9 DESILS (extended sample), estimating the cross correlation using \texttt{pymaster} to decouple the mask~\cite{2019MNRAS.484.4127A}.\footnote{The use of the DR9 sample on large scales instead of the DR10 sample on large scales is the main difference with respect to {Ref.}~\cite{McCarthy_kSZ2_25} and allows us to use the wider sky area available in DR9 compared to DR10.} We use the same redshift binning scheme defined in {Ref.}~\cite{McCarthy_kSZ2_25} with 16 redshift bins, and (following Ref.~\cite{McCarthy_kSZ2_25}) estimate the covariance of the datapoints using the analytic mask-decoupling procedure implemented in  \texttt{pymaster}. For this analytic covariance calculation, we assume the \textit{measured} power of the galaxy overdensity maps and the \textit{analytic} reconstruction noise prediction for the kSZ velocity reconstruction, which was found to be close to the truth in {Ref.}~\cite{McCarthy_kSZ2_25}.

The full {$C_L^{\hat{v}^\alpha g^\beta}$} matrix that we measure has $N_\mathrm{bins}\times N_\mathrm{bins}$ elements at each $L$, where $N_\mathrm{bins}=16$. We focus only on the first off-diagonal, $C_L^{\hat{v}^\alpha g^\beta}$ with $|\alpha-\beta|=1$ (i.e., the cross correlation between the velocities with the galaxies in neighboring bins), where the signal-to-noise is concentrated. In all fits, we use the same $L$ range as {Ref.}~\cite{McCarthy_kSZ2_25}, which included four binned multipoles at $6\leq L\leq 53$ in the analysis.

\begin{figure}\centering
\includegraphics[width=\textwidth]{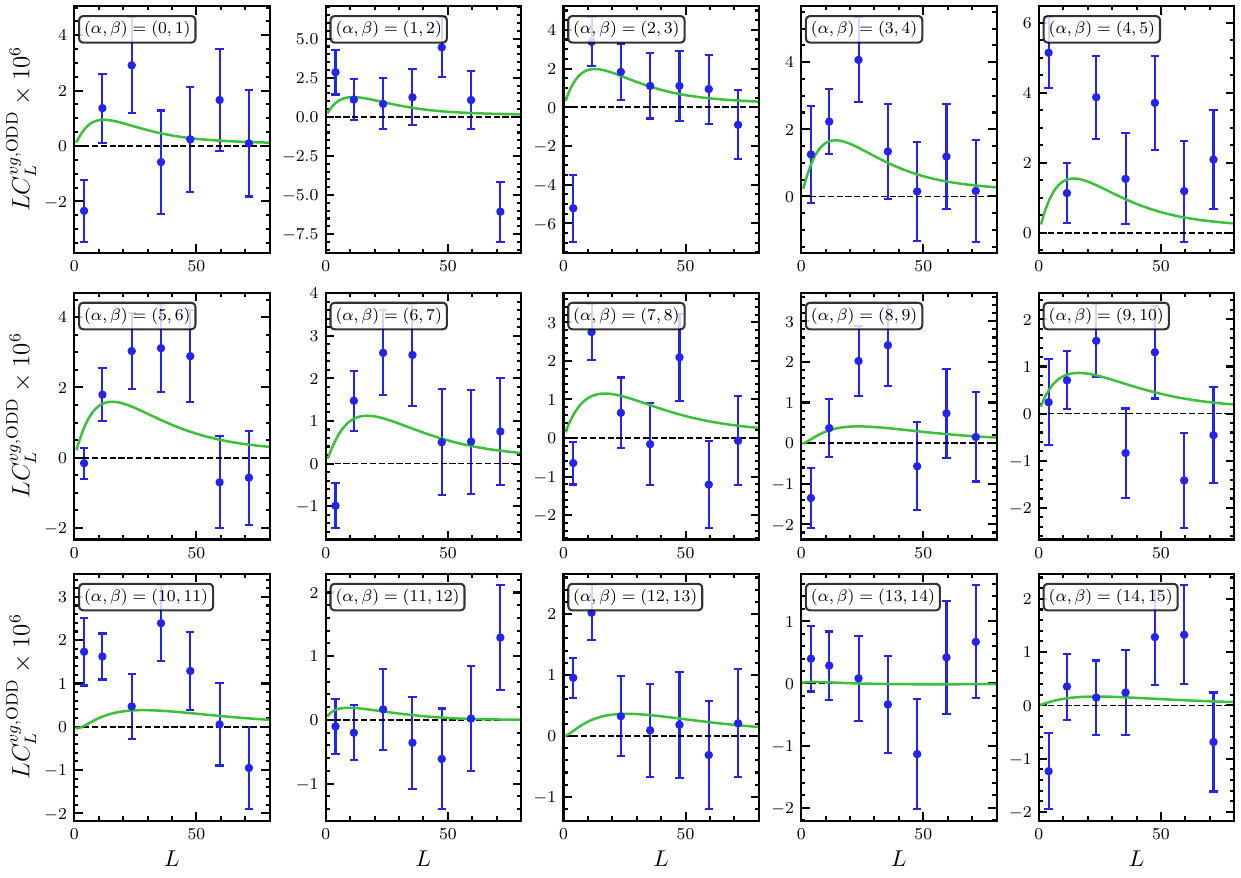}
\includegraphics[width=0.7\textwidth]{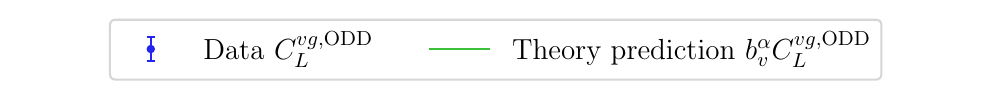}
\caption{Odd-parity $C_L^{\hat{v} g,\mathrm{ODD}^{\alpha,\beta}}$  datapoints measured for $|\alpha-\beta|=1$ compared with the theory prediction (green lines) without foregrounds. In this theory prediction,  we use the best-fit $b_v^\alpha$ from our fit to these data at $6\le L\le53$ (as described in the text). This odd-parity combination contains the kSZ signal with foregrounds suppressed, as explained in the text. The PTE of this best-fitting model with respect to the datapoints is 0.46.}\label{fig:clvg_data_odd}
\end{figure}

We show the $C_L^{\hat{v}^\alpha g^\beta}$ datapoints in Fig.~\ref{fig:clvg_data_new} and $C_L^{\hat{v} g,\mathrm{ODD}^{\alpha,\beta}}$ in Fig.~\ref{fig:clvg_data_odd}. We consider two models: one with only a single, redshift-independent velocity-bias parameter $b_v$, and one with a velocity-bias parameter per redshift bin $b_v^\alpha$. When we fit our model for $C_L^{vg}$ to the {$C_L^{\hat{v}^\alpha g^\beta}$} datapoints (using the model and likelihood described in {Ref.}~\cite{McCarthy_kSZ2_25}), we recover an amplitude for $b_v$ that is consistent with what was found in {Ref.~\cite{McCarthy_kSZ2_25}} using the cross correlation with the continuity-equation-reconstructed velocity $C_L^{\hat{v}\, \hat{v}^{\mathrm{cont}}}$, but we do not have a good fit to the data: our probability-to-exceed (PTE) is $3\times10^{-11}$ for the $b_v$ model and  $8\times10^{-7}$ for the $b_v^\alpha$ model, both clearly unacceptable; the PTEs are summarized in Table~\ref{tab:PTEs}.

\begin{figure}\centering
\includegraphics[width=\textwidth]{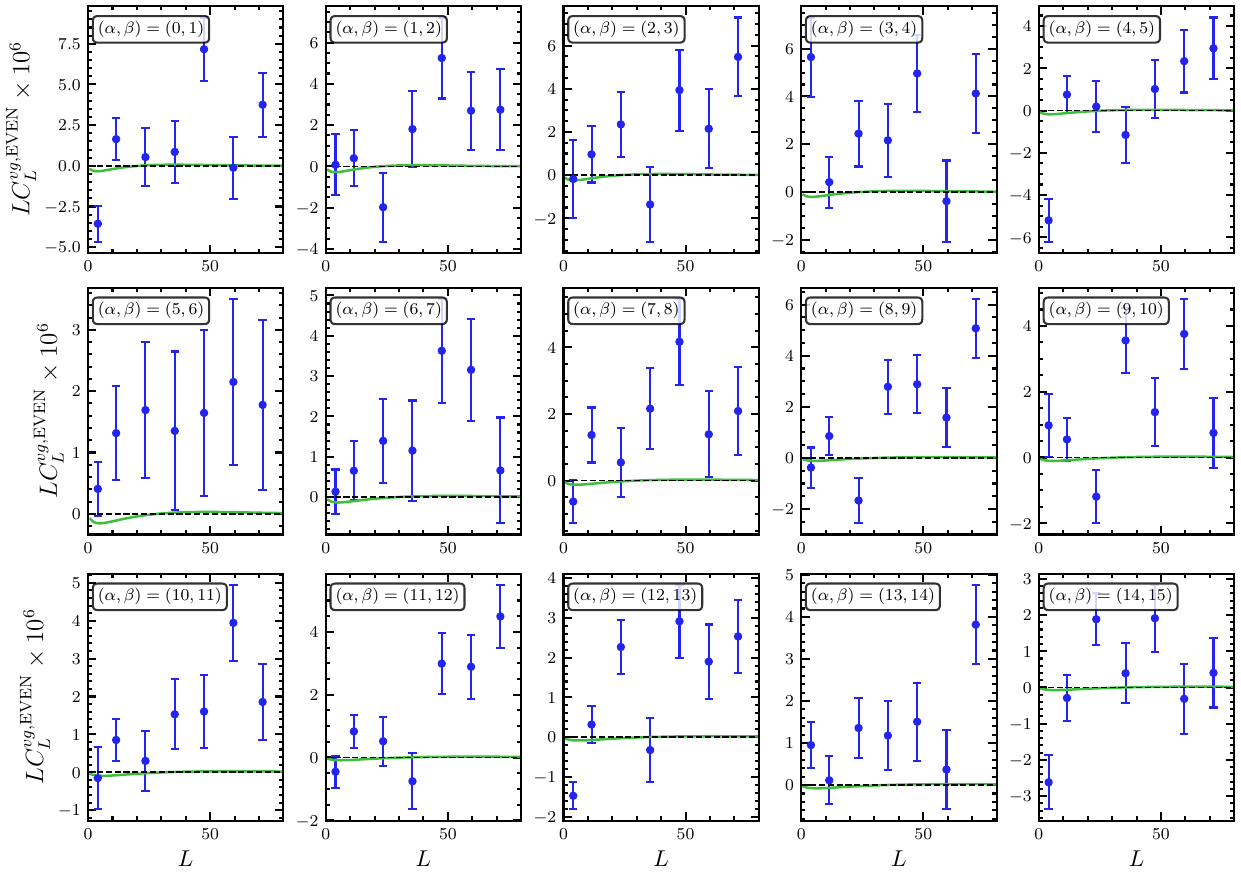}
\includegraphics[width=0.7\textwidth]{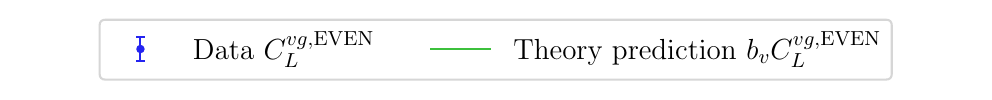}
\caption{Even-parity $C_L^{\hat{v} g,\mathrm{EVEN}^{\alpha,\beta}}$  datapoints measured for $|\alpha-\beta|=1$. The PTE of the $\chi^2$ with respect to zero for these datapoints (with $6<L<55$) is $2\times10^{-9}$. The theory prediction without foregrounds (green lines) is close to zero.}\label{fig:clvg_data_even}
\end{figure}

However, when we fit the same models to the combinations $C_L^{\hat{v} g,\mathrm{ODD}^{\alpha,\beta}}$ (focusing only on the 15 combinations we can make with the first off-diagonal), our goodness-of-fit improves dramatically, to PTEs of $0.001$ for the $b_v$ model and  $0.46$ for the $b_v^\alpha$ model (again, these are summarized in Table~\ref{tab:PTEs}).\footnote{The central value and the errorbar of $b_v$ is unchanged, as the theory that we fit is antisymmetric and is therefore already only sensitive to antisymmetric structure in the data.} The improvement in PTE can also be understood as a manifestation of the fact that the \textit{symmetric} combination $C_L^{\hat{v} g,\mathrm{EVEN}^{\alpha,\beta}}\equiv C_L^{\hat{v}^\alpha g^\beta}+C_L^{\hat{v}^\beta g^\alpha}$ is inconsistent with zero: indeed, the PTE for this combination with respect to zero is $2\times10^{-9}$, corresponding to a $\sim6\sigma$ detection of excess foreground power (assuming all excess power can be attributed to foregrounds). Without foregrounds, the theory prediction for $C_L^{\hat{v} g,\mathrm{EVEN}^{\alpha,\beta}}$ would be close to zero. We show this combination in Fig.~\ref{fig:clvg_data_even}.

\begin{figure}\centering
\includegraphics[width=0.8\textwidth]{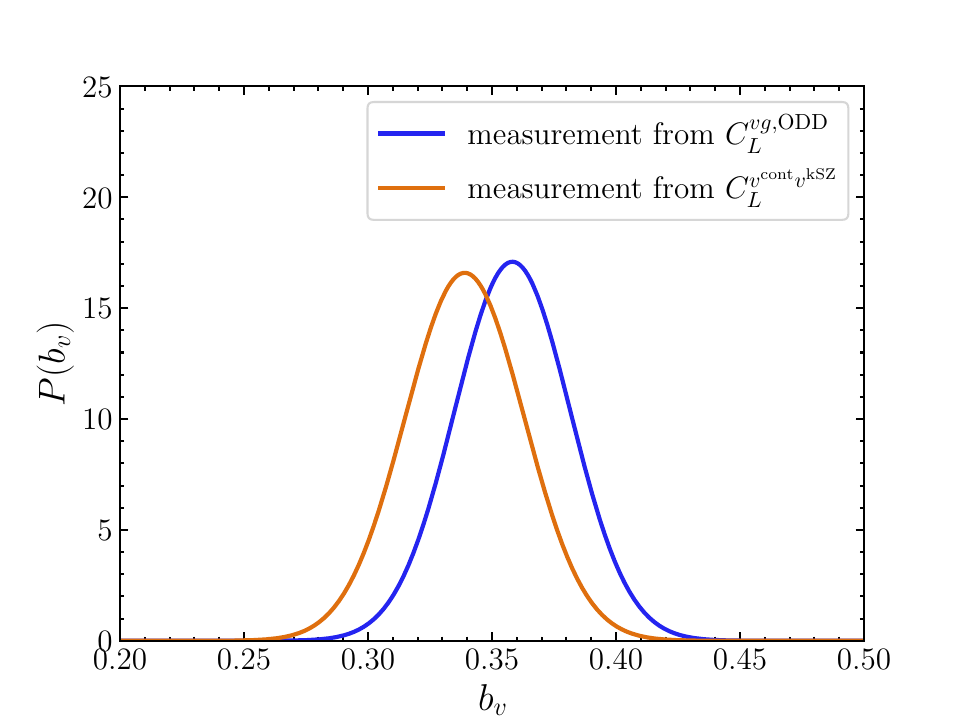}
\caption{Constraint on $b_v$, along with a comparison to the constraint from~\cite{McCarthy_kSZ2_25}, which used a very different method to mitigate foregrounds. These constraints are highly consistent.}\label{fig:bvconstraint}
\end{figure}

We show the constraint on $b_v$ in Fig.~\ref{fig:bvconstraint} along with a comparison to the $b_v$ found in {Ref.}~\cite{McCarthy_kSZ2_25}; these are highly consistent, with only $0.4\sigma$ shift (we note that the data combinations are slightly different). We find that $b_v=0.358\pm0.033$, compared to  $b_v=0.339\pm0.034$ from {Ref.~\cite{McCarthy_kSZ2_25}}. Overall, we have detected the kSZ signal in $C_L^{\hat vg}$ using the presented parity-odd estimator at $11\sigma$ and found a very good fit to our model. 

\begin{table}\centering
\begin{tabular}
{cccccccc}\hline\hline
dataset &  $\chi^2_{\mathrm{min}} (b_v)$ & DOF & PTE & $\chi^2_{\mathrm{min}} (b_v^\alpha)$ & DOF & PTE \\ \hline
$C_L^{v^\alpha g^\beta}$ & 249 &$120-1$&$ 3\times10^{-11}$& 207 &$120-16$&$ 8\times10^{-7}$\\
$C_L^{vg,\mathrm{ODD}}$ & 99 &$60-1$&0.001& 54 &$60-16$&0.46 \\ \hline

\end{tabular}\caption{Summary of the goodness-of-fit of the $b_v$ and $b_v^\alpha$ models to the data. As a reminder, the $b_v$ model is where we have one velocity-bias parameter that is the same for all redshift bins, while the $b_v^\alpha$ model has 16 free bias parameters (one in each redshift bin). The best-fits in the standard $C_L^{v^\alpha g^\beta}$  case are clearly unacceptable, while for the $C_L^{vg,\mathrm{ODD}}$ case we find a perfectly acceptable fit with the $b_v^\alpha$ model.}\label{tab:PTEs}
\end{table}

\section{Summary and conclusions}

As the sensitivity of current experiments continues to improve, kSZ velocity-reconstruction measurements are rapidly evolving from first detections into a probe of precision cosmology. Now, systematic effects that were previously subdominant must be carefully understood in order to extract robust cosmological information. In this work we present a dedicated study of extragalactic foreground contamination present in CMB temperature maps to kSZ velocity-reconstruction measurements. In particular, we focus on the bias induced on the velocity--galaxy angular cross-spectrum at the large angular scales relevant for constraints on primordial non-Gaussianity.

We study this bias analytically by developing a formalism that allows the velocity-reconstruction mode-coupling integrals to be evaluated directly within the \texttt{CosmoBLENDER} framework~\cite{baleato_2025}. We implement the appropriate mode-coupling weights for the kSZ velocity-reconstruction estimator and make this extension publicly available as a new branch of the code. We identify the foreground contribution to the reconstructed signal with the squeezed foreground--galaxy--galaxy bispectrum and model its contributions using the halo model. In particular, we describe the contamination from both the tSZ effect and the CIB, restricting our analysis to the 1-halo and 2-halo terms. While the precise predictions depend on details of the halo-model prescription, such as the galaxy HOD or the assumed electron pressure profile, our analytical approach provides a clear description of the behavior of the different foreground contributions.

We also interpret the foreground contribution within the response formalism. Foregrounds induce an additional contribution to the expectation value of the reconstructed velocity field averaged over small-scale modes. On the large scales of interest here, this arises from the response of the small-scale foreground--galaxy cross-power spectrum to the presence of a long-wavelength matter density perturbation. As a result, we show that the presence of foregrounds generates an additional contribution to the reconstructed velocity field that is directly proportional to the large-scale matter field.

Within the halo-model framework, we first characterize the foreground biases in the same-redshift-bin configuration, where the cosmological velocity--galaxy signal is expected to be negligible, which allows us to isolate the foreground contamination contribution. We find that, at 90 and 150\,GHz, the tSZ and CIB contributions have opposite signs as the tSZ appears as decrement to the CMB maps at these observing frequencies. This leads to a partial cancellation in the total foreground bias. We also show explicitly that, as expected, on the large angular scales relevant for \(f_{\mathrm{NL}}\) analyses, the contamination is dominated by the 2-halo term. Studying the redshift dependence of these contributions shows that the tSZ dominates the foreground contamination for DESI-like redshifts, while the CIB becomes increasingly important at higher redshifts, and can dominate at redshifts close to the epoch of peak star formation. Then, we study the neighboring-redshift-bin configuration, which contributes the most to the true cosmological signal. Comparing the foreground bias to the expected kSZ velocity--galaxy signal, we find that the foreground contamination can reach the level of the signal across the range of scales relevant for $f_{\text{NL}}$ analyses.

{We compare our analytic predictions using two different electron pressure profile prescriptions (Battaglia and Amodeo) to measurements obtained using ACT DR6 single-frequency maps and DESI Legacy galaxies processed with the reconstruction pipeline of Ref.~\cite{McCarthy_kSZ2_25}. At $150\,\mathrm{GHz}$, we find qualitative agreement between the analytic predictions and the observed residual contamination, as well as a moderate dependence on the electron pressure profile used. At $90\,\mathrm{GHz}$, we find the agreement with data to be slightly worse, especially in the case of the Battaglia profile. At this frequency, our predictions are even more sensitive to the modeling of the electron pressure profile and therefore to the high-mass tail of the galaxy HOD, while still being sensitive to CIB modeling. Given that the parameters in the Amodeo pressure profile were calibrated in a different galaxy sample, this model is not expected to provide a high-precision description of our measurement. We note that, at this frequency, the Amodeo profile predicts a lower signal more in agreement with the data than the Battaglia profile. We stress, however, that this conclusion depends on details of the high-mass end of the galaxy HOD and no strong statements about the distribution of electrons within halos should be made from this analysis. In the future, it would be interesting to explore this directly through a $\left<y g\right>$ cross-correlation measurement.}

We present a parity-odd estimator (first proposed in Ref.~\cite{McCarthy_kSZ2_25}), that exploits the different symmetry properties of the cosmological signal compared to the foreground contributions. It is constructed from an antisymmetric combination of the velocity--galaxy cross-correlation: $C_L^{\hat{v}^\alpha g^\beta} - C_L^{\hat{v}^\beta g^\alpha}$, where $\alpha$ and $\beta$ refer to different redshift bins ($\alpha\neq \beta$). We show analytically that, under the Limber approximation and assuming the fields do not evolve significantly with redshift, this estimator cancels the foreground contribution entirely while preserving the full cosmological signal. Applying this estimator directly to the data significantly improves the goodness of fit to the model, while maintaining the full constraining power compared to the previous analysis. 

{We note that all of our foreground-bias calculations were performed within the Limber approximation, which is not fully accurate on the very large scales of interest for kSZ velocity measurements. It would be of great interest to extend the foreground--bias calculations to a beyond-Limber calculation, to investigate how much extremely large-scale correlations can break the symmetry arguments argued here (which can break down due to evolution of the fields along the light cone). In particular, it would be of interest to investigate the leakage into the parity-odd estimator, or---to avoid the complex calculations involved in the halo model---to use the response formalism to quantify and marginalize over any remaining bias in the cross-correlation. Finally, we note that we have focused only on the {cross-correlation} of the kSZ velocity with galaxies; it would additionally be of great interest to understand the foreground bias to the kSZ velocity autospectrum, which is a trispectrum at the level of the data.} 

Our analysis demonstrates the need for robust foreground mitigation techniques in kSZ velocity reconstruction measurements. To that end, the presented parity-odd estimator provides a powerful approach to mitigating foreground contamination in projected velocity--galaxy cross-correlation measurements, analogous to the dipole estimator used in three-dimensional kSZ velocity reconstruction, paving the way for the use of kSZ velocity reconstruction measurements as a probe for precision cosmology.

\acknowledgments

\input{acknowledgements.tex}

\appendix

\section{Halo model expressions for bispectrum biases} \label{app:bispectra}

We present the explicit expressions for the 3D bispectra that produce a foreground bias to the cross-power spectrum of the  kSZ-reconstructed velocity and galaxy overdensity. As in the main text, we assume that the small-scale galaxies entering the kSZ velocity reconstruction and the large-scale ones with which the reconstructed velocity is correlated belong to the same galaxy population, and so we describe them with the same HOD.

First, to describe the foregrounds arising from the tSZ effect, we show the 1-halo and 2-halo terms of the $\left< ygg\right>$ bispectrum: 
\begin{equation}
\begin{split}
    B_{y g g}^{\text{1h}}(\boldsymbol{k}_1,\boldsymbol{k}_2,\boldsymbol{k}_3;z) 
    = \int dM \, n(M,z) 
    \frac{\langle N_{\text{sat}} \rangle \left[2 \langle N_{\text{cen}}\rangle + \langle N_{\text{sat}} \rangle \right]} {\bar{n}_{\text{g}}^2} \\
    \times \, y_{\text{3D}}(k_1,M,z) 
    u(k_2,M,z)
    u(k_3,M,z) \; ,
\end{split}
\end{equation}
and
\begin{equation}
\begin{split}
B_{y g g}^{\text{2h}}(\boldsymbol{k}_1,\boldsymbol{k}_2,\boldsymbol{k}_3;z) 
=  P_{\text{lin}}(k_1,z)\int dM' n(M',z) b(M',z) y_{\text{3D}}(k_1,M',z)
\\
\times \int dM \, n(M,z) b(M,z)
\frac{\langle N_{\text{sat}} \rangle \left[2 \langle N_{\text{cen}}\rangle + \langle N_{\text{sat}} \rangle \right]} {\bar{n}_{\text{g}}^2}
u(k_2,M,z)
u(k_3,M,z) \; 
\\ 
+ \; P_{\text{lin}}(k_2,z)\int dM' n(M',z) b(M',z) \frac{\left< N_{\text{gal}} \right>}{\bar{n}_g} u(k_2,M',z)
\\
\times \int dM n(M,z) b(M,z) y_{\text{3D}}(k_1,M,z) \frac{\left< N_{\text{gal}} \right>}{\bar{n}_g} u(k_3,M,z)
\\
+ \; P_{\text{lin}}(k_3,z)\int dM' n(M',z) b(M',z) \frac{\left< N_{\text{gal}} \right>}{\bar{n}_g} u(k_3,M',z)
\\
\times \int dM n(M,z) b(M,z) y_{\text{3D}}(k_1,M,z) \frac{\left< N_{\text{gal}} \right>}{\bar{n}_g} u(k_2,M,z) \; ,
\end{split}
\end{equation}
where all the elements entering these expression have been previously defined in Sec.~\ref{sec:halo_model}.

For the biases induced by the CIB, the corresponding expressions are:
\begin{equation}
\begin{split}
B_{I g g}^{\text{1h}}(\boldsymbol{k}_1,\boldsymbol{k}_2,\boldsymbol{k}_3;z) 
= \int dM \, n(M,z) 
\frac{\langle N_{\text{sat}} \rangle \left[2 \langle N_{\text{cen}}\rangle + \langle N_{\text{sat}} \rangle \right]} {\bar{n}_{\text{g}}^2}
u(k_2,M,z)
u(k_3,M,z) \; 
\\
\times u(k_1,M,z)\left[\frac{f^{\text{cen}}(M,z)}{u(k_1,M,z)} + f^{\text{sat}}(M,z) \right] \; ,
\label{eq:1h-ygg}
\end{split}
\end{equation}
and
\begin{equation}
\begin{split}
B_{I g g}^{\text{2h}}(\boldsymbol{k}_1,\boldsymbol{k}_2,\boldsymbol{k}_3;z) 
=  P_{\text{lin}}(k_1,z)\int dM' n(M',z) b(M',z) u(k_1,M',z) \left[\frac{f^{\text{cen}}(M',z)}{u(k_1,M',z)} + f^{\text{sat}}(M',z) \right]
\\
\times \int dM \, n(M,z) b(M,z)
\frac{\langle N_{\text{sat}} \rangle \left[2 \langle N_{\text{cen}}\rangle + \langle N_{\text{sat}} \rangle \right]} {\bar{n}_{\text{g}}^2}
u(k_2,M,z)
u(k_3,M,z) \; 
\\ 
+ \; P_{\text{lin}}(k_2,z)\int dM' n(M',z) b(M',z) \frac{\left< N_{\text{gal}} \right>}{\bar{n}_g} u(k_2,M',z)
\\
\times \int dM n(M,z) b(M,z) u(k_1,M,z)\left[\frac{f^{\text{cen}}(M,z)}{u(k_1,M,z)} + f^{\text{sat}}(M,z) \right] \frac{\left< N_{\text{gal}} \right>}{\bar{n}_g} u(k_3,M,z)
\\
+ \; P_{\text{lin}}(k_3,z)\int dM' n(M',z) b(M',z) \frac{\left< N_{\text{gal}} \right>}{\bar{n}_g} u(k_3,M',z)
\\
\times \int dM n(M,z) b(M,z) u(k_1,M,z)\left[\frac{f^{\text{cen}}(M,z)}{u(k_1,M,z)} + f^{\text{sat}}(M,z) \right] \frac{\left< N_{\text{gal}} \right>}{\bar{n}_g} u(k_2,M,z) \; .
\end{split}
\end{equation}

\section{Shot-noise terms} \label{app:shot_noise}

To accurately describe our signal with the halo model, we also need to take into account contributions from the shot-noise terms. These terms appear whenever there is more than one discrete tracer in the same halo, and they correspond to the terms in which two legs from the correlator are in the same galaxy. Therefore, this will be present both in the tSZ and CIB contributions as well as 1-halo and 2-halo terms:
\begin{equation}
    B_{y g g}^{\text{1h-SN}}(\boldsymbol{k}_1,\boldsymbol{k}_2,\boldsymbol{k}_3;z) 
    = \int dM \, n(M,z) 
    \frac{1} {\bar{n}_{\text{g}}^2} \; 
    y_{\text{3D}}(k_1,M,z)
\end{equation}
and
\begin{equation}
\begin{split}
B_{y g g}^{\text{2h-SN}}(\boldsymbol{k}_1,\boldsymbol{k}_2,\boldsymbol{k}_3;z) 
=  P_{\text{lin}}(k_1,z)\int dM' n(M',z) b(M',z) y_{\text{3D}}(k_1,M',z)
\\
\times \int dM \, n(M,z) b(M,z)
\frac{1} {\bar{n}_{\text{g}}^2} \; .
\end{split}
\end{equation}
The equivalent expressions for the CIB  simply replace the $y_{\text{3D}}$ profile for the appropriate CIB term, defined in Eq.~\eqref{eq:CIB_profile}. We note that the CIB is also composed of discrete tracers within halos, but that we are assuming a different population to the galaxies used in the reconstruction. In particular, as discussed in Sec.~\ref{sec:cib_modeling}, we adopt the CIB model presented in Ref.~\cite{Shang:2011mh}.

\newpage
\bibliographystyle{JHEP}
\bibliography{biblio}
\appendix

\end{document}

%% file: authors_cev.tex
\author[1,2]{Carmen Embil Villagra\,\orcidlink{0009-0001-3987-7104}}
\author[1,2]{Fiona McCarthy}
\author[3,4]{Antón Baleato Lizancos\,\orcidlink{0000-0002-0232-6480}}
\author[1,2]{Blake~D.~Sherwin}
\author[5,1,2]{Anthony Challinor\,\orcidlink{0000-0003-3479-7823}}

\affiliation[1]{DAMTP, Centre for Mathematical Sciences, University of Cambridge, Wilberforce Road, Cambridge CB3 OWA, UK}
\affiliation[2]{Kavli Institute for Cosmology Cambridge, Madingley Road, Cambridge CB3 0HA, UK}
\affiliation[3]{Berkeley Center for Cosmological Physics, Department of Physics, University of California, Berkeley, CA 94720, USA}
\affiliation[4]{Lawrence Berkeley National Laboratory, One Cyclotron Road, Berkeley, CA 94720, USA}
\affiliation[5]{Institute of Astronomy, University of Cambridge, Madingley Road, Cambridge CB3 0HA, UK}

%% file: acknowledgements.tex
We thank Edmond Chaussidon, Thomas Colas, Simone Ferraro, Boryana Hadzhiyska, Selim Hotinli, Matthew Johnson, Fabian Schmidt, and Kendrick Smith for useful conversations. We thank Ed Wollack for providing comments on the draft. AC acknowledges support from STFC (grant reference UKRI1164). BDS, FMcC and CEV acknowledge support from the European Research Council (ERC) under the European Union’s Horizon 2020 research and innovation programme (Grant agreement No. 851274).

%% file: biblio.bib
@article{Smith:2007rg,
    author = "Smith, Kendrick M. and Zahn, Oliver and Dore, Olivier",
    title = "{Detection of Gravitational Lensing in the Cosmic Microwave Background}",
    eprint = "0705.3980",
    archivePrefix = "arXiv",
    primaryClass = "astro-ph",
    doi = "10.1103/PhysRevD.76.043510",
    journal = "Phys. Rev. D",
    volume = "76",
    pages = "043510",
    year = "2007"
}

@article{McCarthy_kSZ2_25,
    author = "McCarthy, Fiona and others",
    title = "{The Atacama Cosmology Telescope: Cross-correlation of kSZ and continuity equation velocity reconstruction with photometric DESI LRGs}",
    eprint = "2511.15701",
    archivePrefix = "arXiv",
    primaryClass = "astro-ph.CO",
    month = "11",
    year = "2025"
}

@article{ksz_bispectrum,
    author = {Smith, Kendrick M. and Madhavacheril, Mathew S. and M{\"u}nchmeyer, Moritz and Ferraro, Simone and Giri, Utkarsh and Johnson, Matthew C.},
    title = "{KSZ tomography and the bispectrum}",
    eprint = "1810.13423",
    archivePrefix = "arXiv",
    primaryClass = "astro-ph.CO",
    month = "10",
    year = "2018"
}

@ARTICLE{2019MNRAS.484.4127A,
       author = {{Alonso}, David and {Sanchez}, Javier and {Slosar}, An{\v{z}}e and {LSST Dark Energy Science Collaboration}},
        title = "{A unified pseudo-C$_{{\ensuremath{\ell}}}$ framework}",
      journal = {\mnras},
         year = 2019,
        month = apr,
       volume = {484},
       number = {3},
        pages = {4127-4151},
          doi = {10.1093/mnras/stz093},
archivePrefix = {arXiv},
       eprint = {1809.09603},
 primaryClass = {astro-ph.CO},
       adsurl = {https://ui.adsabs.harvard.edu/abs/2019MNRAS.484.4127A}
}

@article{lai_ksz_25,
    author = {Lai, Anderson C. M. and Kvasiuk, Yurii and M{\"u}nchmeyer, Moritz},
    title = "{KSZ Velocity Reconstruction with ACT and DESI-LS using a Tomographic QML Power Spectrum Estimator}",
    eprint = "2506.21684",
    archivePrefix = "arXiv",
    primaryClass = "astro-ph.CO",
    month = "6",
    year = "2025"
}

@article{Munchmeyer_ksz_forecast_18,
    author = {M{\"u}nchmeyer, Moritz and Madhavacheril, Mathew S. and Ferraro, Simone and Johnson, Matthew C. and Smith, Kendrick M.},
    title = "{Constraining local non-Gaussianities with kinetic Sunyaev-Zel{\textquoteright}dovich tomography}",
    eprint = "1810.13424",
    archivePrefix = "arXiv",
    primaryClass = "astro-ph.CO",
    doi = "10.1103/PhysRevD.100.083508",
    journal = "Phys. Rev. D",
    volume = "100",
    number = "8",
    pages = "083508",
    year = "2019"
}

@article{Kvasiuk_ksz_23,
    author = {Kvasiuk, Yurii and M{\"u}nchmeyer, Moritz},
    title = "{Autodifferentiable likelihood pipeline for the cross-correlation of CMB and large-scale structure due to the kinetic Sunyaev-Zeldovich effect}",
    eprint = "2305.08903",
    archivePrefix = "arXiv",
    primaryClass = "astro-ph.CO",
    doi = "10.1103/PhysRevD.109.083515",
    journal = "Phys. Rev. D",
    volume = "109",
    number = "8",
    pages = "083515",
    year = "2024"
}

@article{Ried_ksz_25,
    author = "Ried Guachalla, Bernardita and others",
    title = "{Backlighting extended gas halos around luminous red galaxies: Kinematic Sunyaev-Zel{\textquoteright}dovich effect from DESI Y1 and ACT data}",
    eprint = "2503.19870",
    archivePrefix = "arXiv",
    primaryClass = "astro-ph.GA",
    reportNumber = "FERMILAB-PUB-25-0196-PPD",
    doi = "10.1103/lqbj-wcqj",
    journal = "Phys. Rev. D",
    volume = "112",
    number = "10",
    pages = "103512",
    year = "2025"
}

@article{McCarthy_ksz_2024,
    author = "McCarthy, Fiona and others",
    title = "{The Atacama Cosmology Telescope: Large-scale velocity reconstruction with the kinematic Sunyaev-Zel'dovich effect and DESI LRGs}",
    eprint = "2410.06229",
    archivePrefix = "arXiv",
    primaryClass = "astro-ph.CO",
    doi = "10.1088/1475-7516/2025/05/057",
    journal = "JCAP",
    volume = "05",
    pages = "057",
    year = "2025"
}

@article{Hadzhiyska_ksz_baryon_24,
    author = "Hadzhiyska, B. and others",
    title = "{Evidence for large baryonic feedback at low and intermediate redshifts from kinematic Sunyaev-Zel{\textquoteright}dovich observations with ACT and DESI photometric galaxies}",
    eprint = "2407.07152",
    archivePrefix = "arXiv",
    primaryClass = "astro-ph.CO",
    reportNumber = "FERMILAB-PUB-24-0383-PPD",
    doi = "10.1103/kclp-x5j1",
    journal = "Phys. Rev. D",
    volume = "112",
    number = "8",
    pages = "083509",
    year = "2025"
}

@article{Gong:2025ffw,
    author = "Gong, Yulin and others",
    title = "{Detection of the pairwise kinematic Sunyaev-Zel{\textquoteright}dovich effect and pairwise velocity with DESI DR1 galaxies and ACT DR6 and Planck CMB data}",
    eprint = "2511.23417",
    archivePrefix = "arXiv",
    primaryClass = "astro-ph.CO",
    reportNumber = "FERMILAB-PUB-25-0868-PPD",
    doi = "10.1103/kkrh-1svk",
    journal = "Phys. Rev. D",
    volume = "113",
    number = "6",
    pages = "063538",
    year = "2026"
}

@article{Hadzhiyska:2025egz,
    author = "Hadzhiyska, B. and others",
    title = "{Probing cosmic velocities with the pairwise kinematic Sunyaev-Zel'dovich signal in DESI Bright Galaxy Sample DR1 and ACT DR6}",
    eprint = "2510.14135",
    archivePrefix = "arXiv",
    primaryClass = "astro-ph.CO",
    reportNumber = "FERMILAB-PUB-25-0758-PPD",
    doi = "10.5281/zenodo.17307202",
    month = "10",
    year = "2025"
}

@article{Deutsch_ksz_17,
    author = {Deutsch, Anne-Sylvie and Dimastrogiovanni, Emanuela and Johnson, Matthew C. and M{\"u}nchmeyer, Moritz and Terrana, Alexandra},
    title = "{Reconstruction of the remote dipole and quadrupole fields from the kinetic Sunyaev Zel{\textquoteright}dovich and polarized Sunyaev Zel{\textquoteright}dovich effects}",
    eprint = "1707.08129",
    archivePrefix = "arXiv",
    primaryClass = "astro-ph.CO",
    doi = "10.1103/PhysRevD.98.123501",
    journal = "Phys. Rev. D",
    volume = "98",
    number = "12",
    pages = "123501",
    year = "2018"
}

@article{Kamionkowski:1997na,
    author = "Kamionkowski, Marc and Loeb, Abraham",
    title = "{Getting around cosmic variance}",
    eprint = "astro-ph/9703118",
    archivePrefix = "arXiv",
    reportNumber = "CU-TP-822, CAL-632, CFA-4578",
    doi = "10.1103/PhysRevD.56.4511",
    journal = "Phys. Rev. D",
    volume = "56",
    pages = "4511--4513",
    year = "1997"
}

@article{Giri:2020pkk,
    author = "Giri, Utkarsh and Smith, Kendrick M.",
    title = "{Exploring KSZ velocity reconstruction with N-body simulations and the halo~model}",
    eprint = "2010.07193",
    archivePrefix = "arXiv",
    primaryClass = "astro-ph.CO",
    doi = "10.1088/1475-7516/2022/09/028",
    journal = "JCAP",
    volume = "09",
    pages = "028",
    year = "2022"
}

@ARTICLE{2024PhRvD.109f3530C,
       author = {{Coulton}, William and {Madhavacheril}, Mathew S. and {Duivenvoorden}, Adriaan J. and {Hill}, J. Colin and {Abril-Cabezas}, Irene and {Ade}, Peter A.~R. and {Aiola}, Simone and {Alford}, Tommy and {Amiri}, Mandana and {Amodeo}, Stefania and {An}, Rui and {Atkins}, Zachary and {Austermann}, Jason E. and {Battaglia}, Nicholas and {Battistelli}, Elia Stefano and {Beall}, James A. and {Bean}, Rachel and {Beringue}, Benjamin and {Bhandarkar}, Tanay and {Biermann}, Emily and {Bolliet}, Boris and {Bond}, J. Richard and {Cai}, Hongbo and {Calabrese}, Erminia and {Calafut}, Victoria and {Capalbo}, Valentina and {Carrero}, Felipe and {Chesmore}, Grace E. and {Cho}, Hsiao-mei and {Choi}, Steve K. and {Clark}, Susan E. and {Rosado}, Rodrigo C{\'o}rdova and {Cothard}, Nicholas F. and {Coughlin}, Kevin and {Crowley}, Kevin T. and {Devlin}, Mark J. and {Dicker}, Simon and {Doze}, Peter and {Duell}, Cody J. and {Duff}, Shannon M. and {Dunkley}, Jo and {D{\"u}nner}, Rolando and {Fanfani}, Valentina and {Fankhanel}, Max and {Farren}, Gerrit and {Ferraro}, Simone and {Freundt}, Rodrigo and {Fuzia}, Brittany and {Gallardo}, Patricio A. and {Garrido}, Xavier and {Givans}, Jahmour and {Gluscevic}, Vera and {Golec}, Joseph E. and {Guan}, Yilun and {Halpern}, Mark and {Han}, Dongwon and {Hasselfield}, Matthew and {Healy}, Erin and {Henderson}, Shawn and {Hensley}, Brandon and {Herv{\'\i}as-Caimapo}, Carlos and {Hilton}, Gene C. and {Hilton}, Matt and {Hincks}, Adam D. and {Hlo{\v{z}}ek}, Ren{\'e}e and {Ho}, Shuay-Pwu Patty and {Huber}, Zachary B. and {Hubmayr}, Johannes and {Huffenberger}, Kevin M. and {Hughes}, John P. and {Irwin}, Kent and {Isopi}, Giovanni and {Jense}, Hidde T. and {Keller}, Ben and {Kim}, Joshua and {Knowles}, Kenda and {Koopman}, Brian J. and {Kosowsky}, Arthur and {Kramer}, Darby and {Kusiak}, Aleksandra and {La Posta}, Adrien and {Lakey}, Victoria and {Lee}, Eunseong and {Li}, Zack and {Li}, Yaqiong and {Limon}, Michele and {Lokken}, Martine and {Louis}, Thibaut and {Lungu}, Marius and {MacCrann}, Niall and {MacInnis}, Amanda and {Maldonado}, Diego and {Maldonado}, Felipe and {Mallaby-Kay}, Maya and {Marques}, Gabriela A. and {van Marrewijk}, Joshiwa and {McCarthy}, Fiona and {McMahon}, Jeff and {Mehta}, Yogesh and {Menanteau}, Felipe and {Moodley}, Kavilan and {Morris}, Thomas W. and {Mroczkowski}, Tony and {Naess}, Sigurd and {Namikawa}, Toshiya and {Nati}, Federico and {Newburgh}, Laura and {Nicola}, Andrina and {Niemack}, Michael D. and {Nolta}, Michael R. and {Orlowski-Scherer}, John and {Page}, Lyman A. and {Pandey}, Shivam and {Partridge}, Bruce and {Prince}, Heather and {Puddu}, Roberto and {Qu}, Frank J. and {Radiconi}, Federico and {Robertson}, Naomi and {Rojas}, Felipe and {Sakuma}, Tai and {Salatino}, Maria and {Schaan}, Emmanuel and {Schmitt}, Benjamin L. and {Sehgal}, Neelima and {Shaikh}, Shabbir and {Sherwin}, Blake D. and {Sierra}, Carlos and {Sievers}, Jon and {Sif{\'o}n}, Crist{\'o}bal and {Simon}, Sara and {Sonka}, Rita and {Spergel}, David N. and {Staggs}, Suzanne T. and {Storer}, Emilie and {Switzer}, Eric R. and {Tampier}, Niklas and {Thornton}, Robert and {Trac}, Hy and {Treu}, Jesse and {Tucker}, Carole and {Ullom}, Joel and {Vale}, Leila R. and {Van Engelen}, Alexander and {Van Lanen}, Jeff and {Vargas}, Cristian and {Vavagiakis}, Eve M. and {Wagoner}, Kasey and {Wang}, Yuhan and {Wenzl}, Lukas and {Wollack}, Edward J. and {Xu}, Zhilei and {Zago}, Fernando and {Zheng}, Kaiwen},
        title = "{Atacama Cosmology Telescope: High-resolution component-separated maps across one third of the sky}",
      journal = {\prd},
         year = 2024,
        month = mar,
       volume = {109},
       number = {6},
          eid = {063530},
        pages = {063530},
          doi = {10.1103/PhysRevD.109.063530},
archivePrefix = {arXiv},
       eprint = {2307.01258},
 primaryClass = {astro-ph.CO},
       adsurl = {https://ui.adsabs.harvard.edu/abs/2024PhRvD.109f3530C}
}

@ARTICLE{2025JCAP...11..061N,
       author = {{Naess}, Sigurd and {Guan}, Yilun and {Duivenvoorden}, Adriaan J. and {Hasselfield}, Matthew and {Wang}, Yuhan and {Abril-Cabezas}, Irene and {Addison}, Graeme E. and {Ade}, Peter A.~R. and {Aiola}, Simone and {Alford}, Tommy and et al.},
        title = "{The Atacama Cosmology Telescope: DR6 maps}",
      journal = {\jcap},
         year = 2025,
        month = nov,
       volume = {2025},
       number = {11},
          eid = {061},
        pages = {061},
          doi = {10.1088/1475-7516/2025/11/061},
archivePrefix = {arXiv},
       eprint = {2503.14451},
 primaryClass = {astro-ph.CO},
       adsurl = {https://ui.adsabs.harvard.edu/abs/2025JCAP...11..061N}
}

@ARTICLE{2023JCAP...02..051C,
       author = {{Cayuso}, Juan and {Bloch}, Richard and {Hotinli}, Selim C. and {Johnson}, Matthew C. and {McCarthy}, Fiona},
        title = "{Velocity reconstruction with the cosmic microwave background and galaxy surveys}",
      journal = {\jcap},
         year = 2023,
        month = feb,
       volume = {2023},
       number = {2},
          eid = {051},
        pages = {051},
          doi = {10.1088/1475-7516/2023/02/051},
archivePrefix = {arXiv},
       eprint = {2111.11526},
 primaryClass = {astro-ph.CO},
       adsurl = {https://ui.adsabs.harvard.edu/abs/2023JCAP...02..051C}
}

@article{Cayuso:2019hen,
    author = "Cayuso, Juan I. and Johnson, Matthew C.",
    title = "{Towards testing CMB anomalies using the kinetic and polarized Sunyaev-Zel{\textquoteright}dovich effects}",
    eprint = "1904.10981",
    archivePrefix = "arXiv",
    primaryClass = "astro-ph.CO",
    doi = "10.1103/PhysRevD.101.123508",
    journal = "Phys. Rev. D",
    volume = "101",
    number = "12",
    pages = "123508",
    year = "2020"
}

@article{AnilKumar:2022flx,
    author = "Anil Kumar, Neha and Sato-Polito, Gabriela and Kamionkowski, Marc and Hotinli, Selim C.",
    title = "{Primordial trispectrum from kinetic Sunyaev-Zel{\textquoteright}dovich tomography}",
    eprint = "2205.03423",
    archivePrefix = "arXiv",
    primaryClass = "astro-ph.CO",
    doi = "10.1103/PhysRevD.106.063533",
    journal = "Phys. Rev. D",
    volume = "106",
    number = "6",
    pages = "063533",
    year = "2022"
}

@article{Kumar:2022bly,
    author = "Kumar, Neha Anil and Hotinli, Selim C. and Kamionkowski, Marc",
    title = "{Uncorrelated compensated isocurvature perturbations from kinetic Sunyaev-Zeldovich tomography}",
    eprint = "2208.02829",
    archivePrefix = "arXiv",
    primaryClass = "astro-ph.CO",
    doi = "10.1103/PhysRevD.107.043504",
    journal = "Phys. Rev. D",
    volume = "107",
    number = "4",
    pages = "043504",
    year = "2023"
}

@article{Hotinli:2019wdp,
    author = "Hotinli, Selim C. and Mertens, James B. and Johnson, Matthew C. and Kamionkowski, Marc",
    title = "{Probing correlated compensated isocurvature perturbations using scale-dependent galaxy bias}",
    eprint = "1908.08953",
    archivePrefix = "arXiv",
    primaryClass = "astro-ph.CO",
    doi = "10.1103/PhysRevD.100.103528",
    journal = "Phys. Rev. D",
    volume = "100",
    number = "10",
    pages = "103528",
    year = "2019"
}

@article{Adolff:2025fli,
    author = "Adolff, Julius and Hotinli, Selim and Dalal, Neal",
    title = "{Probing Dark Energy Microphysics with kSZ Tomography}",
    eprint = "2511.05653",
    archivePrefix = "arXiv",
    primaryClass = "astro-ph.CO",
    month = "11",
    year = "2025"
}

@article{Patki:2025oyp,
    author = "Patki, Raagini and Battaglia, Nicholas and Bean, Rachel",
    title = "{Probing gravity at large scales with kinematic Sunyaev-Zel{\textquoteright}dovich-reconstructed velocities and CMB lensing}",
    eprint = "2510.27605",
    archivePrefix = "arXiv",
    primaryClass = "astro-ph.CO",
    doi = "10.1103/5d6p-4bys",
    journal = "Phys. Rev. D",
    volume = "113",
    number = "2",
    pages = "023527",
    year = "2026"
}

@article{Adshead:2024paa,
    author = "Adshead, Peter and Tishue, Avery J.",
    title = "{Probing beyond local-type non-Gaussianity with kinematic Sunyaev-Zeldovich tomography}",
    eprint = "2407.21094",
    archivePrefix = "arXiv",
    primaryClass = "astro-ph.CO",
    doi = "10.1103/PhysRevD.110.103549",
    journal = "Phys. Rev. D",
    volume = "110",
    number = "10",
    pages = "103549",
    year = "2024"
}

@article{Bloch:2024kpn,
    author = "Bloch, Richard and Johnson, Matthew C.",
    title = "{Kinetic Sunyaev Zel'dovich velocity reconstruction from Planck and unWISE}",
    eprint = "2405.00809",
    archivePrefix = "arXiv",
    primaryClass = "astro-ph.CO",
    month = "5",
    year = "2024"
}

@ARTICLE{2023JCAP...11..097Z,
       author = {{Zhou}, Rongpu and {Ferraro}, Simone and {White}, Martin and {DeRose}, Joseph and {Sailer}, Noah and {Aguilar}, Jessica and {Ahlen}, Steven and {Bailey}, Stephen and {Brooks}, David and {Claybaugh}, Todd and {Dawson}, Kyle and {de la Macorra}, Axel and {Dey}, Biprateep and {Doel}, Peter and {Font-Ribera}, Andreu and {Forero-Romero}, Jaime E. and {Gontcho A Gontcho}, Satya and {Guy}, Julien and {Kremin}, Anthony and {Lambert}, Andrew and {Le Guillou}, Laurent and {Levi}, Michael and {Magneville}, Christophe and {Manera}, Marc and {Meisner}, Aaron and {Miquel}, Ramon and {Moustakas}, John and {Myers}, Adam D. and {Newman}, Jeffrey A. and {Nie}, Jundan and {Percival}, Will and {Rezaie}, Mehdi and {Rossi}, Graziano and {Sanchez}, Eusebio and {Schlegel}, David and {Schubnell}, Michael and {Seo}, Hee-Jong and {Tarl{\'e}}, Gregory and {Zhou}, Zhimin},
        title = "{DESI luminous red galaxy samples for cross-correlations}",
      journal = {\jcap},
         year = 2023,
        month = nov,
       volume = {2023},
       number = {11},
          eid = {097},
        pages = {097},
          doi = {10.1088/1475-7516/2023/11/097},
archivePrefix = {arXiv},
       eprint = {2309.06443},
 primaryClass = {astro-ph.CO},
       adsurl = {https://ui.adsabs.harvard.edu/abs/2023JCAP...11..097Z}
}

@article{thesimonsobservatoryacollaboration2025simonsobservatorysciencegoals,
    author = "Abitbol, M. and others",
    collaboration = "Simons Observatory",
    title = "{The Simons Observatory: science goals and forecasts for the enhanced Large Aperture Telescope}",
    eprint = "2503.00636",
    archivePrefix = "arXiv",
    primaryClass = "astro-ph.IM",
    reportNumber = "FERMILAB-PUB-25-0188-PPD",
    doi = "10.1088/1475-7516/2025/08/034",
    journal = "JCAP",
    volume = "08",
    pages = "034",
    year = "2025"
}

@article{sehgal2019cmbhdultradeephighresolutionmillimeterwave,
    author = "Sehgal, Neelima and others",
    title = "{CMB-HD: An Ultra-Deep, High-Resolution Millimeter-Wave Survey Over Half the Sky}",
    eprint = "1906.10134",
    archivePrefix = "arXiv",
    primaryClass = "astro-ph.CO",
    journal = "Bull. Am. Astron. Soc.",
    volume = "51",
    number = "7",
    pages = "1--23",
    year = "2019"
}

@ARTICLE{2011PASP..123..568C,
       author = {{Carlstrom}, J.~E. and {Ade}, P.~A.~R. and {Aird}, K.~A. and {Benson}, B.~A. and {Bleem}, L.~E. and {Busetti}, S. and {Chang}, C.~L. and {Chauvin}, E. and {Cho}, H.-M. and {Crawford}, T.~M. and {Crites}, A.~T. and {Dobbs}, M.~A. and {Halverson}, N.~W. and {Heimsath}, S. and {Holzapfel}, W.~L. and {Hrubes}, J.~D. and {Joy}, M. and {Keisler}, R. and {Lanting}, T.~M. and {Lee}, A.~T. and {Leitch}, E.~M. and {Leong}, J. and {Lu}, W. and {Lueker}, M. and {Luong-Van}, D. and {McMahon}, J.~J. and {Mehl}, J. and {Meyer}, S.~S. and {Mohr}, J.~J. and {Montroy}, T.~E. and {Padin}, S. and {Plagge}, T. and {Pryke}, C. and {Ruhl}, J.~E. and {Schaffer}, K.~K. and {Schwan}, D. and {Shirokoff}, E. and {Spieler}, H.~G. and {Staniszewski}, Z. and {Stark}, A.~A. and {Tucker}, C. and {Vanderlinde}, K. and {Vieira}, J.~D. and {Williamson}, R.},
        title = "{The 10 Meter South Pole Telescope}",
      journal = {\pasp},
         year = 2011,
        month = may,
       volume = {123},
       number = {903},
        pages = {568},
          doi = {10.1086/659879},
archivePrefix = {arXiv},
       eprint = {0907.4445},
 primaryClass = {astro-ph.IM},
       adsurl = {https://ui.adsabs.harvard.edu/abs/2011PASP..123..568C}
}

@ARTICLE{2011ApJS..194...41S,
       author = {{Swetz}, D.~S. and {Ade}, P.~A.~R. and {Amiri}, M. and {Appel}, J.~W. and {Battistelli}, E.~S. and {Burger}, B. and {Chervenak}, J. and {Devlin}, M.~J. and {Dicker}, S.~R. and {Doriese}, W.~B. and {D{\"u}nner}, R. and {Essinger-Hileman}, T. and {Fisher}, R.~P. and {Fowler}, J.~W. and {Halpern}, M. and {Hasselfield}, M. and {Hilton}, G.~C. and {Hincks}, A.~D. and {Irwin}, K.~D. and {Jarosik}, N. and {Kaul}, M. and {Klein}, J. and {Lau}, J.~M. and {Limon}, M. and {Marriage}, T.~A. and {Marsden}, D. and {Martocci}, K. and {Mauskopf}, P. and {Moseley}, H. and {Netterfield}, C.~B. and {Niemack}, M.~D. and {Nolta}, M.~R. and {Page}, L.~A. and {Parker}, L. and {Staggs}, S.~T. and {Stryzak}, O. and {Switzer}, E.~R. and {Thornton}, R. and {Tucker}, C. and {Wollack}, E. and {Zhao}, Y.},
        title = "{Overview of the Atacama Cosmology Telescope: Receiver, Instrumentation, and Telescope Systems}",
      journal = {\apjs},
         year = 2011,
        month = jun,
       volume = {194},
       number = {2},
          eid = {41},
        pages = {41},
          doi = {10.1088/0067-0049/194/2/41},
archivePrefix = {arXiv},
       eprint = {1007.0290},
 primaryClass = {astro-ph.IM},
       adsurl = {https://ui.adsabs.harvard.edu/abs/2011ApJS..194...41S}
}

@article{ACTPol:2015teu,
    author = "Schaan, Emmanuel and others",
    collaboration = "ACTPol",
    title = "{Evidence for the kinematic Sunyaev-Zel{\textquoteright}dovich effect with the Atacama Cosmology Telescope and velocity reconstruction from the Baryon Oscillation Spectroscopic Survey}",
    eprint = "1510.06442",
    archivePrefix = "arXiv",
    primaryClass = "astro-ph.CO",
    doi = "10.1103/PhysRevD.93.082002",
    journal = "Phys. Rev. D",
    volume = "93",
    number = "8",
    pages = "082002",
    year = "2016"
}

@article{sunyaev,
	author = {Sunyaev, R. A. and Zeldovich, Ya. B.},
	date = {1970/04/01},
	doi = {10.1007/BF00653471},
	id = {Sunyaev1970},
	isbn = {1572-946X},
	journal = {Astrophysics and Space Science},
	number = {1},
	pages = {3--19},
	title = {Small-scale fluctuations of relic radiation},
	url = {https://doi.org/10.1007/BF00653471},
	volume = {7},
	year = {1970}}

@article{Chiang:2025ujk,
    author = "Chiang, Yi-Kuan and Makiya, Ryu and M{\'e}nard, Brice",
    title = "{Cosmic Infrared Background Tomography and a Census of Cosmic Dust and Star Formation}",
    eprint = "2504.05384",
    archivePrefix = "arXiv",
    primaryClass = "astro-ph.CO",
    doi = "10.3847/1538-4357/adfb6a",
    journal = "Astrophys. J.",
    volume = "992",
    number = "1",
    pages = "65",
    year = "2025"
}

@article{Jego:2022eqo,
    author = "Jego, Baptiste and Ruiz-Zapatero, Jaime and Garc{\'\i}a-Garc{\'\i}a, Carlos and Koukoufilippas, Nick and Alonso, David",
    title = "{The star-formation history in the last 10 billion years from CIB cross-correlations}",
    eprint = "2206.15394",
    archivePrefix = "arXiv",
    primaryClass = "astro-ph.GA",
    doi = "10.1093/mnras/stad213",
    journal = "Mon. Not. Roy. Astron. Soc.",
    volume = "520",
    number = "2",
    pages = "1895--1912",
    year = "2023"
}

@article{zeldovich,
	author = {Zeldovich, Ya. B. and Sunyaev, R. A.},
	date = {1969/07/01},
	doi = {10.1007/BF00661821},
	id = {Zeldovich1969},
	isbn = {1572-946X},
	journal = {Astrophysics and Space Science},
	number = {3},
	pages = {301--316},
	title = {The interaction of matter and radiation in a hot-model universe},
	url = {https://doi.org/10.1007/BF00661821},
	volume = {4},
	year = {1969}}

@article{Hotinli:2025tul,
    author = "Hotinli, Selim C. and Smith, Kendrick M. and Ferraro, Simone",
    title = "{Velocity Reconstruction from KSZ: Measuring $f_{NL}$ with ACT and DESILS}",
    eprint = "2506.21657",
    archivePrefix = "arXiv",
    primaryClass = "astro-ph.CO",
    month = "6",
    year = "2025"
}

@article{Hadzhiyska:2025mvt,
    author = "Hadzhiyska, Boryana and Ferraro, Simone and Farren, Gerrit S. and Sailer, Noah and Zhou, Rongpu",
    title = "{Missing baryons recovered: A measurement of the gas fraction in galaxies and groups with the kinematic Sunyaev-Zel{\textquoteright}dovich effect and CMB lensing}",
    eprint = "2507.14136",
    archivePrefix = "arXiv",
    primaryClass = "astro-ph.CO",
    doi = "10.1103/mdhz-fgj8",
    journal = "Phys. Rev. D",
    volume = "112",
    number = "12",
    pages = "123507",
    year = "2025"
}

@article{krywonos2024constraintscosmologylambdacdmkinetic,
    author = "Krywonos, Jordan and Hotinli, Selim C. and Johnson, Matthew C.",
    title = "{Constraints on cosmology beyond $\Lambda$CDM with kinetic Sunyaev Zel'dovich velocity reconstruction}",
    eprint = "2408.05264",
    archivePrefix = "arXiv",
    primaryClass = "astro-ph.CO",
    month = "8",
    year = "2024"
}

@ARTICLE{2017JCAP...02..040T,
       author = {{Terrana}, Alexandra and {Harris}, Mary-Jean and {Johnson}, Matthew C.},
        title = "{Analyzing the cosmic variance limit of remote dipole measurements of the cosmic microwave background using the large-scale kinetic Sunyaev Zel'dovich effect}",
      journal = {\jcap},
         year = 2017,
        month = feb,
       volume = {2017},
       number = {2},
          eid = {040},
        pages = {040},
          doi = {10.1088/1475-7516/2017/02/040},
archivePrefix = {arXiv},
       eprint = {1610.06919},
 primaryClass = {astro-ph.CO},
       adsurl = {https://ui.adsabs.harvard.edu/abs/2017JCAP...02..040T}
}

@article{zhang_v,
   title={The dark flow induced small-scale kinetic Sunyaev–Zel’dovich effect},
   volume={407},
   ISSN={1745-3933},
   url={http://dx.doi.org/10.1111/j.1745-3933.2010.00899.x},
   DOI={10.1111/j.1745-3933.2010.00899.x},
   number={1},
   journal={Monthly Notices of the Royal Astronomical Society: Letters},
   publisher={Oxford University Press (OUP)},
   author={Zhang, Pengjie},
   year={2010},
   month=sep, pages={L36–L40} }

@ARTICLE{Hand2012,
       author = {{Hand}, Nick and {Addison}, Graeme E. and {Aubourg}, Eric and {Battaglia}, Nick and {Battistelli}, Elia S. and {Bizyaev}, Dmitry and {Bond}, J. Richard and {Brewington}, Howard and {Brinkmann}, Jon and {Brown}, Benjamin R. and {Das}, Sudeep and {Dawson}, Kyle S. and {Devlin}, Mark J. and {Dunkley}, Joanna and {Dunner}, Rolando and {Eisenstein}, Daniel J. and {Fowler}, Joseph W. and {Gralla}, Megan B. and {Hajian}, Amir and {Halpern}, Mark and {Hilton}, Matt and {Hincks}, Adam D. and {Hlozek}, Ren{\'e}e and {Hughes}, John P. and {Infante}, Leopoldo and {Irwin}, Kent D. and {Kosowsky}, Arthur and {Lin}, Yen-Ting and {Malanushenko}, Elena and {Malanushenko}, Viktor and {Marriage}, Tobias A. and {Marsden}, Danica and {Menanteau}, Felipe and {Moodley}, Kavilan and {Niemack}, Michael D. and {Nolta}, Michael R. and {Oravetz}, Daniel and {Page}, Lyman A. and {Palanque-Delabrouille}, Nathalie and {Pan}, Kaike and {Reese}, Erik D. and {Schlegel}, David J. and {Schneider}, Donald P. and {Sehgal}, Neelima and {Shelden}, Alaina and {Sievers}, Jon and {Sif{\'o}n}, Crist{\'o}bal and {Simmons}, Audrey and {Snedden}, Stephanie and {Spergel}, David N. and {Staggs}, Suzanne T. and {Swetz}, Daniel S. and {Switzer}, Eric R. and {Trac}, Hy and {Weaver}, Benjamin A. and {Wollack}, Edward J. and {Yeche}, Christophe and {Zunckel}, Caroline},
        title = "{Evidence of Galaxy Cluster Motions with the Kinematic Sunyaev-Zel'dovich Effect}",
      journal = {\prl},
         year = 2012,
        month = jul,
       volume = {109},
       number = {4},
          eid = {041101},
        pages = {041101},
          doi = {10.1103/PhysRevLett.109.041101},
archivePrefix = {arXiv},
       eprint = {1203.4219},
 primaryClass = {astro-ph.CO},
       adsurl = {https://ui.adsabs.harvard.edu/abs/2012PhRvL.109d1101H}
}

@article{Planck2013_kSZ,
    author = "Ade, P. A. R. and others",
    collaboration = "Planck",
    title = "{Planck intermediate results. XIII. Constraints on peculiar velocities}",
    eprint = "1303.5090",
    archivePrefix = "arXiv",
    primaryClass = "astro-ph.CO",
    doi = "10.1051/0004-6361/201321299",
    journal = "Astron. Astrophys.",
    volume = "561",
    pages = "A97",
    year = "2014"
}

@article{Soergel2016,
    author = "Soergel, B. and others",
    collaboration = "DES, SPT",
    title = "{Detection of the kinematic Sunyaev{\textendash}Zel'dovich effect with DES Year 1 and SPT}",
    eprint = "1603.03904",
    archivePrefix = "arXiv",
    primaryClass = "astro-ph.CO",
    reportNumber = "DES-2015-0154, FERMILAB-PUB-16-036-AE",
    doi = "10.1093/mnras/stw1455",
    journal = "Mon. Not. Roy. Astron. Soc.",
    volume = "461",
    number = "3",
    pages = "3172--3193",
    year = "2016"
}

@article{Hill2016,
    author = "Hill, J. Colin and Ferraro, Simone and Battaglia, Nick and Liu, Jia and Spergel, David N.",
    title = "{Kinematic Sunyaev-Zel{\textquoteright}dovich Effect with Projected Fields: A Novel Probe of the Baryon Distribution with Planck, WMAP, and WISE Data}",
    eprint = "1603.01608",
    archivePrefix = "arXiv",
    primaryClass = "astro-ph.CO",
    doi = "10.1103/PhysRevLett.117.051301",
    journal = "Phys. Rev. Lett.",
    volume = "117",
    number = "5",
    pages = "051301",
    year = "2016"
}

@article{Kusiak2021,
    author = "Kusiak, Aleksandra and Bolliet, Boris and Ferraro, Simone and Hill, J. Colin and Krolewski, Alex",
    title = "{Constraining the baryon abundance with the kinematic Sunyaev-Zel{\textquoteright}dovich effect: Projected-field detection using Planck, WMAP, and unWISE}",
    eprint = "2102.01068",
    archivePrefix = "arXiv",
    primaryClass = "astro-ph.CO",
    doi = "10.1103/PhysRevD.104.043518",
    journal = "Phys. Rev. D",
    volume = "104",
    number = "4",
    pages = "043518",
    year = "2021"
}

@article{Tanimura2021,
    author = "Tanimura, Hideki and Zaroubi, Saleem and Aghanim, Nabila",
    title = "{Direct detection of the kinetic Sunyaev-Zel'dovich effect in galaxy clusters}",
    eprint = "2007.02952",
    archivePrefix = "arXiv",
    primaryClass = "astro-ph.CO",
    doi = "10.1051/0004-6361/202038846",
    journal = "Astron. Astrophys.",
    volume = "645",
    pages = "A112",
    year = "2021"
}

@article{Calafut2021,
    author = "Calafut, Victoria and others",
    title = "{The Atacama Cosmology Telescope: Detection of the pairwise kinematic Sunyaev-Zel{\textquoteright}dovich effect with SDSS DR15 galaxies}",
    eprint = "2101.08374",
    archivePrefix = "arXiv",
    primaryClass = "astro-ph.CO",
    doi = "10.1103/PhysRevD.104.043502",
    journal = "Phys. Rev. D",
    volume = "104",
    number = "4",
    pages = "043502",
    year = "2021"
}

@article{Lague2024,
    author = {Lagu{\"e}, Alex and Madhavacheril, Mathew S. and Smith, Kendrick M. and Ferraro, Simone and Schaan, Emmanuel},
    title = "{Constraints on Local Primordial Non-Gaussianity with 3D Velocity Reconstruction from the Kinetic Sunyaev-Zeldovich Effect}",
    eprint = "2411.08240",
    archivePrefix = "arXiv",
    primaryClass = "astro-ph.CO",
    doi = "10.1103/PhysRevLett.134.151003",
    journal = "Phys. Rev. Lett.",
    volume = "134",
    number = "15",
    pages = "151003",
    year = "2025"
}

@article{Roper2025,
    author = "Roper, Finn A. and Cai, Yan-Chuan and Peacock, John A.",
    title = "{Mass dependence of halo baryon fractions from the kinetic Sunyaev-Zeldovich effect}",
    eprint = "2510.12553",
    archivePrefix = "arXiv",
    primaryClass = "astro-ph.CO",
    month = "10",
    year = "2025"
}

@article{Seljak_2009,
   title={Extracting Primordial Non-Gaussianity without Cosmic Variance},
   volume={102},
   ISSN={1079-7114},
   url={http://dx.doi.org/10.1103/PhysRevLett.102.021302},
   DOI={10.1103/physrevlett.102.021302},
   number={2},
   journal={Physical Review Letters},
   publisher={American Physical Society (APS)},
   author={Seljak, Uroš},
   year={2009},
   month=jan }

@article{Dalal:2007cu,
    author = "Dalal, Neal and Dore, Olivier and Huterer, Dragan and Shirokov, Alexander",
    title = "{The imprints of primordial non-gaussianities on large-scale structure: scale dependent bias and abundance of virialized objects}",
    eprint = "0710.4560",
    archivePrefix = "arXiv",
    primaryClass = "astro-ph",
    doi = "10.1103/PhysRevD.77.123514",
    journal = "Phys. Rev. D",
    volume = "77",
    pages = "123514",
    year = "2008"
}

@article{baleato_2025,
    author = "Baleato Lizancos, A. and Coulton, W. and Challinor, A. and Sherwin, B. D. and Mehta, Y.",
    title = "{A halo model of extragalactic contamination to CMB lensing, delensing, and cross-correlations}",
    eprint = "2507.03859",
    archivePrefix = "arXiv",
    primaryClass = "astro-ph.CO",
    doi = "10.1088/1475-7516/2025/11/031",
    journal = "JCAP",
    volume = "11",
    pages = "031",
    year = "2025"
}

@article{Planck:2013wqd,
    author = "Ade, P. A. R. and others",
    collaboration = "Planck",
    title = "{Planck 2013 results. XXX. Cosmic infrared background measurements and implications for star formation}",
    eprint = "1309.0382",
    archivePrefix = "arXiv",
    primaryClass = "astro-ph.CO",
    doi = "10.1051/0004-6361/201322093",
    journal = "Astron. Astrophys.",
    volume = "571",
    pages = "A30",
    year = "2014"
}

@article{Crawford:2013uka,
    author = "Crawford, T. M. and others",
    title = "{A Measurement of the Secondary-CMB and Millimeter-wave-foreground Bispectrum using 800 deg$^2$ of South Pole Telescope Data}",
    eprint = "1303.3535",
    archivePrefix = "arXiv",
    primaryClass = "astro-ph.CO",
    doi = "10.1088/0004-637X/784/2/143",
    journal = "Astrophys. J.",
    volume = "784",
    pages = "143",
    year = "2014"
}

@article{Scoccimarro:2000sn,
    author = "Scoccimarro, Roman",
    title = "{The bispectrum: from theory to observations}",
    eprint = "astro-ph/0004086",
    archivePrefix = "arXiv",
    doi = "10.1086/317248",
    journal = "Astrophys. J.",
    volume = "544",
    pages = "597",
    year = "2000"
}

@ARTICLE{2021PhRvD.103f3513S,
       author = {{Schaan}, Emmanuel and {Ferraro}, Simone and {Amodeo}, Stefania and {Battaglia}, Nicholas and {Aiola}, Simone and {Austermann}, Jason E. and {Beall}, James A. and {Bean}, Rachel and {Becker}, Daniel T. and {Bond}, Richard J. and {Calabrese}, Erminia and {Calafut}, Victoria and {Choi}, Steve K. and {Denison}, Edward V. and {Devlin}, Mark J. and {Duff}, Shannon M. and {Duivenvoorden}, Adriaan J. and {Dunkley}, Jo and {D{\"u}nner}, Rolando and {Gallardo}, Patricio A. and {Guan}, Yilun and {Han}, Dongwon and {Hill}, J. Colin and {Hilton}, Gene C. and {Hilton}, Matt and {Hlo{\v{z}}ek}, Ren{\'e}e and {Hubmayr}, Johannes and {Huffenberger}, Kevin M. and {Hughes}, John P. and {Koopman}, Brian J. and {MacInnis}, Amanda and {McMahon}, Jeff and {Madhavacheril}, Mathew S. and {Moodley}, Kavilan and {Mroczkowski}, Tony and {Naess}, Sigurd and {Nati}, Federico and {Newburgh}, Laura B. and {Niemack}, Michael D. and {Page}, Lyman A. and {Partridge}, Bruce and {Salatino}, Maria and {Sehgal}, Neelima and {Schillaci}, Alessandro and {Sif{\'o}n}, Crist{\'o}bal and {Smith}, Kendrick M. and {Spergel}, David N. and {Staggs}, Suzanne and {Storer}, Emilie R. and {Trac}, Hy and {Ullom}, Joel N. and {Van Lanen}, Jeff and {Vale}, Leila R. and {van Engelen}, Alexander and {Maga{\~n}a}, Mariana Vargas and {Vavagiakis}, Eve M. and {Wollack}, Edward J. and {Xu}, Zhilei and {Atacama Cosmology Telescope Collaboration}},
        title = "{Atacama Cosmology Telescope: Combined kinematic and thermal Sunyaev-Zel'dovich measurements from BOSS CMASS and LOWZ halos}",
      journal = {\prd},
         year = 2021,
        month = mar,
       volume = {103},
       number = {6},
          eid = {063513},
        pages = {063513},
          doi = {10.1103/PhysRevD.103.063513},
archivePrefix = {arXiv},
       eprint = {2009.05557},
 primaryClass = {astro-ph.CO},
       adsurl = {https://ui.adsabs.harvard.edu/abs/2021PhRvD.103f3513S}
}

@article{Serra:2014pva,
    author = "Serra, Paolo and Lagache, Guilaine and Dor{\'e}, Olivier and Pullen, Anthony and White, Martin",
    title = "{Cross-correlation of cosmic far-infrared background anisotropies with large scale structures}",
    eprint = "1404.1933",
    archivePrefix = "arXiv",
    primaryClass = "astro-ph.CO",
    doi = "10.1051/0004-6361/201423958",
    journal = "Astron. Astrophys.",
    volume = "570",
    pages = "A98",
    year = "2014"
}

@article{Liu:2025zqo,
    author = "Liu, R. Henry and others",
    title = "{Measurements of the thermal Sunyaev-Zel{\textquoteright}dovich effect with ACT and DESI luminous red galaxies}",
    eprint = "2502.08850",
    archivePrefix = "arXiv",
    primaryClass = "astro-ph.CO",
    reportNumber = "FERMILAB-PUB-25-0094-PPD",
    doi = "10.1103/jqn8-19gx",
    journal = "Phys. Rev. D",
    volume = "112",
    number = "8",
    pages = "083561",
    year = "2025"
}

@article{LaPosta:2026cek,
    author = "La Posta, Adrien and Alonso, David and Garc{\'\i}a-Garc{\'\i}a, Carlos and Maleubre, Sara",
    title = "{Joint tomographic measurement of thermal Sunyaev Zeldovich and the cosmic infrared background}",
    eprint = "2603.04269",
    archivePrefix = "arXiv",
    primaryClass = "astro-ph.CO",
    month = "3",
    year = "2026"
}

@ARTICLE{1953ApJ...117..134L,
       author = {{Limber}, D. Nelson},
        title = "{The Analysis of Counts of the Extragalactic Nebulae in Terms of a Fluctuating Density Field.}",
      journal = {\apj},
         year = 1953,
        month = jan,
       volume = {117},
        pages = {134},
          doi = {10.1086/145672},
       adsurl = {https://ui.adsabs.harvard.edu/abs/1953ApJ...117..134L}
}

@article{Lemos:2017arq,
    author = "Lemos, Pablo and Challinor, Anthony and Efstathiou, George",
    title = "{The effect of Limber and flat-sky approximations on galaxy weak lensing}",
    eprint = "1704.01054",
    archivePrefix = "arXiv",
    primaryClass = "astro-ph.CO",
    doi = "10.1088/1475-7516/2017/05/014",
    journal = "JCAP",
    volume = "05",
    pages = "014",
    year = "2017"
}

@article{Barreira:2017sqa,
    author = "Barreira, Alexandre and Schmidt, Fabian",
    title = "{Responses in Large-Scale Structure}",
    eprint = "1703.09212",
    archivePrefix = "arXiv",
    primaryClass = "astro-ph.CO",
    doi = "10.1088/1475-7516/2017/06/053",
    journal = "JCAP",
    volume = "06",
    pages = "053",
    year = "2017"
}

@article{Barreira:2017fjz,
    author = "Barreira, Alexandre and Krause, Elisabeth and Schmidt, Fabian",
    title = "{Complete super-sample lensing covariance in the response approach}",
    eprint = "1711.07467",
    archivePrefix = "arXiv",
    primaryClass = "astro-ph.CO",
    doi = "10.1088/1475-7516/2018/06/015",
    journal = "JCAP",
    volume = "06",
    pages = "015",
    year = "2018"
}

@article{Planck:2018vyg,
    author = "Aghanim, N. and others",
    collaboration = "Planck",
    title = "{Planck 2018 results. VI. Cosmological parameters}",
    eprint = "1807.06209",
    archivePrefix = "arXiv",
    primaryClass = "astro-ph.CO",
    doi = "10.1051/0004-6361/201833910",
    journal = "Astron. Astrophys.",
    volume = "641",
    pages = "A6",
    year = "2020",
    note = "[Erratum: Astron.Astrophys. 652, C4 (2021)]"
}

@article{Maksimova_2021,
    author = "Maksimova, Nina A. and Garrison, Lehman H. and Eisenstein, Daniel J. and Hadzhiyska, Boryana and Bose, Sownak and Satterthwaite, Thomas P.",
    title = "{AbacusSummit: a massive set of high-accuracy, high-resolution N-body simulations}",
    eprint = "2110.11398",
    archivePrefix = "arXiv",
    primaryClass = "astro-ph.CO",
    doi = "10.1093/mnras/stab2484",
    journal = "Mon. Not. Roy. Astron. Soc.",
    volume = "508",
    number = "3",
    pages = "4017--4037",
    year = "2021"
}

@article{Hadzhiyska_2021,
    author = "Hadzhiyska, Boryana and Garrison, Lehman H. and Eisenstein, Daniel and Bose, Sownak",
    title = "{The halo light-cone catalogues of AbacusSummit}",
    eprint = "2110.11413",
    archivePrefix = "arXiv",
    primaryClass = "astro-ph.CO",
    doi = "10.1093/mnras/stab3066",
    journal = "Mon. Not. Roy. Astron. Soc.",
    volume = "509",
    number = "2",
    pages = "2194--2208",
    year = "2021"
}

@article{nelson2021illustristngsimulationspublicdata,
    author = "Nelson, Dylan and others",
    title = "{The IllustrisTNG simulations: public data release}",
    eprint = "1812.05609",
    archivePrefix = "arXiv",
    primaryClass = "astro-ph.GA",
    doi = "10.1186/s40668-019-0028-x",
    journal = "Comput. Astrophys. Cosmol.",
    volume = "6",
    number = "1",
    pages = "2",
    year = "2019"
}

@article{Shao:2010md,
    author = "Shao, Jiawei and Zhang, Pengjie and Lin, Weipeng and Jing, Yipeng and Pan, Jun",
    title = "{The kinetic SZ tomography with spectroscopic redshift surveys}",
    eprint = "1004.1301",
    archivePrefix = "arXiv",
    primaryClass = "astro-ph.CO",
    doi = "10.1111/j.1365-2966.2011.18166.x",
    journal = "Mon. Not. Roy. Astron. Soc.",
    volume = "413",
    pages = "628--642",
    year = "2011"
}

@ARTICLE{2009arXiv0903.2845H,
       author = {{Ho}, Shirley and {Dedeo}, Simon and {Spergel}, David},
        title = "{Finding the Missing Baryons Using CMB as a Backlight}",
      journal = {arXiv e-prints},
         year = 2009,
        month = mar,
          eid = {arXiv:0903.2845},
        pages = {arXiv:0903.2845},
          doi = {10.48550/arXiv.0903.2845},
archivePrefix = {arXiv},
       eprint = {0903.2845},
 primaryClass = {astro-ph.CO},
       adsurl = {https://ui.adsabs.harvard.edu/abs/2009arXiv0903.2845H}
}

@ARTICLE{2019JCAP...10..024C,
       author = {{Contreras}, Dagoberto and {Johnson}, Matthew C. and {Mertens}, James B.},
        title = "{Towards detection of relativistic effects in galaxy number counts using kSZ tomography}",
      journal = {\jcap},
         year = 2019,
        month = oct,
       volume = {2019},
       number = {10},
          eid = {024},
        pages = {024},
          doi = {10.1088/1475-7516/2019/10/024},
archivePrefix = {arXiv},
       eprint = {1904.10033},
 primaryClass = {astro-ph.CO},
       adsurl = {https://ui.adsabs.harvard.edu/abs/2019JCAP...10..024C}
}

@article{tishue2025kszopticaldepthdegeneracy,
    author = "Tishue, Avery J. and Shiveshwarkar, Charuhas and Holder, Gilbert",
    title = "{The kSZ optical depth degeneracy and future constraints on local primordial non-Gaussianity}",
    eprint = "2510.25821",
    archivePrefix = "arXiv",
    primaryClass = "astro-ph.CO",
    month = "10",
    year = "2025"
}

@article{Puget:1996fx,
    author = "Puget, J. L. and Abergel, A. and Bernard, J. P. and Boulanger, F. and Burton, W. B. and Desert, F. X. and Hartmann, D.",
    title = "{Tentative detection of a cosmic far - infrared background with COBE}",
    journal = "Astron. Astrophys.",
    volume = "308",
    pages = "L5",
    year = "1996"
}

@article{HOD,
doi = {10.1086/341469},
url = {https://doi.org/10.1086/341469},
year = {2002},
month = {aug},
publisher = {},
volume = {575},
number = {2},
pages = {587},
author = {Berlind, Andreas A. and Weinberg, David H.},
title = {The Halo Occupation Distribution: Toward an Empirical Determination of the Relation between Galaxies and Mass},
journal = {The Astrophysical Journal}
}

@ARTICLE{2012ApJ...758...75B,
       author = {{Battaglia}, N. and {Bond}, J.~R. and {Pfrommer}, C. and {Sievers}, J.~L.},
        title = "{On the Cluster Physics of Sunyaev-Zel'dovich and X-Ray Surveys. II. Deconstructing the Thermal SZ Power Spectrum}",
      journal = {\apj},
         year = 2012,
        month = oct,
       volume = {758},
       number = {2},
          eid = {75},
        pages = {75},
          doi = {10.1088/0004-637X/758/2/75},
archivePrefix = {arXiv},
       eprint = {1109.3711},
 primaryClass = {astro-ph.CO},
       adsurl = {https://ui.adsabs.harvard.edu/abs/2012ApJ...758...75B}
}

@ARTICLE{2021PhRvD.103f3514A,
       author = {{Amodeo}, Stefania and {Battaglia}, Nicholas and {Schaan}, Emmanuel and {Ferraro}, Simone and {Moser}, Emily and {Aiola}, Simone and {Austermann}, Jason E. and {Beall}, James A. and {Bean}, Rachel and {Becker}, Daniel T. and {Bond}, Richard J. and {Calabrese}, Erminia and {Calafut}, Victoria and {Choi}, Steve K. and {Denison}, Edward V. and {Devlin}, Mark and {Duff}, Shannon M. and {Duivenvoorden}, Adriaan J. and {Dunkley}, Jo and {D{\"u}nner}, Rolando and {Gallardo}, Patricio A. and {Hall}, Kirsten R. and {Han}, Dongwon and {Hill}, J. Colin and {Hilton}, Gene C. and {Hilton}, Matt and {Hlo{\v{z}}ek}, Ren{\'e}e and {Hubmayr}, Johannes and {Huffenberger}, Kevin M. and {Hughes}, John P. and {Koopman}, Brian J. and {MacInnis}, Amanda and {McMahon}, Jeff and {Madhavacheril}, Mathew S. and {Moodley}, Kavilan and {Mroczkowski}, Tony and {Naess}, Sigurd and {Nati}, Federico and {Newburgh}, Laura B. and {Niemack}, Michael D. and {Page}, Lyman A. and {Partridge}, Bruce and {Schillaci}, Alessandro and {Sehgal}, Neelima and {Sif{\'o}n}, Crist{\'o}bal and {Spergel}, David N. and {Staggs}, Suzanne and {Storer}, Emilie R. and {Ullom}, Joel N. and {Vale}, Leila R. and {van Engelen}, Alexander and {Van Lanen}, Jeff and {Vavagiakis}, Eve M. and {Wollack}, Edward J. and {Xu}, Zhilei},
        title = "{Atacama Cosmology Telescope: Modeling the gas thermodynamics in BOSS CMASS galaxies from kinematic and thermal Sunyaev-Zel'dovich measurements}",
      journal = {\prd},
         year = 2021,
        month = mar,
       volume = {103},
       number = {6},
          eid = {063514},
        pages = {063514},
          doi = {10.1103/PhysRevD.103.063514},
archivePrefix = {arXiv},
       eprint = {2009.05558},
 primaryClass = {astro-ph.CO},
       adsurl = {https://ui.adsabs.harvard.edu/abs/2021PhRvD.103f3514A}
}

@ARTICLE{2014ApJS..211...17A,
       author = {{Ahn}, Christopher P. and {Alexandroff}, Rachael and {Allende Prieto}, Carlos and {Anders}, Friedrich and {Anderson}, Scott F. and {Anderton}, Timothy and {Andrews}, Brett H. and {Aubourg}, {\'E}ric and {Bailey}, Stephen and {Bastien}, Fabienne A. and {Bautista}, Julian E. and {Beers}, Timothy C. and {Beifiori}, Alessandra and {Bender}, Chad F. and {Berlind}, Andreas A. and {Beutler}, Florian and {Bhardwaj}, Vaishali and {Bird}, Jonathan C. and {Bizyaev}, Dmitry and {Blake}, Cullen H. and {Blanton}, Michael R. and {Blomqvist}, Michael and {Bochanski}, John J. and {Bolton}, Adam S. and {Borde}, Arnaud and {Bovy}, Jo and {Shelden Bradley}, Alaina and {Brandt}, W.~N. and {Brauer}, Doroth{\'e}e and {Brinkmann}, J. and {Brownstein}, Joel R. and {Busca}, Nicol{\'a}s G. and {Carithers}, William and {Carlberg}, Joleen K. and {Carnero}, Aurelio R. and {Carr}, Michael A. and {Chiappini}, Cristina and {Chojnowski}, S. Drew and {Chuang}, Chia-Hsun and {Comparat}, Johan and {Crepp}, Justin R. and {Cristiani}, Stefano and {Croft}, Rupert A.~C. and {Cuesta}, Antonio J. and {Cunha}, Katia and {da Costa}, Luiz N. and {Dawson}, Kyle S. and {De Lee}, Nathan and {Dean}, Janice D.~R. and {Delubac}, Timoth{\'e}e and {Deshpande}, Rohit and {Dhital}, Saurav and {Ealet}, Anne and {Ebelke}, Garrett L. and {Edmondson}, Edward M. and {Eisenstein}, Daniel J. and {Epstein}, Courtney R. and {Escoffier}, Stephanie and {Esposito}, Massimiliano and {Evans}, Michael L. and {Fabbian}, D. and {Fan}, Xiaohui and {Favole}, Ginevra and {Femen{\'\i}a Castell{\'a}}, Bruno and {Fern{\'a}ndez Alvar}, Emma and {Feuillet}, Diane and {Filiz Ak}, Nurten and {Finley}, Hayley and {Fleming}, Scott W. and {Font-Ribera}, Andreu and {Frinchaboy}, Peter M. and {Galbraith-Frew}, J.~G. and {Garc{\'\i}a-Hern{\'a}ndez}, D.~A. and {Garc{\'\i}a P{\'e}rez}, Ana E. and {Ge}, Jian and {G{\'e}nova-Santos}, R. and {Gillespie}, Bruce A. and {Girardi}, L{\'e}o and {Gonz{\'a}lez Hern{\'a}ndez}, Jonay I. and {Gott}, III, J. Richard and {Gunn}, James E. and {Guo}, Hong and {Halverson}, Samuel and {Harding}, Paul and {Harris}, David W. and {Hasselquist}, Sten and {Hawley}, Suzanne L. and {Hayden}, Michael and {Hearty}, Frederick R. and {Herrero Dav{\'o}}, Artemio and {Ho}, Shirley and {Hogg}, David W. and {Holtzman}, Jon A. and {Honscheid}, Klaus and {Huehnerhoff}, Joseph and {Ivans}, Inese I. and {Jackson}, Kelly M. and {Jiang}, Peng and {Johnson}, Jennifer A. and {Kinemuchi}, K. and {Kirkby}, David and {Klaene}, Mark A. and {Kneib}, Jean-Paul and {Koesterke}, Lars and {Lan}, Ting-Wen and {Lang}, Dustin and {Le Goff}, Jean-Marc and {Leauthaud}, Alexie and {Lee}, Khee-Gan and {Lee}, Young Sun and {Long}, Daniel C. and {Loomis}, Craig P. and {Lucatello}, Sara and {Lupton}, Robert H. and {Ma}, Bo and {Mack}, III, Claude E. and {Mahadevan}, Suvrath and {Maia}, Marcio A.~G. and {Majewski}, Steven R. and {Malanushenko}, Elena and {Malanushenko}, Viktor and {Manchado}, A. and {Manera}, Marc and {Maraston}, Claudia and {Margala}, Daniel and {Martell}, Sarah L. and {Masters}, Karen L. and {McBride}, Cameron K. and {McGreer}, Ian D. and {McMahon}, Richard G. and {M{\'e}nard}, Brice and {M{\'e}sz{\'a}ros}, Sz. and {Miralda-Escud{\'e}}, Jordi and {Miyatake}, Hironao and {Montero-Dorta}, Antonio D. and {Montesano}, Francesco and {More}, Surhud and {Morrison}, Heather L. and {Muna}, Demitri and {Munn}, Jeffrey A. and {Myers}, Adam D. and {Nguyen}, Duy Cuong and {Nichol}, Robert C. and {Nidever}, David L. and {Noterdaeme}, Pasquier and {Nuza}, Sebasti{\'a}n E. and {O'Connell}, Julia E. and {O'Connell}, Robert W. and {O'Connell}, Ross and {Olmstead}, Matthew D. and {Oravetz}, Daniel J. and {Owen}, Russell and {Padmanabhan}, Nikhil and {Palanque-Delabrouille}, Nathalie and {Pan}, Kaike and {Parejko}, John K. and {Parihar}, Prachi and {P{\^a}ris}, Isabelle and {Pepper}, Joshua and {Percival}, Will J. and {P{\'e}rez-R{\`a}fols}, Ignasi and {Dotto Perottoni}, H{\'e}lio and {Petitjean}, Patrick and {Pieri}, Matthew M. and {Pinsonneault}, M.~H. and {Prada}, Francisco and {Price-Whelan}, Adrian M. and {Raddick}, M. Jordan and {Rahman}, Mubdi and {Rebolo}, Rafael and {Reid}, Beth A. and {Richards}, Jonathan C. and {Riffel}, Rog{\'e}rio and {Robin}, Annie C. and {Rocha-Pinto}, H.~J. and {Rockosi}, Constance M. and {Roe}, Natalie A. and {Ross}, Ashley J. and {Ross}, Nicholas P. and {Rossi}, Graziano and {Roy}, Arpita and {Rubi{\~n}o-Martin}, J.~A. and {Sabiu}, Cristiano G. and {S{\'a}nchez}, Ariel G. and {Santiago}, Bas{\'\i}lio and {Sayres}, Conor and {Schiavon}, Ricardo P. and {Schlegel}, David J. and {Schlesinger}, Katharine J. and {Schmidt}, Sarah J. and {Schneider}, Donald P. and {Schultheis}, Mathias and {Sellgren}, Kris and {Seo}, Hee-Jong and {Shen}, Yue and {Shetrone}, Matthew and {Shu}, Yiping and {Simmons}, Audrey E. and {Skrutskie}, M.~F. and {Slosar}, An{\v{z}}e},
        title = "{The Tenth Data Release of the Sloan Digital Sky Survey: First Spectroscopic Data from the SDSS-III Apache Point Observatory Galactic Evolution Experiment}",
      journal = {\apjs},
         year = 2014,
        month = apr,
       volume = {211},
       number = {2},
          eid = {17},
        pages = {17},
          doi = {10.1088/0067-0049/211/2/17},
archivePrefix = {arXiv},
       eprint = {1307.7735},
 primaryClass = {astro-ph.IM},
       adsurl = {https://ui.adsabs.harvard.edu/abs/2014ApJS..211...17A}
}

@ARTICLE{2002ARA&A..40..643C,
       author = {{Carlstrom}, John E. and {Holder}, Gilbert P. and {Reese}, Erik D.},
        title = "{Cosmology with the Sunyaev-Zel'dovich Effect}",
      journal = {\araa},
         year = 2002,
        month = jan,
       volume = {40},
        pages = {643-680},
          doi = {10.1146/annurev.astro.40.060401.093803},
archivePrefix = {arXiv},
       eprint = {astro-ph/0208192},
 primaryClass = {astro-ph},
       adsurl = {https://ui.adsabs.harvard.edu/abs/2002ARA&A..40..643C}
}

@ARTICLE{2013MNRAS.431..609N,
       author = {{Namikawa}, Toshiya and {Hanson}, Duncan and {Takahashi}, Ryuichi},
        title = "{Bias-hardened CMB lensing}",
      journal = {\mnras},
         year = 2013,
        month = may,
       volume = {431},
       number = {1},
        pages = {609-620},
          doi = {10.1093/mnras/stt195},
archivePrefix = {arXiv},
       eprint = {1209.0091},
 primaryClass = {astro-ph.CO},
       adsurl = {https://ui.adsabs.harvard.edu/abs/2013MNRAS.431..609N}
}

@ARTICLE{2014JCAP...03..024O,
       author = {{Osborne}, Stephen J. and {Hanson}, Duncan and {Dor{\'e}}, Olivier},
        title = "{Extragalactic foreground contamination in temperature-based CMB lens reconstruction}",
      journal = {\jcap},
         year = 2014,
        month = mar,
       volume = {2014},
       number = {3},
          eid = {024},
        pages = {024},
          doi = {10.1088/1475-7516/2014/03/024},
archivePrefix = {arXiv},
       eprint = {1310.7547},
 primaryClass = {astro-ph.CO},
       adsurl = {https://ui.adsabs.harvard.edu/abs/2014JCAP...03..024O}
}

@ARTICLE{2020PhRvD.102f3517S,
       author = {{Sailer}, Noah and {Schaan}, Emmanuel and {Ferraro}, Simone},
        title = "{Lower bias, lower noise CMB lensing with foreground-hardened estimators}",
      journal = {\prd},
         year = 2020,
        month = sep,
       volume = {102},
       number = {6},
          eid = {063517},
        pages = {063517},
          doi = {10.1103/PhysRevD.102.063517},
archivePrefix = {arXiv},
       eprint = {2007.04325},
 primaryClass = {astro-ph.CO},
       adsurl = {https://ui.adsabs.harvard.edu/abs/2020PhRvD.102f3517S}
}

@article{AtacamaCosmologyTelescope:2025vnj,
    author = "Naess, Sigurd and others",
    collaboration = "Atacama Cosmology Telescope",
    title = "{The Atacama Cosmology Telescope: DR6 maps}",
    eprint = "2503.14451",
    archivePrefix = "arXiv",
    primaryClass = "astro-ph.CO",
    reportNumber = "FERMILAB-PUB-25-0160-PPD",
    doi = "10.1088/1475-7516/2025/11/061",
    journal = "JCAP",
    volume = "11",
    pages = "061",
    year = "2025"
}

@article{AtacamaCosmologyTelescope:2025blo,
    author = "Louis, Thibaut and others",
    collaboration = "Atacama Cosmology Telescope",
    title = "{The Atacama Cosmology Telescope: DR6 power spectra, likelihoods and {\ensuremath{\Lambda}}CDM parameters}",
    eprint = "2503.14452",
    archivePrefix = "arXiv",
    primaryClass = "astro-ph.CO",
    reportNumber = "FERMILAB-PUB-25-0071-PPD",
    doi = "10.1088/1475-7516/2025/11/062",
    journal = "JCAP",
    volume = "11",
    pages = "062",
    year = "2025"
}

@article{AtacamaCosmologyTelescope:2025nti,
    author = "Calabrese, Erminia and others",
    collaboration = "Atacama Cosmology Telescope",
    title = "{The Atacama Cosmology Telescope: DR6 constraints on extended cosmological models}",
    eprint = "2503.14454",
    archivePrefix = "arXiv",
    primaryClass = "astro-ph.CO",
    reportNumber = "FERMILAB-PUB-25-0157-PPD",
    doi = "10.1088/1475-7516/2025/11/063",
    journal = "JCAP",
    volume = "11",
    pages = "063",
    year = "2025"
}

@ARTICLE{2023AJ....165...58Z,
       author = {{Zhou}, Rongpu and {Dey}, Biprateep and {Newman}, Jeffrey A. and {Eisenstein}, Daniel J. and {Dawson}, K. and {Bailey}, S. and {Berti}, A. and {Guy}, J. and {Lan}, Ting-Wen and {Zou}, H. and {Aguilar}, J. and {Ahlen}, S. and {Alam}, Shadab and {Brooks}, D. and {de la Macorra}, A. and {Dey}, A. and {Dhungana}, G. and {Fanning}, K. and {Font-Ribera}, A. and {Gontcho}, S. Gontcho A. and {Honscheid}, K. and {Ishak}, Mustapha and {Kisner}, T. and {Kov{\'a}cs}, A. and {Kremin}, A. and {Landriau}, M. and {Levi}, Michael E. and {Magneville}, C. and {Manera}, Marc and {Martini}, P. and {Meisner}, Aaron M. and {Miquel}, R. and {Moustakas}, J. and {Myers}, Adam D. and {Nie}, Jundan and {Palanque-Delabrouille}, N. and {Percival}, W.~J. and {Poppett}, C. and {Prada}, F. and {Raichoor}, A. and {Ross}, A.~J. and {Schlafly}, E. and {Schlegel}, D. and {Schubnell}, M. and {Tarl{\'e}}, Gregory and {Weaver}, B.~A. and {Wechsler}, R.~H. and {Y{\'e}che}, Christophe and {Zhou}, Zhimin},
        title = "{Target Selection and Validation of DESI Luminous Red Galaxies}",
      journal = {\aj},
         year = 2023,
        month = feb,
       volume = {165},
       number = {2},
          eid = {58},
        pages = {58},
          doi = {10.3847/1538-3881/aca5fb},
archivePrefix = {arXiv},
       eprint = {2208.08515},
 primaryClass = {astro-ph.CO},
       adsurl = {https://ui.adsabs.harvard.edu/abs/2023AJ....165...58Z}
}

@ARTICLE{2019AJ....157..168D,
       author = {{Dey}, Arjun and {Schlegel}, David J. and {Lang}, Dustin and {Blum}, Robert and {Burleigh}, Kaylan and {Fan}, Xiaohui and {Findlay}, Joseph R. and {Finkbeiner}, Doug and {Herrera}, David and {Juneau}, St{\'e}phanie and {Landriau}, Martin and {Levi}, Michael and {McGreer}, Ian and {Meisner}, Aaron and {Myers}, Adam D. and {Moustakas}, John and {Nugent}, Peter and {Patej}, Anna and {Schlafly}, Edward F. and {Walker}, Alistair R. and {Valdes}, Francisco and {Weaver}, Benjamin A. and {Y{\`e}che}, Christophe and {Zou}, Hu and {Zhou}, Xu and {Abareshi}, Behzad and {Abbott}, T.~M.~C. and {Abolfathi}, Bela and {Aguilera}, C. and {Alam}, Shadab and {Allen}, Lori and {Alvarez}, A. and {Annis}, James and {Ansarinejad}, Behzad and {Aubert}, Marie and {Beechert}, Jacqueline and {Bell}, Eric F. and {BenZvi}, Segev Y. and {Beutler}, Florian and {Bielby}, Richard M. and {Bolton}, Adam S. and {Brice{\~n}o}, C{\'e}sar and {Buckley-Geer}, Elizabeth J. and {Butler}, Karen and {Calamida}, Annalisa and {Carlberg}, Raymond G. and {Carter}, Paul and {Casas}, Ricard and {Castander}, Francisco J. and {Choi}, Yumi and {Comparat}, Johan and {Cukanovaite}, Elena and {Delubac}, Timoth{\'e}e and {DeVries}, Kaitlin and {Dey}, Sharmila and {Dhungana}, Govinda and {Dickinson}, Mark and {Ding}, Zhejie and {Donaldson}, John B. and {Duan}, Yutong and {Duckworth}, Christopher J. and {Eftekharzadeh}, Sarah and {Eisenstein}, Daniel J. and {Etourneau}, Thomas and {Fagrelius}, Parker A. and {Farihi}, Jay and {Fitzpatrick}, Mike and {Font-Ribera}, Andreu and {Fulmer}, Leah and {G{\"a}nsicke}, Boris T. and {Gaztanaga}, Enrique and {George}, Koshy and {Gerdes}, David W. and {Gontcho}, Satya Gontcho A. and {Gorgoni}, Claudio and {Green}, Gregory and {Guy}, Julien and {Harmer}, Diane and {Hernandez}, M. and {Honscheid}, Klaus and {Huang}, Lijuan Wendy and {James}, David J. and {Jannuzi}, Buell T. and {Jiang}, Linhua and {Joyce}, Richard and {Karcher}, Armin and {Karkar}, Sonia and {Kehoe}, Robert and {Kneib}, Jean-Paul and {Kueter-Young}, Andrea and {Lan}, Ting-Wen and {Lauer}, Tod R. and {Le Guillou}, Laurent and {Le Van Suu}, Auguste and {Lee}, Jae Hyeon and {Lesser}, Michael and {Perreault Levasseur}, Laurence and {Li}, Ting S. and {Mann}, Justin L. and {Marshall}, Robert and {Mart{\'\i}nez-V{\'a}zquez}, C.~E. and {Martini}, Paul and {du Mas des Bourboux}, H{\'e}lion and {McManus}, Sean and {Meier}, Tobias Gabriel and {M{\'e}nard}, Brice and {Metcalfe}, Nigel and {Mu{\~n}oz-Guti{\'e}rrez}, Andrea and {Najita}, Joan and {Napier}, Kevin and {Narayan}, Gautham and {Newman}, Jeffrey A. and {Nie}, Jundan and {Nord}, Brian and {Norman}, Dara J. and {Olsen}, Knut A.~G. and {Paat}, Anthony and {Palanque-Delabrouille}, Nathalie and {Peng}, Xiyan and {Poppett}, Claire L. and {Poremba}, Megan R. and {Prakash}, Abhishek and {Rabinowitz}, David and {Raichoor}, Anand and {Rezaie}, Mehdi and {Robertson}, A.~N. and {Roe}, Natalie A. and {Ross}, Ashley J. and {Ross}, Nicholas P. and {Rudnick}, Gregory and {Safonova}, Sasha and {Saha}, Abhijit and {S{\'a}nchez}, F. Javier and {Savary}, Elodie and {Schweiker}, Heidi and {Scott}, Adam and {Seo}, Hee-Jong and {Shan}, Huanyuan and {Silva}, David R. and {Slepian}, Zachary and {Soto}, Christian and {Sprayberry}, David and {Staten}, Ryan and {Stillman}, Coley M. and {Stupak}, Robert J. and {Summers}, David L. and {Sien Tie}, Suk and {Tirado}, H. and {Vargas-Maga{\~n}a}, Mariana and {Vivas}, A. Katherina and {Wechsler}, Risa H. and {Williams}, Doug and {Yang}, Jinyi and {Yang}, Qian and {Yapici}, Tolga and {Zaritsky}, Dennis and {Zenteno}, A. and {Zhang}, Kai and {Zhang}, Tianmeng and {Zhou}, Rongpu and {Zhou}, Zhimin},
        title = "{Overview of the DESI Legacy Imaging Surveys}",
      journal = {\aj},
         year = 2019,
        month = may,
       volume = {157},
       number = {5},
          eid = {168},
        pages = {168},
          doi = {10.3847/1538-3881/ab089d},
archivePrefix = {arXiv},
       eprint = {1804.08657},
 primaryClass = {astro-ph.IM},
       adsurl = {https://ui.adsabs.harvard.edu/abs/2019AJ....157..168D}
}

@ARTICLE{bennett_92,
       author = {{Bennett}, C.~L. and {Smoot}, G.~F. and {Hinshaw}, G. and {Wright}, E.~L. and {Kogut}, A. and {de Amici}, G. and {Meyer}, S.~S. and {Weiss}, R. and {Wilkinson}, D.~T. and {Gulkis}, S. and {Janssen}, M. and {Boggess}, N.~W. and {Cheng}, E.~S. and {Hauser}, M.~G. and {Kelsall}, T. and {Mather}, J.~C. and {Moseley}, Jr., S.~H. and {Murdock}, T.~L. and {Silverberg}, R.~F.},
        title = "{Preliminary Separation of Galactic and Cosmic Microwave Emission for the COBE Differential Microwave Radiometers}",
      journal = {\apjl},
         year = 1992,
        month = sep,
       volume = {396},
        pages = {L7},
          doi = {10.1086/186505},
       adsurl = {https://ui.adsabs.harvard.edu/abs/1992ApJ...396L...7B}
}

@ARTICLE{2009A&A...493..835D,
       author = {{Delabrouille}, J. and {Cardoso}, J.-F. and {Le Jeune}, M. and {Betoule}, M. and {Fay}, G. and {Guilloux}, F.},
        title = "{A full sky, low foreground, high resolution CMB map from WMAP}",
      journal = {\aap},
         year = 2009,
        month = jan,
       volume = {493},
       number = {3},
        pages = {835-857},
          doi = {10.1051/0004-6361:200810514},
archivePrefix = {arXiv},
       eprint = {0807.0773},
 primaryClass = {astro-ph},
       adsurl = {https://ui.adsabs.harvard.edu/abs/2009A&A...493..835D}
}

@article{Sheth-Torman,
   title={Large-scale bias and the peak background split},
   volume={308},
   ISSN={1365-2966},
   url={http://dx.doi.org/10.1046/j.1365-8711.1999.02692.x},
   DOI={10.1046/j.1365-8711.1999.02692.x},
   number={1},
   journal={Monthly Notices of the Royal Astronomical Society},
   publisher={Oxford University Press (OUP)},
   author={Sheth, Ravi K. and Tormen, Giuseppe},
   year={1999},
   month=sep, pages={119–126} }

@ARTICLE{2002MNRAS.329...61S,
       author = {{Sheth}, Ravi K. and {Tormen}, Giuseppe},
        title = "{An excursion set model of hierarchical clustering: ellipsoidal collapse and the moving barrier}",
      journal = {\mnras},
         year = 2002,
        month = jan,
       volume = {329},
       number = {1},
        pages = {61-75},
          doi = {10.1046/j.1365-8711.2002.04950.x},
archivePrefix = {arXiv},
       eprint = {astro-ph/0105113},
 primaryClass = {astro-ph},
       adsurl = {https://ui.adsabs.harvard.edu/abs/2002MNRAS.329...61S}
}

@article{SimonsObservatory:2018koc,
    author = "Ade, Peter and others",
    collaboration = "Simons Observatory",
    title = "{The Simons Observatory: Science goals and forecasts}",
    eprint = "1808.07445",
    archivePrefix = "arXiv",
    primaryClass = "astro-ph.CO",
    doi = "10.1088/1475-7516/2019/02/056",
    journal = "JCAP",
    volume = "02",
    pages = "056",
    year = "2019"
}

@article{Shang:2011mh,
    author = "Shang, Cien and Haiman, Zoltan and Knox, Lloyd and Oh, S. Peng",
    title = "{Improved Models for Cosmic Infrared Background Anisotropies: New Constraints on the IR Galaxy Population}",
    eprint = "1109.1522",
    archivePrefix = "arXiv",
    primaryClass = "astro-ph.CO",
    doi = "10.1111/j.1365-2966.2012.20510.x",
    journal = "Mon. Not. Roy. Astron. Soc.",
    volume = "421",
    pages = "2832",
    year = "2012"
}

@article{Sunyaev:1980nv,
    author = "Sunyaev, R. A. and Zeldovich, Ya. B.",
    title = "{The Velocity of clusters of galaxies relative to the microwave background. The Possibility of its measurement}",
    journal = "Mon. Not. Roy. Astron. Soc.",
    volume = "190",
    pages = "413--420",
    year = "1980"
}

@article{Contreras:2022zdz,
    author = "Contreras, Dagoberto and McCarthy, Fiona and Johnson, Matthew C.",
    title = "{Maximum likelihood kinetic Sunyaev-Zel{\textquoteright}dovich velocity reconstruction}",
    eprint = "2205.15779",
    archivePrefix = "arXiv",
    primaryClass = "astro-ph.CO",
    doi = "10.1103/PhysRevD.107.023521",
    journal = "Phys. Rev. D",
    volume = "107",
    number = "2",
    pages = "023521",
    year = "2023"
}

@article{Fixsen:1996nj,
    author = "Fixsen, D. J. and Cheng, E. S. and Gales, J. M. and Mather, John C. and Shafer, R. A. and Wright, E. L.",
    title = "{The Cosmic Microwave Background spectrum from the full COBE FIRAS data set}",
    eprint = "astro-ph/9605054",
    archivePrefix = "arXiv",
    doi = "10.1086/178173",
    journal = "Astrophys. J.",
    volume = "473",
    pages = "576",
    year = "1996"
}

@article{Mather:1998gm,
    author = "Mather, John C. and Fixsen, D. J. and Shafer, R. A. and Mosier, C. and Wilkinson, D. T.",
    title = "{Calibrator design for the COBE far infrared absolute spectrophotometer (FIRAS)}",
    eprint = "astro-ph/9810373",
    archivePrefix = "arXiv",
    doi = "10.1086/306805",
    journal = "Astrophys. J.",
    volume = "512",
    pages = "511--520",
    year = "1999"
}

@article{Fixsen:2009ug,
    author = "Fixsen, D. J.",
    title = "{The Temperature of the Cosmic Microwave Background}",
    eprint = "0911.1955",
    archivePrefix = "arXiv",
    primaryClass = "astro-ph.CO",
    doi = "10.1088/0004-637X/707/2/916",
    journal = "Astrophys. J.",
    volume = "707",
    pages = "916--920",
    year = "2009"
}

@article{Lewis:1999bs,
    author = "Lewis, Antony and Challinor, Anthony and Lasenby, Anthony",
    title = "{Efficient computation of CMB anisotropies in closed FRW models}",
    eprint = "astro-ph/9911177",
    archivePrefix = "arXiv",
    doi = "10.1086/309179",
    journal = "Astrophys. J.",
    volume = "538",
    pages = "473--476",
    year = "2000"
}

@ARTICLE{2002PhR...372....1C,
       author = {{Cooray}, Asantha and {Sheth}, Ravi},
        title = "{Halo models of large scale structure}",
      journal = {\physrep},
         year = 2002,
        month = dec,
       volume = {372},
       number = {1},
        pages = {1-129},
          doi = {10.1016/S0370-1573(02)00276-4},
archivePrefix = {arXiv},
       eprint = {astro-ph/0206508},
 primaryClass = {astro-ph},
       adsurl = {https://ui.adsabs.harvard.edu/abs/2002PhR...372....1C}
}

@article{PhysRevD.101.103522,
  title = {Amending the halo model to satisfy cosmological conservation laws},
  author = {Chen, Alice Y. and Afshordi, Niayesh},
  journal = {Phys. Rev. D},
  volume = {101},
  issue = {10},
  pages = {103522},
  numpages = {9},
  year = {2020},
  month = {May},
  publisher = {American Physical Society},
  doi = {10.1103/PhysRevD.101.103522},
  url = {https://link.aps.org/doi/10.1103/PhysRevD.101.103522}
}

@ARTICLE{2015MNRAS.454.1958M,
       author = {{Mead}, A.~J. and {Peacock}, J.~A. and {Heymans}, C. and {Joudaki}, S. and {Heavens}, A.~F.},
        title = "{An accurate halo model for fitting non-linear cosmological power spectra and baryonic feedback models}",
      journal = {\mnras},
         year = 2015,
        month = dec,
       volume = {454},
       number = {2},
        pages = {1958-1975},
          doi = {10.1093/mnras/stv2036},
archivePrefix = {arXiv},
       eprint = {1505.07833},
 primaryClass = {astro-ph.CO},
       adsurl = {https://ui.adsabs.harvard.edu/abs/2015MNRAS.454.1958M}
}

@article{DES:2021sgf,
    author = "Pandey, S. and others",
    collaboration = "DES, ACT",
    title = "{Cross-correlation of Dark Energy Survey Year 3 lensing data with ACT and Planck thermal Sunyaev-Zel{\textquoteright}dovich effect observations. II. Modeling and constraints on halo pressure profiles}",
    eprint = "2108.01601",
    archivePrefix = "arXiv",
    primaryClass = "astro-ph.CO",
    reportNumber = "DES-2021-0636, FERMILAB-PUB-21-331-AE",
    doi = "10.1103/PhysRevD.105.123526",
    journal = "Phys. Rev. D",
    volume = "105",
    number = "12",
    pages = "123526",
    year = "2022"
}

@ARTICLE{1996ApJ...462..563N,
       author = {{Navarro}, Julio F. and {Frenk}, Carlos S. and {White}, Simon D.~M.},
        title = "{The Structure of Cold Dark Matter Halos}",
      journal = {\apj},
         year = 1996,
        month = may,
       volume = {462},
        pages = {563},
          doi = {10.1086/177173},
archivePrefix = {arXiv},
       eprint = {astro-ph/9508025},
 primaryClass = {astro-ph},
       adsurl = {https://ui.adsabs.harvard.edu/abs/1996ApJ...462..563N}
}

@ARTICLE{2014MNRAS.439..123L,
       author = {{Lacasa}, F. and {P{\'e}nin}, A. and {Aghanim}, N.},
        title = "{Non-Gaussianity of the cosmic infrared background anisotropies - I. Diagrammatic formalism and application to the angular bispectrum}",
      journal = {\mnras},
         year = 2014,
        month = mar,
       volume = {439},
       number = {1},
        pages = {123-142},
          doi = {10.1093/mnras/stt2373},
archivePrefix = {arXiv},
       eprint = {1312.1251},
 primaryClass = {astro-ph.CO},
       adsurl = {https://ui.adsabs.harvard.edu/abs/2014MNRAS.439..123L}
}

@ARTICLE{2005ApJ...633..791Z,
       author = {{Zheng}, Zheng and {Berlind}, Andreas A. and {Weinberg}, David H. and {Benson}, Andrew J. and {Baugh}, Carlton M. and {Cole}, Shaun and {Dav{\'e}}, Romeel and {Frenk}, Carlos S. and {Katz}, Neal and {Lacey}, Cedric G.},
        title = "{Theoretical Models of the Halo Occupation Distribution: Separating Central and Satellite Galaxies}",
      journal = {\apj},
         year = 2005,
        month = nov,
       volume = {633},
       number = {2},
        pages = {791-809},
          doi = {10.1086/466510},
archivePrefix = {arXiv},
       eprint = {astro-ph/0408564},
 primaryClass = {astro-ph},
       adsurl = {https://ui.adsabs.harvard.edu/abs/2005ApJ...633..791Z}
}

@ARTICLE{2004ApJ...609...35K,
       author = {{Kravtsov}, Andrey V. and {Berlind}, Andreas A. and {Wechsler}, Risa H. and {Klypin}, Anatoly A. and {Gottl{\"o}ber}, Stefan and {Allgood}, Brandon and {Primack}, Joel R.},
        title = "{The Dark Side of the Halo Occupation Distribution}",
      journal = {\apj},
         year = 2004,
        month = jul,
       volume = {609},
       number = {1},
        pages = {35-49},
          doi = {10.1086/420959},
archivePrefix = {arXiv},
       eprint = {astro-ph/0308519},
 primaryClass = {astro-ph},
       adsurl = {https://ui.adsabs.harvard.edu/abs/2004ApJ...609...35K}
}

@ARTICLE{1998ApJ...495...80B,
       author = {{Bryan}, Greg L. and {Norman}, Michael L.},
        title = "{Statistical Properties of X-Ray Clusters: Analytic and Numerical Comparisons}",
      journal = {\apj},
         year = 1998,
        month = mar,
       volume = {495},
       number = {1},
        pages = {80-99},
          doi = {10.1086/305262},
archivePrefix = {arXiv},
       eprint = {astro-ph/9710107},
 primaryClass = {astro-ph},
       adsurl = {https://ui.adsabs.harvard.edu/abs/1998ApJ...495...80B}
}

@ARTICLE{2020JCAP...05..010F,
       author = {{Fang}, Xiao and {Krause}, Elisabeth and {Eifler}, Tim and {MacCrann}, Niall},
        title = "{Beyond Limber: efficient computation of angular power spectra for galaxy clustering and weak lensing}",
      journal = {\jcap},
         year = 2020,
        month = may,
       volume = {2020},
       number = {5},
          eid = {010},
        pages = {010},
          doi = {10.1088/1475-7516/2020/05/010},
archivePrefix = {arXiv},
       eprint = {1911.11947},
 primaryClass = {astro-ph.CO},
       adsurl = {https://ui.adsabs.harvard.edu/abs/2020JCAP...05..010F}
}

@ARTICLE{2019PhRvD.100h3522P,
       author = {{Pan}, Zhen and {Johnson}, Matthew C.},
        title = "{Forecasted constraints on modified gravity from Sunyaev-Zel'dovich tomography}",
      journal = {\prd},
         year = 2019,
        month = oct,
       volume = {100},
       number = {8},
          eid = {083522},
        pages = {083522},
          doi = {10.1103/PhysRevD.100.083522},
archivePrefix = {arXiv},
       eprint = {1906.04208},
 primaryClass = {astro-ph.CO},
       adsurl = {https://ui.adsabs.harvard.edu/abs/2019PhRvD.100h3522P}
}

@ARTICLE{2020PhRvD.102d3520M,
       author = {{McCarthy}, Fiona and {Johnson}, Matthew C.},
        title = "{Remote dipole field reconstruction with dusty galaxies}",
      journal = {\prd},
         year = 2020,
        month = aug,
       volume = {102},
       number = {4},
          eid = {043520},
        pages = {043520},
          doi = {10.1103/PhysRevD.102.043520},
archivePrefix = {arXiv},
       eprint = {1907.06678},
 primaryClass = {astro-ph.CO},
       adsurl = {https://ui.adsabs.harvard.edu/abs/2020PhRvD.102d3520M}
}

@ARTICLE{2003PhRvD..67h3002O,
       author = {{Okamoto}, Takemi and {Hu}, Wayne},
        title = "{Cosmic microwave background lensing reconstruction on the full sky}",
      journal = {\prd},
         year = 2003,
        month = apr,
       volume = {67},
       number = {8},
          eid = {083002},
        pages = {083002},
          doi = {10.1103/PhysRevD.67.083002},
archivePrefix = {arXiv},
       eprint = {astro-ph/0301031},
 primaryClass = {astro-ph},
       adsurl = {https://ui.adsabs.harvard.edu/abs/2003PhRvD..67h3002O}
}

@ARTICLE{2021PhRvD.103j3515M,
       author = {{McCarthy}, Fiona and {Madhavacheril}, Mathew S.},
        title = "{Improving models of the cosmic infrared background using CMB lensing mass maps}",
      journal = {\prd},
         year = 2021,
        month = may,
       volume = {103},
       number = {10},
          eid = {103515},
        pages = {103515},
          doi = {10.1103/PhysRevD.103.103515},
archivePrefix = {arXiv},
       eprint = {2010.16405},
 primaryClass = {astro-ph.CO},
       adsurl = {https://ui.adsabs.harvard.edu/abs/2021PhRvD.103j3515M}
}

@ARTICLE{2025arXiv250411794L,
       author = {{Liu}, R. Henry and {Hadzhiyska}, Boryana and {Ferraro}, Simone and {Bose}, Sownak and {Hern{\'a}ndez-Aguayo}, C{\'e}sar},
        title = "{Fast Baryonic Field Painting for Sunyaev-Zel'dovich Analyses: Transfer Function vs. Hybrid Effective Field Theory}",
      journal = {arXiv e-prints},
         year = 2025,
        month = apr,
          eid = {arXiv:2504.11794},
        pages = {arXiv:2504.11794},
          doi = {10.48550/arXiv.2504.11794},
archivePrefix = {arXiv},
       eprint = {2504.11794},
 primaryClass = {astro-ph.CO},
       adsurl = {https://ui.adsabs.harvard.edu/abs/2025arXiv250411794L}
}

@ARTICLE{2020JCAP...12..047A,
       author = {{Aiola}, Simone and {Calabrese}, Erminia and {Maurin}, Lo{\"\i}c and {Naess}, Sigurd and {Schmitt}, Benjamin L. and {Abitbol}, Maximilian H. and {Addison}, Graeme E. and {Ade}, Peter A.~R. and {Alonso}, David and {Amiri}, Mandana and {Amodeo}, Stefania and {Angile}, Elio and {Austermann}, Jason E. and {Baildon}, Taylor and {Battaglia}, Nick and {Beall}, James A. and {Bean}, Rachel and {Becker}, Daniel T. and {Bond}, J. Richard and {Bruno}, Sarah Marie and {Calafut}, Victoria and {Campusano}, Luis E. and {Carrero}, Felipe and {Chesmore}, Grace E. and {Cho}, Hsiao-mei and {Choi}, Steve K. and {Clark}, Susan E. and {Cothard}, Nicholas F. and {Crichton}, Devin and {Crowley}, Kevin T. and {Darwish}, Omar and {Datta}, Rahul and {Denison}, Edward V. and {Devlin}, Mark J. and {Duell}, Cody J. and {Duff}, Shannon M. and {Duivenvoorden}, Adriaan J. and {Dunkley}, Jo and {D{\"u}nner}, Rolando and {Essinger-Hileman}, Thomas and {Fankhanel}, Max and {Ferraro}, Simone and {Fox}, Anna E. and {Fuzia}, Brittany and {Gallardo}, Patricio A. and {Gluscevic}, Vera and {Golec}, Joseph E. and {Grace}, Emily and {Gralla}, Megan and {Guan}, Yilun and {Hall}, Kirsten and {Halpern}, Mark and {Han}, Dongwon and {Hargrave}, Peter and {Hasselfield}, Matthew and {Helton}, Jakob M. and {Henderson}, Shawn and {Hensley}, Brandon and {Hill}, J. Colin and {Hilton}, Gene C. and {Hilton}, Matt and {Hincks}, Adam D. and {Hlo{\v{z}}ek}, Ren{\'e}e and {Ho}, Shuay-Pwu Patty and {Hubmayr}, Johannes and {Huffenberger}, Kevin M. and {Hughes}, John P. and {Infante}, Leopoldo and {Irwin}, Kent and {Jackson}, Rebecca and {Klein}, Jeff and {Knowles}, Kenda and {Koopman}, Brian and {Kosowsky}, Arthur and {Lakey}, Vincent and {Li}, Dale and {Li}, Yaqiong and {Li}, Zack and {Lokken}, Martine and {Louis}, Thibaut and {Lungu}, Marius and {MacInnis}, Amanda and {Madhavacheril}, Mathew and {Maldonado}, Felipe and {Mallaby-Kay}, Maya and {Marsden}, Danica and {McMahon}, Jeff and {Menanteau}, Felipe and {Moodley}, Kavilan and {Morton}, Tim and {Namikawa}, Toshiya and {Nati}, Federico and {Newburgh}, Laura and {Nibarger}, John P. and {Nicola}, Andrina and {Niemack}, Michael D. and {Nolta}, Michael R. and {Orlowski-Sherer}, John and {Page}, Lyman A. and {Pappas}, Christine G. and {Partridge}, Bruce and {Phakathi}, Phumlani and {Pisano}, Giampaolo and {Prince}, Heather and {Puddu}, Roberto and {Qu}, Frank J. and {Rivera}, Jesus and {Robertson}, Naomi and {Rojas}, Felipe and {Salatino}, Maria and {Schaan}, Emmanuel and {Schillaci}, Alessandro and {Sehgal}, Neelima and {Sherwin}, Blake D. and {Sierra}, Carlos and {Sievers}, Jon and {Sifon}, Cristobal and {Sikhosana}, Precious and {Simon}, Sara and {Spergel}, David N. and {Staggs}, Suzanne T. and {Stevens}, Jason and {Storer}, Emilie and {Sunder}, Dhaneshwar D. and {Switzer}, Eric R. and {Thorne}, Ben and {Thornton}, Robert and {Trac}, Hy and {Treu}, Jesse and {Tucker}, Carole and {Vale}, Leila R. and {Van Engelen}, Alexander and {Van Lanen}, Jeff and {Vavagiakis}, Eve M. and {Wagoner}, Kasey and {Wang}, Yuhan and {Ward}, Jonathan T. and {Wollack}, Edward J. and {Xu}, Zhilei and {Zago}, Fernando and {Zhu}, Ningfeng},
        title = "{The Atacama Cosmology Telescope: DR4 maps and cosmological parameters}",
      journal = {\jcap},
         year = 2020,
        month = dec,
       volume = {2020},
       number = {12},
          eid = {047},
        pages = {047},
          doi = {10.1088/1475-7516/2020/12/047},
archivePrefix = {arXiv},
       eprint = {2007.07288},
 primaryClass = {astro-ph.CO},
       adsurl = {https://ui.adsabs.harvard.edu/abs/2020JCAP...12..047A}
}

@ARTICLE{2021ApJS..253....3H,
       author = {{Hilton}, M. and {Sif{\'o}n}, C. and {Naess}, S. and {Madhavacheril}, M. and {Oguri}, M. and {Rozo}, E. and {Rykoff}, E. and {Abbott}, T.~M.~C. and {Adhikari}, S. and {Aguena}, M. and {Aiola}, S. and {Allam}, S. and {Amodeo}, S. and {Amon}, A. and {Annis}, J. and {Ansarinejad}, B. and {Aros-Bunster}, C. and {Austermann}, J.~E. and {Avila}, S. and {Bacon}, D. and {Battaglia}, N. and {Beall}, J.~A. and {Becker}, D.~T. and {Bernstein}, G.~M. and {Bertin}, E. and {Bhandarkar}, T. and {Bhargava}, S. and {Bond}, J.~R. and {Brooks}, D. and {Burke}, D.~L. and {Calabrese}, E. and {Carrasco Kind}, M. and {Carretero}, J. and {Choi}, S.~K. and {Choi}, A. and {Conselice}, C. and {da Costa}, L.~N. and {Costanzi}, M. and {Crichton}, D. and {Crowley}, K.~T. and {D{\"u}nner}, R. and {Denison}, E.~V. and {Devlin}, M.~J. and {Dicker}, S.~R. and {Diehl}, H.~T. and {Dietrich}, J.~P. and {Doel}, P. and {Duff}, S.~M. and {Duivenvoorden}, A.~J. and {Dunkley}, J. and {Everett}, S. and {Ferraro}, S. and {Ferrero}, I. and {Fert{\'e}}, A. and {Flaugher}, B. and {Frieman}, J. and {Gallardo}, P.~A. and {Garc{\'\i}a-Bellido}, J. and {Gaztanaga}, E. and {Gerdes}, D.~W. and {Giles}, P. and {Golec}, J.~E. and {Gralla}, M.~B. and {Grandis}, S. and {Gruen}, D. and {Gruendl}, R.~A. and {Gschwend}, J. and {Gutierrez}, G. and {Han}, D. and {Hartley}, W.~G. and {Hasselfield}, M. and {Hill}, J.~C. and {Hilton}, G.~C. and {Hincks}, A.~D. and {Hinton}, S.~R. and {Ho}, S.-P.~P. and {Honscheid}, K. and {Hoyle}, B. and {Hubmayr}, J. and {Huffenberger}, K.~M. and {Hughes}, J.~P. and {Jaelani}, A.~T. and {Jain}, B. and {James}, D.~J. and {Jeltema}, T. and {Kent}, S. and {Knowles}, K. and {Koopman}, B.~J. and {Kuehn}, K. and {Lahav}, O. and {Lima}, M. and {Lin}, Y.-T. and {Lokken}, M. and {Loubser}, S.~I. and {MacCrann}, N. and {Maia}, M.~A.~G. and {Marriage}, T.~A. and {Martin}, J. and {McMahon}, J. and {Melchior}, P. and {Menanteau}, F. and {Miquel}, R. and {Miyatake}, H. and {Moodley}, K. and {Morgan}, R. and {Mroczkowski}, T. and {Nati}, F. and {Newburgh}, L.~B. and {Niemack}, M.~D. and {Nishizawa}, A.~J. and {Ogando}, R.~L.~C. and {Orlowski-Scherer}, J. and {Page}, L.~A. and {Palmese}, A. and {Partridge}, B. and {Paz-Chinch{\'o}n}, F. and {Phakathi}, P. and {Plazas}, A.~A. and {Robertson}, N.~C. and {Romer}, A.~K. and {Carnero Rosell}, A. and {Salatino}, M. and {Sanchez}, E. and {Schaan}, E. and {Schillaci}, A. and {Sehgal}, N. and {Serrano}, S. and {Shin}, T. and {Simon}, S.~M. and {Smith}, M. and {Soares-Santos}, M. and {Spergel}, D.~N. and {Staggs}, S.~T. and {Storer}, E.~R. and {Suchyta}, E. and {Swanson}, M.~E.~C. and {Tarle}, G. and {Thomas}, D. and {To}, C. and {Trac}, H. and {Ullom}, J.~N. and {Vale}, L.~R. and {Van Lanen}, J. and {Vavagiakis}, E.~M. and {De Vicente}, J. and {Wilkinson}, R.~D. and {Wollack}, E.~J. and {Xu}, Z. and {Zhang}, Y.},
        title = "{The Atacama Cosmology Telescope: A Catalog of >4000 Sunyaev-Zel{\textquoteright}dovich Galaxy Clusters}",
      journal = {\apjs},
         year = 2021,
        month = mar,
       volume = {253},
       number = {1},
          eid = {3},
        pages = {3},
          doi = {10.3847/1538-4365/abd023},
archivePrefix = {arXiv},
       eprint = {2009.11043},
 primaryClass = {astro-ph.CO},
       adsurl = {https://ui.adsabs.harvard.edu/abs/2021ApJS..253....3H}
}

@ARTICLE{2016JCAP...08..058B,
       author = {{Battaglia}, N.},
        title = "{The tau of galaxy clusters}",
      journal = {\jcap},
         year = 2016,
        month = aug,
       volume = {2016},
       number = {8},
          eid = {058},
        pages = {058},
          doi = {10.1088/1475-7516/2016/08/058},
archivePrefix = {arXiv},
       eprint = {1607.02442},
 primaryClass = {astro-ph.CO},
       adsurl = {https://ui.adsabs.harvard.edu/abs/2016JCAP...08..058B}
}

@ARTICLE{2018PhRvD..98f3502C,
       author = {{Cayuso}, Juan I. and {Johnson}, Matthew C. and {Mertens}, James B.},
        title = "{Simulated reconstruction of the remote dipole field using the kinetic Sunyaev Zel'dovich effect}",
      journal = {\prd},
         year = 2018,
        month = sep,
       volume = {98},
       number = {6},
          eid = {063502},
        pages = {063502},
          doi = {10.1103/PhysRevD.98.063502},
archivePrefix = {arXiv},
       eprint = {1806.01290},
 primaryClass = {astro-ph.CO},
       adsurl = {https://ui.adsabs.harvard.edu/abs/2018PhRvD..98f3502C}
}

@ARTICLE{2024PhRvD.109j3533R,
       author = {{Ried Guachalla}, Bernardita and {Schaan}, Emmanuel and {Hadzhiyska}, Boryana and {Ferraro}, Simone},
        title = "{Velocity reconstruction in the era of DESI and Rubin/LSST. I. Exploring spectroscopic, photometric, and hybrid samples}",
      journal = {\prd},
         year = 2024,
        month = may,
       volume = {109},
       number = {10},
          eid = {103533},
        pages = {103533},
          doi = {10.1103/PhysRevD.109.103533},
archivePrefix = {arXiv},
       eprint = {2312.12435},
 primaryClass = {astro-ph.CO},
       adsurl = {https://ui.adsabs.harvard.edu/abs/2024PhRvD.109j3533R}
}

@ARTICLE{2024PhRvD.109j3534H,
       author = {{Hadzhiyska}, Boryana and {Ferraro}, Simone and {Ried Guachalla}, Bernardita and {Schaan}, Emmanuel},
        title = "{Velocity reconstruction in the era of DESI and Rubin/LSST. II. Realistic samples on the light cone}",
      journal = {\prd},
         year = 2024,
        month = may,
       volume = {109},
       number = {10},
          eid = {103534},
        pages = {103534},
          doi = {10.1103/PhysRevD.109.103534},
archivePrefix = {arXiv},
       eprint = {2312.12434},
 primaryClass = {astro-ph.CO},
       adsurl = {https://ui.adsabs.harvard.edu/abs/2024PhRvD.109j3534H}
}

@ARTICLE{2015JCAP...11..007S,
       author = {{Senatore}, Leonardo},
        title = "{Bias in the effective field theory of large scale structures}",
      journal = {\jcap},
         year = 2015,
        month = nov,
       volume = {2015},
       number = {11},
        pages = {007-007},
          doi = {10.1088/1475-7516/2015/11/007},
archivePrefix = {arXiv},
       eprint = {1406.7843},
 primaryClass = {astro-ph.CO},
       adsurl = {https://ui.adsabs.harvard.edu/abs/2015JCAP...11..007S}
}

@ARTICLE{2015JCAP...07..030M,
       author = {{Mirbabayi}, Mehrdad and {Schmidt}, Fabian and {Zaldarriaga}, Matias},
        title = "{Biased tracers and time evolution}",
      journal = {\jcap},
         year = 2015,
        month = jul,
       volume = {2015},
       number = {7},
        pages = {030-030},
          doi = {10.1088/1475-7516/2015/07/030},
archivePrefix = {arXiv},
       eprint = {1412.5169},
 primaryClass = {astro-ph.CO},
       adsurl = {https://ui.adsabs.harvard.edu/abs/2015JCAP...07..030M}
}

@ARTICLE{2021JCAP...05..069V,
       author = {{Voivodic}, Rodrigo and {Barreira}, Alexandre},
        title = "{Responses of Halo Occupation Distributions: a new ingredient in the halo model \& the impact on galaxy bias}",
      journal = {\jcap},
         year = 2021,
        month = may,
       volume = {2021},
       number = {5},
          eid = {069},
        pages = {069},
          doi = {10.1088/1475-7516/2021/05/069},
archivePrefix = {arXiv},
       eprint = {2012.04637},
 primaryClass = {astro-ph.CO},
       adsurl = {https://ui.adsabs.harvard.edu/abs/2021JCAP...05..069V}
}
